\documentclass[journal,twoside,print]{ieeecolor}
\usepackage[table,dvipsnames]{xcolor}
\usepackage{generic}
\usepackage{cite}
\usepackage{amsmath,amssymb,amsfonts}
\usepackage{algorithmic}
\usepackage{graphicx}
\usepackage[hidelinks]{hyperref}
\usepackage{textcomp}
\usepackage{xcolor}
\def\BibTeX{{\rm B\kern-.05em{\sc i\kern-.025em b}\kern-.08em
    T\kern-.1667em\lower.7ex\hbox{E}\kern-.125emX}}
\newcommand{\Reals}{\mathbb{R}}

\newcommand{\OfferingAgent}{A}
\newcommand{\respondingAgent}{B}

\newcommand{\OfferingState}{S_{\OfferingAgent}} %
\newcommand{\respondingState}{S_{\respondingAgent}}
\newcommand{\OfferingUtility}[1]{f^{\OfferingAgent}(#1)}
\newcommand{\respondingUtility}[1]{f^{\respondingAgent}(#1)}
\newcommand{\respondingUtilityWoArguments}{f^{\respondingAgent}}
\newcommand{\OfferingUtilityWoArguments}{f^{\OfferingAgent}}
\newcommand{\Trade}{T}
\newcommand{\TradeItem}[1]{t_{#1}}

\newcommand{\ResponseFunction}[1]{R_{\respondingAgent}(#1)}

\newcommand{\Quadrant}{Q}

\newcommand{\Cone}{C}
\newcommand{\ConeCenter}{\tau}
\newcommand{\OpeningAngle}{\theta}
\newcommand{\AngleIncrease}{\gamma}
\newcommand{\UnacceptedOffers}{k}

\newcommand{\Smoothness}{\beta}
\newcommand{\SamplingDistance}{d}
\newcommand{\GradientThreshold}{\kappa}
\newcommand{\NumItems}{n}

\newcommand{\Circle}{\mathbb{B}}
\newcommand{\CircleCenter}{c}
\newcommand{\CircleRadius}{r}
\newcommand{\polytope}{P}

\newcommand{\BasisSet}{V_{b}}
\newcommand{\HyperplaneA}{a}
\newcommand{\HyperplaneB}{b}
\newcommand{\ClosePoint}{c_{close}}
\newcommand{\FarPoint}{c_{far}}

\newcommand{\UtilityA}{Q}
\newcommand{\Utilityb}{u}

\usepackage{pgfplots}
\pgfplotsset{compat=1.18}
\usepackage{lipsum}

\usepackage{amsmath,amsfonts,amssymb}
\usepackage{amsthm}
\newtheorem{theorem}{Theorem}
\newtheorem{remark}{Remark}
\newtheorem{corollary}{Corollary}[theorem]
\newtheorem{lemma}[theorem]{Lemma}
\usepackage[capitalize]{cleveref}
\usepackage{graphicx}
\DeclareMathOperator*{\argmax}{arg\,max}
\DeclareMathOperator*{\argmin}{arg\,min}
\usepackage{subcaption}
\usepackage{lineno}
\usepackage{tikz}
\usepackage{tikz-3dplot}
\usepackage{amsmath}
\usepackage{float}
\usepackage{tabularx}
\usepackage{lipsum}
\usepackage{nicefrac}
\usepackage{caption}  %
\usepackage[norelsize, linesnumbered, ruled, lined, boxed, noend, commentsnumbered]{algorithm2e}
\usetikzlibrary{shapes}
\usetikzlibrary{shapes.geometric}
\usepackage{tcolorbox}
\tcbuselibrary{listingsutf8}
\newtcolorbox{gptbox}[1][]{%
    colback=white, %
    colframe=black, %
    fonttitle=\bfseries, %
    boxrule=0.5mm, %
    arc=3mm, %
    width=\linewidth, %
    before=\footnotesize, %
    listing only,
    listing options={basicstyle=\ttfamily, breaklines=true, breakindent=0pt},
    title=GPT Prompt, %
    #1
}

\usepackage[T1]{fontenc}
\usepackage{newtxtext}
\usepackage{verbatim}
\usepackage{pifont}

\usepackage{listings}
\definecolor{computer}{RGB}{25,25,110}
\definecolor{user}{RGB}{120,25,25}
\definecolor{system}{RGB}{25,110,25}
\definecolor{computer_internal}{RGB}{209,114,15}
\definecolor{backgroundgray}{RGB}{252,252,252}
\lstdefinelanguage{trade}{
    basicstyle=\fontfamily{ptm}\fontsize{8pt}{9pt}\selectfont,
    showstringspaces=false,
    keywordstyle=\color{blue},
    commentstyle=\color{gray},
    morekeywords={}
}

\begin{document}
\title{Sequential Resource Trading Using Comparison-Based Gradient Estimation}
\author{
Surya Murthy\textsuperscript{1,*}%
\thanks{\textsuperscript{1} The University of Texas at Austin.}\thanks{\textsuperscript{*} Corresponding author: surya.murthy@utexas.edu.}, 
Mustafa O. Karabag\textsuperscript{1}, 
and Ufuk Topcu\textsuperscript{1}
}

\maketitle

\begin{abstract}
We study sequential multi-issue trading between two greedily rational agents who exchange resources from a finite set of categories. 
Each agent’s utility depends on its allocation, but the offering agent does not know the responding agent’s utility function and receives only accept/reject feedback.
We propose a comparison-based algorithm that interprets acceptance and rejection responses as pairwise state comparisons, allowing the offering agent to iteratively estimate the responding agent’s gradient. Rejected offers prune the space of feasible gradient directions, enabling systematic refinement of possibly mutually beneficial trades.
The algorithm guarantees that each accepted trade strictly improves both agents’ utilities and, after finitely many rejected offers, either identifies a mutually beneficial trade or certifies that the current allocation is $\epsilon$-weakly Pareto optimal. We further show that the sequence of accepted trades asymptotically converges to the Pareto front under mild assumptions.
We evaluate the method against standard baselines and show that it achieves higher societal benefit with fewer offers across multiple trading settings. We further validate the approach in a user study, demonstrating strong performance in scenarios with substantial resource conflict.
\end{abstract}


\section{Introduction}

Multi-agent systems require autonomous agents to coordinate their decisions while respecting individual preferences and objectives \cite{shoham2008multiagent, wooldridge2009introduction}. 
A fundamental form of coordination arises in resource trading, where agents exchange items across multiple categories to reach mutually beneficial allocations. 
In principle, identifying such trades is straightforward when each agent’s utility function is known. However, in realistic settings, agents typically cannot access or share full utility information due to privacy or computational constraints \cite{hadfield2016cooperative, dragan2013policy, blum2004preference}. 
Instead, they receive only limited feedback, such as acceptance or rejection of proposed trades. 
This restricted feedback makes inferring the other agent’s preferences the central challenge, since mutually beneficial trades depend on identifying directions that improve both utilities without explicit preference access. 
To address this challenge, we introduce Sequential Trading with Cone Refinements (ST-CR), a comparison-based algorithm that leverages binary acceptance and rejection responses to estimate the responding agent’s utility gradient and identify Pareto-improving trades.

In detail, we consider a sequential multi-issue trading problem where two greedily rational agents allocate resources from a finite set of categories between themselves.
Each agent has a continuous state defined by the current amount of resources it possesses in each category. Each agent has a utility function that depends on the agent’s state and is not necessarily separable with respect to the resource categories.
We consider a sequential trading setting, where agents repeatedly exchange a limited amount of resources with each trade.
The offering agent presents a trade offer to improve its utility, and the responding agent accepts offers that improve its utility. 
Upon acceptance, the agents trade resources and continue trading from a new state.

ST-CR treats the acceptance and rejection responses of the responding agent as comparisons between the current state and the post-trade state.
We utilize binary comparisons to estimate the gradient direction of the responding agent’s utility function and, thereby, find an accepted offer.
Building on \cite{karabag2021smooth}, we model the space of potential gradients as a cone and treat rejected offers as halfspace constraints to refine the space. If enough offers get rejected, ST-CR creates a new cone to compactly represent the potential gradients. 
While generating offers, ST-CR chooses offers that are orthogonal to the direction of the cone and previously rejected offers to remove half of the potential gradient directions after every rejected offer and efficiently elicit the responding agent’s utility function gradient.
A key challenge in this procedure is to ensure the inclusion of the gradient in the newly created cone as the halfspace constraints may be incorrect due to first-order gradient approximation. ST-CR addresses this challenge by adequately expanding the cones proposed in \cite{karabag2021smooth}.

We show that, for smooth utility functions, after a finite number of consecutively rejected offers generated by ST-CR, the responding agent is at a near-optimal point or the preferences of the agents are closely aligned; that is, the gradients of the agents’ utility functions are approximately in the same direction. In both cases, the resulting state is \(\epsilon\)\nobreakdash-weakly Pareto optimal under additional assumptions of concavity and Lipschitz continuity. We further show that, when utilities are strongly convex and the accepted trades maintain a bounded angle with at least one agent’s gradient, ST-CR converges to a weak Pareto-optimal point under a diminishing trade-magnitude schedule as the number of accepted offers grows.

We test the proposed ST-CR algorithm in resource allocation scenarios against random trading algorithms and the greedy concession algorithm \cite{baarslag2015optimal}. 
ST-CR achieves a higher societal benefit with fewer offers compared to the baselines. 
We show that ST-CR outperforms the baselines across different numbers of resource categories and different levels of utility function alignment between agents.
We also demonstrate that heuristics utilizing information from previously accepted trades further improve ST-CR’s performance.

To assess the benefits and limitations of ST-CR in bargaining settings with humans, we conduct a user study in which human participants interact with ST-CR and baseline algorithms. We observe that ST-CR produces more accepted trades with fewer offers and achieves higher societal benefit in scenarios where agent goals are highly aligned—i.e., when competition over resources is strongest—highlighting its effectiveness in structured human-agent negotiation.

We make the following contributions. (i) We propose ST-CR, a comparison-based approach to finding mutually beneficial trades for the two-agent, sequential multi-issue trading problem. ST-CR uses acceptance and rejection responses from the responding agent. This algorithm extends a comparison-based optimization approach \cite{karabag2021smooth} to two-agent trading scenarios while ensuring mutual benefit. (ii) We show that, after ST-CR makes a finite number of consecutively rejected offers, either the responding agent is near-optimal or the agents’ preferences are closely aligned, resulting in an \(\epsilon\)\nobreakdash-weakly Pareto optimal state. Additionally, we show that, for a sequence of accepted trades, under mild assumptions, ST-CR will asymptotically converge to a Pareto-optimal point. (iii) We empirically evaluate ST-CR in both numerical experiments and a user study, demonstrating that it efficiently elicits preferences and achieves high societal benefit with few offers, particularly in scenarios with strong preference alignment where existing greedy approaches perform poorly.

\subsection{Related Works}

The sequential trading setting we consider lies at the intersection of automated negotiation, comparison-based optimization, and multi-objective optimization. It combines the challenge of eliciting preferences through limited feedback with the goal of identifying Pareto-optimal outcomes across multiple utility functions.

\textbf{Automated Negotiation.}
Automated negotiation studies algorithms that reach agreements through iterative offer exchange \cite{Faratin_1}. We consider a multi-issue bilateral bargaining problem in which agents trade resources across multiple categories, with utilities defined over these allocations. This can be viewed as a variant of the cutting-pie problem \cite{rubinstein1982perfect}, but unlike the classical single-resource model with time-based incentives, we consider multiple categories without explicit deadlines. Agents are assumed to be greedily rational, accepting any strictly utility-improving allocation, similar to hill-climbing agents in mediated search settings \cite{Klein_1}. Our objective is to reach a Pareto-optimal allocation through a sequence of mutually beneficial trades.

Negotiation algorithms employ offering strategies that generate proposals according to the agent’s objective \cite{baarslag2014decoupling}. In mediated settings, a central mechanism proposes allocations to achieve Pareto efficiency or other social objectives \cite{Klein_1, mori2017dependency, chalamish2012automed, fujita2014efficient, fujita2009secure}. In contrast, bilateral offering strategies typically maximize the proposer’s utility using concession-based heuristics or search-based methods \cite{koeman2015fawkes, van2013hardheaded, Faratin_2, zheng2016automated, li2024offer, baarslag2015optimal}. For example, greedy concession algorithms rank discrete offers by expected utility \cite{baarslag2015optimal, mohammad2023optimal}. Although ST-CR operates in a bilateral setting, its objective aligns with mediated approaches: every accepted offer must strictly improve both agents’ utilities, steering the negotiation toward mutual efficiency rather than unilateral gain.

Other key components of negotiation algorithms include opponent modeling and acceptance strategies \cite{baarslag2014decoupling}. Opponent modeling methods estimate the structure of the other agent’s utility from observed offers, including weighted additive models, interdependency graphs, and learning-based approaches \cite{zhang2015bayesian, fujita2010approach, zafari2016popponent}, with refinement techniques such as Bayesian updating, kernel density estimation, and frequency analysis \cite{zhang2015bayesian, coehoorn2004learning, tunali2017rethinking}. Acceptance strategies typically balance utility improvement against remaining time \cite{kawata2020meta, de2022analysis}. In contrast, ST-CR neither reconstructs the opponent’s full utility model nor relies on time-based concessions. Instead, binary acceptance feedback provides local first-order information that guides offer generation while guaranteeing monotonic improvement.

\textbf{Comparison-Based Optimization.} 
Instead of directly accessing an objective function or its gradients, comparison-based optimization algorithms rely on pairwise comparisons \cite{jamieson2012query}. These methods are useful when the objective is unknown or difficult to measure. In our setting, responses to offers induce comparisons between allocations, enabling optimization without explicit function access.

Prior work uses comparisons to estimate gradients or locate optima. Jamieson et al.\ \cite{jamieson2012query} combine comparison-based line search with coordinate descent, while Sign-OPT \cite{cheng2020signopt}, SCOBO \cite{cai2022one}, and Comparison-GDE \cite{zhang2024comparisons} estimate descent directions from pairwise feedback using randomized sampling or cutting-plane updates. Karabag et al.\ \cite{karabag2021smooth} iteratively refine a cone of feasible gradient directions via halfspace cuts derived from comparisons. These approaches focus on single-objective optimization. In contrast, we consider a multi-objective setting in which proposed offers must improve the offering agent’s utility, restricting the comparisons that can be made.

Similar to \cite{karabag2021smooth}, we use comparisons to iteratively shrink a cone of feasible gradient directions. However, Karabag et al.\ \cite{karabag2021smooth} rely on three-point comparisons to obtain valid halfspace cuts. Under greedy rationality, only two-point comparisons are permitted, which may introduce erroneous cuts. We therefore derive a modified cone-angle update rule that provably encloses all feasible gradient directions despite such errors.

\textbf{Multi-Objective Optimization.}
Multi-objective optimization (MOO) studies problems involving multiple competing objectives and typically seeks Pareto-optimal solutions \cite{deb2016multi}. Our negotiation setting can be viewed as a two-objective optimization problem in which each agent’s utility defines an objective. Unlike standard MOO formulations that assume direct access to all objectives, we observe only one utility and obtain information about the other through comparisons.

Common MOO approaches include scalarization methods such as weighted sums and Tchebycheff formulations \cite{Pardalos2017}, evolutionary algorithms \cite{coello2006evolutionary, deb2016multi}, and multi-objective gradient methods \cite{fliege2000steepest, drummond2004projected, desideri2012multiple}. These methods typically assume direct access to objective values or gradients, which is unavailable in our setting. In contrast, ST-CR operates directly on comparison feedback and drives the negotiation toward Pareto-efficient allocations by ensuring that each accepted offer strictly improves both agents’ utilities.

\subsection{Preliminaries and Notation}

The $L^{2}$-norm of a vector \(v\) is \(\|v\|\).
A function $f: \mathbb{R}^n \rightarrow \mathbb{R}$ is \(L\)-Lipschitz continuous if $|f(y) - f(x)| \leq L \lVert x - y \rVert$ for all $x, y \in \mathbb{R}^n$.
A function $f: \mathbb{R}^n \rightarrow \mathbb{R}$ is \(\beta\)-smooth if $|f(y) - f(x) - \langle \nabla f(x), y-x \rangle| \leq \beta \lVert x - y \rVert^2 / 2$ for all $x, y \in \mathbb{R}^n$.
A function $f: \mathbb{R}^n \rightarrow \mathbb{R}$ is concave if $f(y) \leq f(x) + \langle \nabla f(x), y-x \rangle$.
The directional derivative of \(f\) along vector \(v\) is \(\nabla_{v} f(x) = \langle \nabla f(x), v \rangle / \|v\|\).
The angle between $v_1$ and $v_2$ is $\angle(v_1, v_2) = \cos^{-1} \left(\frac{\langle v_1, v_2 \rangle}{\lVert v_1 \rVert \lVert v_2 \rVert}\right)$. 
A right circular cone with direction $\tau$ and semi-vertical angle $\theta$ is \(C(\tau, \theta) = \left\{ x \in \mathbb{R}^n | \angle (x, \tau) \leq \theta \right\}.\)
The floor operator $\lfloor x \rfloor$ denotes the greatest integer less than or equal to $x$.

\section{Sequential Multi-Issue Trading Problem}
We consider a trading setting where an offering agent $\OfferingAgent$ and a responding agent $\respondingAgent$ make sequential trades of resources belonging to $n$ categories (issues). 
Each agent’s state is defined by the non-negative amount of resources they possess in each category. 
The offering and responding agents’ states are $\OfferingState \in \mathbb{R}_{\geq0}^n$ and $\respondingState \in \mathbb{R}_{\geq0}^n$, respectively, where $\mathbb{R}_{\geq0}^n$ is the $n$-dimensional set of non-negative real numbers.
We consider a \textit{continuous} setting, where the agents’ states and trade offers can take on continuous values.
For example, consider a trading problem with coffee and milk as categories. An example agent state is $(x_{1}, x_{2}) =(1.7, 0.3)$ where $x_{1}$ and $x_{2}$ are cups of coffee and milk, respectively, in the agent’s possession.
The offering and responding agents have continuous utility functions $\OfferingUtility{\OfferingState}$ and $\respondingUtility{\respondingState}$, respectively, which reflect their valuations of different states. We assume that these utility functions are continuous and locally differentiable on the interior of the feasible region, which allows us to interpret accept/reject responses to small trades via first-order approximations.
The utility functions are not assumed to be separable across resource categories.
That is, we do not assume that $u(x_1,\dots,x_n) = \sum_{i=1}^n u_i(x_i)$. The marginal value of one resource may depend on the quantities of other resources.
For example, in the coffee–milk setting, a person who prefers a specific ratio may have a utility function
\[
u(x_1,x_2) = -(x_1 - 1.9)^2 - (0.1 x_1 - 1.9x_2)^2 - (x_2 - 0.1)^2,
\]
whose optimal point is $(1.9,0.1)$. $(0.1 x_1 - 1.9x_2)^2$ couples coffee and milk, so the utility is non-separable.

The agents aim to improve their utility by sequentially trading with each other. A trade offer  $\Trade=(\TradeItem{1},\ldots,\TradeItem{n} )$ is the change in the amount of resources in each category, where $\TradeItem{i} \in \mathbb{R}$ is the increase in resource $i$ for the offering agent. An offer is \textit{feasible} if it results in no negative amounts of any resource for either agent after the trade.

When determining whether to propose or accept an offer, the agents consider the benefit associated with the offer. At states $\OfferingState$ and $\respondingState$, the benefits of an offer $\Trade$ are $\OfferingUtility{\OfferingState+\Trade}- \OfferingUtility{\OfferingState}$ and $\respondingUtility{\respondingState - \Trade}-\respondingUtility{\respondingState}$ for the offering and responding agents, respectively. The societal benefit is the sum of the benefits for both agents. An offer is \textit{mutually beneficial} if $\OfferingUtility{\OfferingState+\Trade}- \OfferingUtility{\OfferingState} \geq 0$ and $\respondingUtility{\respondingState - \Trade}-\respondingUtility{\respondingState} \geq 0$. The agents are \textit{greedily rational} and will accept every beneficial offer. Consequently, the offering agent will only offer $\Trade$ if $\OfferingUtility{\OfferingState+\Trade}- \OfferingUtility{\OfferingState} \geq 0$, and the responding agent will only accept $\Trade$ if $\respondingUtility{\respondingState - \Trade}-\respondingUtility{\respondingState} \geq 0$. The responding agent’s response function is 
\begin{equation*}
    \ResponseFunction{\Trade}=
    \begin{cases}
        \text{accept} & \text{if } \respondingUtility{\respondingState - \Trade}-\respondingUtility{\respondingState} \geq 0, \\
        \text{reject} & \text{if } \respondingUtility{\respondingState - \Trade}-\respondingUtility{\respondingState} < 0.
    \end{cases}
\end{equation*}
Depending on the responding agent’s response to the offer $\Trade$, the agents transition to the next states $\OfferingState'$ and $\respondingState'$ where
\begin{equation*}
    \OfferingState', \ \respondingState'=
    \begin{cases}
        \OfferingState+\Trade, \ \respondingState-\Trade & \text{if } \ResponseFunction{\Trade}=\text{accept,} \\
        \OfferingState, \ \respondingState  &   \text{if } \ResponseFunction{\Trade}=\text{reject.}
    \end{cases}
\end{equation*}

\paragraph{Weak Pareto Optimality} A joint state \((\OfferingState^*, \respondingState^*)\) is \emph{weakly Pareto optimal} if there exists no feasible trade \( \Trade \) such that both agents strictly improve. That is, for all feasible \( \Trade\),
\[
f^A(\OfferingState^* + \Trade) - f^A(\OfferingState^*) \leq 0 \quad \text{or}\]
\[ f^B(\respondingState^* - \Trade) - f^B(\respondingState^*) \leq 0.
\]

\paragraph{\(\epsilon\)-Weak Pareto Optimality}
A joint state \((\OfferingState^*, \respondingState^*)\) is \emph{\(\epsilon\)-weakly Pareto optimal} for some \(\epsilon > 0\) if there exists no feasible trade \(\Trade\) that improves both agents by more than \(\epsilon\). That is, for all feasible \(\Trade\),
\[
f^A(\OfferingState^* + \Trade) - f^A(\OfferingState^*) \le \epsilon
\quad \text{or} \quad
f^B(\respondingState^* - \Trade) - f^B(\respondingState^*) \le \epsilon.
\]

\textbf{Problem definition.}
Consider a sequential multi-issue trading problem with two greedily rational agents.
The offering agent knows the initial states $\OfferingState$ and $\respondingState$, as well as its own utility function $\OfferingUtility{\OfferingState}$, but does not know the responding agent’s utility function $\respondingUtility{\respondingState}$ and cannot observe its utility values.
The goal is to reach a weak Pareto-optimal allocation of resources through sequential trades.
At a weak Pareto-optimal state, no further feasible and mutually beneficial trade exists, ensuring that the allocation cannot be improved for one agent without reducing the utility of the other.

\begin{figure}[t]
    \centering
    
    \begin{subfigure}{0.49\columnwidth}
        \centering
          \begin{tikzpicture}
    \begin{axis}[
      width=0.90\linewidth,
      view={-55}{25},
      xmin=-0.6, xmax=1.1,
      ymin=-1.1, ymax=1.1,
      zmin=0,   zmax=1.0,
      xticklabels=\empty,
      yticklabels=\empty,
      zticklabels=\empty,
      scale only axis,
    ]
      \addplot3[
        surf,
        domain=0:0.9,
        y domain=0:360,
        samples=25,
        samples y=45,
        shader=interp,
        draw=orange!60!brown,
        colormap={cone}{color=(orange!85!brown) color=(orange!30!black)},
        opacity=0.9,
      ]
      ({(0.5/0.9)*x*cos(y)}, {(0.5/0.9)*x*sin(y)}, {x});

      \addplot3[->,very thick,black] coordinates {(0,0,0) (0.0,0.0,1.0)};

      \node[anchor=south east] at (axis cs:0.0,0.0,1.0) {$\tau$};
    \end{axis}
  \end{tikzpicture}
        \caption{Initial gradient cone.}
        \label{fig:refinement_1}
    \end{subfigure}\hfill
    \begin{subfigure}{0.49\columnwidth}
        \centering
        \begin{tikzpicture}
  \begin{axis}[
    width=0.90\linewidth,
    view={-55}{25},
    xmin=-0.6, xmax=1.1,
    ymin=-1.1, ymax=1.1,
    zmin=0,   zmax=1.0,
    xticklabels=\empty,
    yticklabels=\empty,
    zticklabels=\empty,
    scale only axis,
  ]
    \addplot3[
      surf,
      domain=0:0.9,
      y domain=90:270,
      samples=25,
      samples y=25,
      shader=interp,
      draw=orange!60!brown,
      colormap={cone}{color=(orange!85!brown) color=(orange!30!black)},
      opacity=0.9,
    ]
    ({-(0.5/0.9)*x*cos(y)}, {(0.5/0.9)*x*sin(y)}, {x});
    
    \addplot3[
      surf,
      domain=-1:1,   
      y domain=0:0.9,
      samples=2,
      samples y=2,
      shader=flat,
      draw=orange!60!brown,
      fill=orange!70!black,
      opacity=0.9,
    ]
    ({0},
     {(0.5/0.9)*y*x},
     {y}); 
    
    \addplot3[
      surf,
      domain=-1:1,
      y domain=0:1,
      samples=2,
      samples y=2,
      shader=flat,
      draw=blue!40,
      fill=blue!40,
      opacity=0.18,
    ]
    ({0}, {x}, {y});

    \addplot3[->,very thick,black] coordinates {(0,0,0) (1,0,0)};
    \addplot3[->,very thick,black] coordinates {(0,0,0) (0.0,0.0,1.0)};

    \node[anchor= south west] at (axis cs:1,0,0) {$T_1$};
    \node[anchor=south east] at (axis cs:0.0,0.0,1.0) {$\tau$};
  \end{axis}
\end{tikzpicture}
        \caption{Rejected offers cut the cone.}
        \label{fig:refinement_2}
    \end{subfigure}

    \vspace{0.5em}

    \begin{subfigure}{0.49\columnwidth}
        \centering
          \begin{tikzpicture}
    \begin{axis}[
      width=0.90\linewidth,
      view={-55}{25},
      xmin=-0.6, xmax=1.1,
      ymin=-1.1, ymax=1.1,
      zmin=0,   zmax=1.0,
      xticklabels=\empty,
      yticklabels=\empty,
      zticklabels=\empty,
      scale only axis,
    ]
      \addplot3[
        surf,
        domain=0:0.9,
        y domain=270:360,
        samples=25,
        samples y=25,
        shader=interp,
        draw=orange!60!brown,
        colormap={cone}{color=(orange!85!brown) color=(orange!30!black)},
        opacity=0.9,
      ]
      ({(0.5/0.9)*x*cos(y)}, {(0.5/0.9)*x*sin(y)}, {x});

      \addplot3[
        surf,
        domain=0:0.9,
        y domain=0:1,
        samples=2,
        samples y=2,
        shader=flat,
        draw=orange!60!brown,
        fill=orange!70!black,
        opacity=0.9,
      ]
      ({(0.5/0.9)*x*y}, {0}, {x});

      \addplot3[
        surf,
        domain=0:0.9,
        y domain=0:1,
        samples=2,
        samples y=2,
        shader=flat,
        draw=orange!60!brown,
        fill=orange!70!black,
        opacity=0.9,
      ]
      ({0}, {-(0.5/0.9)*x*y}, {x});

      \addplot3[
        surf,
        domain=-1:1,
        y domain=0:1,
        samples=2,
        samples y=2,
        shader=flat,
        draw=blue!40,
        fill=blue!40,
        opacity=0.18,
      ]
      ({0}, {x}, {y});

      \addplot3[
        surf,
        domain=-0.6:1.1,
        y domain=0:1,
        samples=2,
        samples y=2,
        shader=flat,
        draw=blue!40,
        fill=blue!40,
        opacity=0.18,
      ]
      ({x}, {0}, {y});

      \addplot3[->,very thick,black] coordinates {(0,0,0) (1,0,0)};
      \addplot3[->,very thick,black] coordinates {(0,0,0) (0.0,-1,0.0)};
      \addplot3[->,very thick,black] coordinates {(0,0,0) (0.0,0.0,1.0)};

      \node[anchor=south west] at (axis cs:1,0,0) {$T_1$};
      \node[anchor=south west] at (axis cs:-0.1,-0.95,0.0) {$T_2$};
      \node[anchor=south east] at (axis cs:0.0,0.0,1.0) {$\tau$};
    \end{axis}
  \end{tikzpicture}
        \caption{Rejected offers cut the cone.}
        \label{fig:refinement_3}
    \end{subfigure}
    \begin{subfigure}{0.49\columnwidth}
        \centering
          \begin{tikzpicture}
    \begin{axis}[
      width=0.90\linewidth,
      view={-55}{25},
      xmin=-0.6, xmax=1.1,
      ymin=-1.1, ymax=1.1,
      zmin=0,   zmax=1.0,
      xticklabels=\empty,
      yticklabels=\empty,
      zticklabels=\empty,
      scale only axis,
    ]
      \pgfmathsetmacro{\betaprime}{18} %

      \addplot3[
        surf,
        domain=0:0.9,
        y domain=270:360,
        samples=25,
        samples y=25,
        shader=interp,
        draw=orange!60!brown,
        colormap={cone}{color=(orange!85!brown) color=(orange!30!black)},
        opacity=0.9,
      ]
      ({(0.5/0.9)*x*cos(y)}, {(0.5/0.9)*x*sin(y)}, {x});

      \addplot3[
        surf,
        domain=0:0.9,
        y domain=0:1,
        samples=2,
        samples y=2,
        shader=flat,
        draw=orange!60!brown,
        fill=orange!70!black,
        opacity=0.9,
      ]
      ({(0.5/0.9)*x*y}, {0}, {x});

      \addplot3[
        surf,
        domain=0:0.9,
        y domain=0:1,
        samples=2,
        samples y=2,
        shader=flat,
        draw=orange!60!brown,
        fill=orange!70!black,
        opacity=0.9,
      ]
      ({0}, {-(0.5/0.9)*x*y}, {x});
      
    \addplot3[
      surf,
      domain=0:1.05,
      y domain=0:360,
      samples=25,
      samples y=45,
      shader=interp,
      colormap={solidblue}{color=(blue!80) color=(blue!80)},
      point meta=0,
      opacity=0.4,
    ]
    ({0.183920196*x + (x*tan(\betaprime))*(cos(y)*0.982957178)},
     {-0.183920196*x + (x*tan(\betaprime))*(cos(y)*0.034417795 + sin(y)*0.982337858)},
     {0.965581029*x + (x*tan(\betaprime))*(cos(y)*(-0.180584058) + sin(y)*0.187115804)});
      \addplot3[->,very thick,black] coordinates {(0,0,0) (1,0,0)};
      \addplot3[->,very thick,black] coordinates {(0,0,0) (0.0,-1,0.0)};
      \addplot3[->,very thick,black] coordinates {(0,0,0) (0.0,0.0,1.0)};

      \node[anchor=south west] at (axis cs:1,0,0) {$T_1$};
      \node[anchor=south west] at (axis cs:-0.1,-0.95,0.0) {$T_2$};
      \node[anchor=south east] at (axis cs:0.0,0.0,1.0) {$\tau$};

      \addplot3[->,very thick,black] coordinates {(0,0,0) (0.20,-0.20,1.05)};
      \node[anchor= west] at (axis cs:0.20,-0.20,1.10) {$\tau'$};
    \end{axis}
  \end{tikzpicture} %
        \caption{New cone encloses gradients.}
        \label{fig:refinement_4}
    \end{subfigure}

    \caption{Illustration of ST-CR cone refinement for three dimensions.}
    \label{fig:refinement}
\end{figure}

\section{Sequential Trading with Cone Refinement}
We propose Sequential Trading with Cone Refinements (ST-CR), an algorithm to find a weak Pareto-optimal point through sequential trades. 
Mutually beneficial trades are offers lying in the intersection of the ascent directions of agents’ utility functions. 
To find these directions, we infer the potential gradients of the responding agent with a cone based on rejected offers.
By the greedy rationality assumption, each rejected offer provides information on the responding agent’s gradient, resulting in a more accurate estimate.

\textbf{Preliminaries on Gradient Direction Estimation.}
Given the sign of the directional derivatives in curated directions, one can approximate the direction of the responding agent’s gradient to find an ascent direction. Building on this observation,
\cite{karabag2021smooth} model the set of potential gradients as a right circular cone. The signs of directional derivatives inferred from comparisons in orthogonal directions are then used as halfspace constraints to refine the cone.
The cone model allows the remaining potential gradients after orthogonal halfspace refinements to be easily enclosed by a new cone.
Similarly, ST-CR maintains a cone $\Cone(\ConeCenter, \OpeningAngle)$ of gradients with direction $\ConeCenter \in \mathbb{R}^{n}$ and semi-vertical angle $\OpeningAngle$. 

To refine the cone of potential gradients, ST-CR utilizes the signs of the directional derivatives. 
By making a linear approximation of each agent’s utility function at the states $\OfferingState$ and $\respondingState$, we use a first-order Taylor approximation to describe beneficial trades in terms of each agent’s utility gradient:
\begin{equation*}
\begin{split}
\OfferingUtility{\OfferingState+\Trade}-\OfferingUtility{\OfferingState} \approx \langle \Trade, \nabla \OfferingUtility{\OfferingState}\rangle\geq0, \\
\respondingUtility{\respondingState-\Trade}-\respondingUtility{\respondingState} \approx \langle- \Trade, \nabla \respondingUtility{\respondingState}\rangle \geq0.
\end{split}
\end{equation*}
By the greedy rationality assumption, the responding agent will reject an offer if $\langle -\Trade, \nabla \respondingUtility{\respondingState}\rangle < 0$, implying that $\nabla_{\Trade} f^{B}(\respondingState) > 0$.  
Therefore, the agent’s responses provide information on the sign of directional derivatives, which ST-CR uses to refine the potential gradient cone.

We detail the procedure in \cref{fig:refinement}. As depicted in \cref{fig:refinement_1}, ST-CR represents all possible gradient directions as a cone. Each rejected trade offer induces a halfspace constraint as discussed above. These halfspace constraints cut the cone and remove infeasible gradient directions as shown in \cref{fig:refinement_2}. After $n-1$ such cuts, the remaining feasible directions, i.e., the intersection of the cone with these halfspaces, is enclosed by a new cone with a smaller opening angle, shown in \cref{fig:refinement_3}. We note that this approach relies on the match between the halfspaces and the signs of the actual directional derivatives, which may not always hold.

Karabag et al.\ \cite{karabag2021smooth} accurately determine the signs of directional derivatives or identify inconclusive directions by making comparisons at three points on a line.
In our setting, making three-point comparisons corresponds to making two opposite offers.
This guarantees that at least one of the offers will not benefit the offering agent, violating the greedy rationality assumption for the offering agent.
Hence, we use two-point comparisons to estimate signs of directional derivatives. 
Two-point comparisons can lead to sign errors, consequently affecting the cone refinement.
As we later show in \cref{thm}, we account for such errors by increasing the cone’s semi-vertical angle to enclose all potential gradients. 

We now overview ST-CR’s two-stage approach. 
The first stage (Algorithm \ref{algo:trade_progression} Lines \ref{algo:stage_1_start}--\ref{algo:stage_1_end}), uses heuristic offers to quickly improve societal benefit. 
The second stage (Algorithm \ref{algo:trade_progression}, Line \ref{algo:cr} Procedure \ref{algo:CR}) uses rejected offers to refine the potential gradient cone. 
Once an offer is accepted, ST-CR transitions to the next state (Algorithm \ref{algo:trade_progression} Lines \ref{algo:transition_1} and \ref{algo:transition_2}).

\textbf{Stage 1: Heuristic Offers (Algorithm 
\ref{algo:trade_progression}, Lines \ref{algo:stage_1_start}--\ref{algo:stage_1_end}).}
When making consecutive trades, information from previous negotiations can inform new offers.
For example, ST-CR begins each new negotiation by proposing the previously accepted offer.
If the responding agent has a smooth utility function, its gradient does not change significantly after accepting an offer, making previously accepted offers likely to be accepted again.
We later show in numerical experiments that this heuristic improves ST-CR’s performance.
As per the greedy rationality assumption, this offer is only presented if it is beneficial to the offering agent.
If the responding agent rejects the offer presented in this stage, we move to stage~2.

\begin{algorithm}[htbp]
    \SetAlgoNlRelativeSize{0}
    \SetAlgoLined
    \setcounter{AlgoLine}{0}
    \KwIn{$\OfferingState \in \mathbb{R}^{n}, \respondingState \in \mathbb{R}^{n}, f^{\OfferingAgent},
    \theta^{*}$ (cone angle threshold),
    $\SamplingDistance$ (offer magnitude),
    $m$ (total offer budget)}
    $q \leftarrow 0$ \;
    
    \While{$q < m$}{
        Offer $\Trade_{\text{prev}}$, $q\leftarrow q + 1$\; \label{algo:stage_1_start}
        
        \If{$\ResponseFunction{\Trade_{\text{prev}}} = \text{accept}$} 
        { %
            $\OfferingState \leftarrow \OfferingState + \Trade_{\text{prev}}$, $\respondingState \leftarrow \respondingState - \Trade_{\text{prev}}$ \; \label{algo:transition_1}
            
            continue \; \label{algo:stage_1_end}
        } 
        \tcp{Find trade using cone refinement}
        $\Trade, q_{\Trade} \leftarrow \text{ CR }(\OfferingState, \respondingState, f^{\OfferingAgent}, \theta^{*}, \SamplingDistance, m - q)$ \; \label{algo:cr} %
        
        $q \leftarrow q + q_{\Trade}$, $\Trade_{\text{prev}} \leftarrow \Trade$ \;
        
        \If{$\Trade \neq -1$}{ %
            $\OfferingState \leftarrow \OfferingState + \Trade$, $\respondingState \leftarrow \respondingState - \Trade$ \;
            \label{algo:transition_2}
            
            continue \; \label{algo:transition}
        }
        break \; %
    }
    \caption{Sequential Trading with Cone Refinement (ST-CR)}
    \label{algo:trade_progression}
    \end{algorithm}
    \renewcommand{\algorithmcfname}{Procedure}
    \begin{algorithm}[htbp]
      \SetAlgoNlRelativeSize{0}
      \SetAlgoLined
      \setcounter{AlgoLine}{13}
      \SetKwRepeat{Do}{do}{while}
    \KwIn{$\OfferingState \in \mathbb{R}^{n}, \respondingState \in \mathbb{R}^{n}, f^{\OfferingAgent},
    \theta^{*}$ (cone angle threshold),
    $\SamplingDistance$ (offer magnitude),
    $m$ (remaining offer budget)}
      \KwOut{$\Trade, q_{\Trade}$}
      
    \tcp{Determine Init. Quadrant}
     
      $q_{\Trade} \leftarrow 0$, $\Quadrant \leftarrow [0, 0, \ldots, 0]$ of size $n$\; \label{algo:stage_2_1_start}

      \For{$i \leftarrow 1$ \KwTo $\NumItems$}{
    
        Set $\Trade \in \mathbb{R}^n \text{ s.t. $\Trade$ is feasible, } \Trade[i] \neq 0, \forall j \neq i,$ $\Trade[j] = 0,  \lVert\Trade\rVert \leq d \text{ and } \langle \nabla \OfferingUtility{\OfferingState}, \Trade \rangle \geq 0$ \; \label{algo:offering_agent_constraint}
        
        Offer $\Trade$, $q_{\Trade} \leftarrow q_{\Trade} + 1$  \; 
        
        \If{$\ResponseFunction{\Trade} = \text{accept}$}
        {return $\Trade, q_{\Trade}$ \; \label{algo:return_quadrant}}
        
        \If{$\ResponseFunction{\Trade} = \text{reject}$}{$\Quadrant \leftarrow \Quadrant + \Trade$ \label{algo:quadrant_refinement}}
        \If{$q_{\Trade} \geq m$}{
            return $-1, q_{\Trade}$ \;
        } 
      }
    
      $\ConeCenter \leftarrow \Quadrant, \OpeningAngle \leftarrow \frac{\pi}{2}, V \leftarrow \{\ConeCenter\}$\; \label{algo:stage_2_1_end}
    
      \Do{$\ResponseFunction{\Trade} \neq \text{accept} \text{ and } \OpeningAngle \geq \theta^{*} \text{ and } q_{\Trade} < m$}{
      
        \tcp{Update cone after \(n-1\) offers} 
        \If{$|V| = n$ }{ \label{algo:stage_2_3_start}
          Set $w_n \leftarrow \ConeCenter / \|\ConeCenter\|$ and for all $i \in \{1, \ldots, n-1\}$, set $w_i \leftarrow \ConeCenter \cos{(\OpeningAngle)}/ \|\ConeCenter\| + \Trade_i \sin{(\OpeningAngle)}/ \|\Trade_{i}\|$\; \label{algo:cone_center_update_1}
          
          $\ConeCenter \leftarrow \sum_{i=1}^{\NumItems} \left(w_i\right) / \left\lVert \sum_{i=1}^{\NumItems} \left(w_i\right) \right\rVert$\;\label{algo:cone_center_update_2}
          
          $\OpeningAngle \leftarrow \sin^{-1}\left(\sin(\OpeningAngle)\sqrt{1 - \nicefrac{1}{2n}}\right)$, $V \leftarrow \{\ConeCenter\}$ \; \label{algo:stage_2_3_end}
        }
        \tcp{Generate Next Trade} \label{algo:stage_2_2_start}
        Set $\Trade$ s.t. $\Trade$ is feasible, $\forall \Trade' \in V$, $\Trade \perp \Trade'$, $ \lVert \Trade \rVert = \SamplingDistance$, and $\langle \nabla \OfferingUtility{\OfferingState}, \Trade \rangle \geq 0$ \; \label{algo:offering_agent_constraint_2}
        
        Offer $\Trade$, $q_{\Trade} \leftarrow q_{\Trade} + 1$, and $V \leftarrow V \cup \{\Trade\}$  \; \label{algo:stage_2_2_end}
      }
    
      \If{$\ResponseFunction{\Trade} \neq \text{accept}$}{return $-1, q_{\Trade}$ \;}
      return $\Trade, q_{\Trade}$ \;
      
      \caption{Cone Refinement (CR)}
      \label{algo:CR}
    \end{algorithm}
    \renewcommand{\algorithmcfname}{Algorithm}

\textbf{Stage 2: Gradient Refinement (Procedure \ref{algo:CR}).}
We use rejected offers to refine the space of potential gradients and identify a mutually beneficial offer. 
If no such offer is found after refinement, the gradients of both agents are closely aligned, indicating a near Pareto-optimal point. 
To refine the space of potential gradients, we represent the potential gradients as an $n$-dimensional cone, $\Cone(\ConeCenter,\OpeningAngle)$.

\textbf{Stage 2.1: Initializing the Cone of Potential Gradients (Procedure \ref{algo:CR} Lines \ref{algo:stage_2_1_start}--\ref{algo:stage_2_1_end}).}
When negotiation begins, the offering agent does not know the responding agent’s utility function \(\respondingUtilityWoArguments\) and, consequently, has no information on $\nabla \respondingUtility{\respondingState}$.
We first determine the $n$-dimensional quadrant of $\nabla \respondingUtility{\respondingState}$ by probing each coordinate direction once.
For each resource category $i \in \{1,\dots,n\}$, ST-CR proposes a trade of the form
\[
\Trade_i = (0,\ldots,0,\pm d,0,\ldots,0),
\]
where the nonzero component is in coordinate $i$ and the sign is chosen so that
$\langle \Trade_i, \nabla \OfferingUtility{\OfferingState} \rangle \geq 0$.
Thus, exactly $n$ offers are made in this stage.
To accurately refine the cone using Taylor approximation and avoid overshooting along mutually beneficial trades, we limit the magnitude of offers by $d$.
If the responding agent accepts any of the offers, ST-CR proceeds to the next state and repeats from stage 1 (Line \ref{algo:return_quadrant}).
If the responding agent rejects all of the offers, the direction of the quadrant is inferred as $\Quadrant[i] = \Trade_i[i]$ (Line \ref{algo:quadrant_refinement}).
The quadrant is enclosed by a cone with direction $\ConeCenter = \Quadrant$ and semi-vertical angle $\OpeningAngle = \pi/2$ (Line \ref{algo:stage_2_1_end}). 
Instead of performing this stage, one can also expand the cone from the previous trade to account for gradient changes and, as a heuristic, use the expanded cone for initialization.

In Procedure~2, the initial set of $n$ orthogonal offers is chosen along the coordinate axes for simplicity and interpretability.
However, we note that the theoretical guarantees of ST-CR presented in Section~\ref{section:theory} do not depend on this specific choice: any orthonormal basis yields the same cone-refinement properties, since each rejected offer induces a valid halfspace constraint on the responding agent’s gradient.
One natural alternative would be to align the first direction with the offering agent’s gradient $\nabla f^A(S_A)$ and choose the remaining $n-1$ directions to be orthogonal to it.
However, this greedy choice can be counterproductive in sequential settings.
If the direction aligned with $\nabla f^A(S_A)$ is rejected, then all remaining orthogonal directions necessarily yield little immediate benefit to the offering agent.
Because stage~2 enforces orthogonality between offers, starting with a maximally beneficial direction can therefore reduce flexibility and slow progress.
The coordinate-axis choice is a naturally interpretable choice that also preserves theoretical guarantees.

\textbf{Stage 2.2: Determining Mutually Beneficial Offers (Procedure \ref{algo:CR} Lines \ref{algo:stage_2_2_start}--\ref{algo:stage_2_2_end}).} Given $\NumItems$ categories and the cone direction $\ConeCenter$, we generate $\NumItems-1$ offers $T_1, \ldots, T_{\NumItems-1}$ that are orthogonal to each other and $\ConeCenter$ (Line \ref{algo:stage_2_2_end}). By the greedy rationality assumption, each offer satisfies $\langle \Trade, \nabla f^{A}(S_{A}) \rangle \geq 0$.
We remark that this set of offers can be sorted in order of decreasing utility for the offering agent to ensure that the offering agent is maximizing its own utility.
If the responding agent accepts an offer, a state transition occurs (Algorithm \ref{algo:trade_progression} Line \ref{algo:transition}), and ST-CR repeats from stage 1. 
After $\NumItems-1$ consecutively rejected offers in stage 2.2, the space of potential gradients reduces to the intersection of the gradient cone and halfspaces generated by the rejected offers.
In stage 2.3, we enclose this space with a smaller cone.

\textbf{Stage 2.3: Updating the Cone of Potential Gradients (Procedure \ref{algo:CR} Lines \ref{algo:stage_2_3_start}--\ref{algo:stage_2_3_end}).}
If the responding agent rejects $\NumItems-1$ offers, we refine the cone using the rejected offers. 
Using the first-order approximation and the greedy rationality assumption, each rejected offer 
$\Trade \in \lbrace T_1, \ldots, T_{\NumItems-1} \rbrace = (V \setminus {\tau})$ implies 
$\langle -\Trade, \nabla \respondingUtility{\respondingState} \rangle < 0$, which is equivalent to 
$\langle \Trade, \nabla \respondingUtility{\respondingState} \rangle > 0$. 
Thus, each rejected offer induces a halfspace constraint 
$\{x \mid \langle \Trade, x \rangle \ge 0\}$, meaning that the true gradient direction must lie on the same side of the hyperplane orthogonal to $\Trade$ as $\Trade$ itself.

We detail this geometric refinement in \cref{fig:refinement}. As depicted in \cref{fig:refinement_1}, ST-CR represents all possible gradient directions as a cone $\Cone(\ConeCenter,\OpeningAngle)$. Each rejected offer induces a halfspace constraint as described above. These halfspaces intersect the cone and remove infeasible gradient directions, effectively slicing the cone by hyperplanes through the origin, as illustrated in \cref{fig:refinement_2}. After $n-1$ such cuts, the remaining feasible directions are given by $\bigcap_{i=1}^{n-1} \{x \mid \langle \Trade_i, x\rangle \geq 0\} \cap \Cone(\tau, \theta)$. This intersection corresponds to the set of gradient directions consistent with all observed rejections. As shown in \cref{fig:refinement_3}, this new region can be enclosed by a new cone with a smaller opening angle.

To maintain the cone representation, ST-CR computes a new enclosing cone $\Cone(\ConeCenter', \OpeningAngle')$ containing the above intersection. The updated cone direction is determined using the prior cone direction together with the extreme boundary directions induced by the halfspace constraints (Procedure \ref{algo:CR}, Lines \ref{algo:cone_center_update_1}--\ref{algo:cone_center_update_2}). Specifically, $\ConeCenter' = {\sum_{i=1}^{\NumItems} w_i}/{\left\lVert \sum_{i=1}^{\NumItems} w_i \right\rVert}$, where $w_n = \ConeCenter/\|\ConeCenter\|$ and for all $i \in \{1, \ldots, n-1\}$, $w_i = \ConeCenter \cos(\OpeningAngle)/\|\ConeCenter\| + \Trade_i \sin(\OpeningAngle)/\|\Trade_i\|$. Each vector $w_i$ represents a boundary direction obtained by rotating the original cone boundary toward the normal direction of the corresponding constraint. The normalized sum therefore provides a central direction that lies within and approximately centers the refined feasible region, yielding the updated cone $\Cone(\ConeCenter', \OpeningAngle')$.

We set a new angle $\OpeningAngle' =  \sin^{-1}\left(\sin(\OpeningAngle)\sqrt{1 - \nicefrac{1}{2n}}\right)$. 
When we explore the theoretical guarantees of ST-CR, we show that this angle suffices to enclose $\cap_{i=1}^{n-1} \{x \mid \langle \Trade, x\rangle \geq 0\} \cap \Cone(\tau, \theta)$.

After updating the cone, ST-CR repeats stages 2.2--2.3 until an offer is accepted, the total number of offers exceeds the budget $m$, or $\OpeningAngle$ is less than a threshold $\OpeningAngle^{*}$.

\begin{remark}[Greedy Rationality Assumption]
Motivated by the first-order Taylor approximation, ST-CR selects offer directions that align with the offering agent's gradient direction, ensuring that \( \langle \Trade, \nabla f^{A}(S_{A})\rangle \geq 0 \). 
For non-linear utility functions, this method can lead to non-beneficial trades for the offering agent if \(\lVert\nabla_{\Trade} f^{A}(\OfferingState)\rVert\) is small, overshooting the optimal trade in the direction of \(\Trade\). 
However, for a \( \Smoothness \)-smooth \(\OfferingUtilityWoArguments\), the potential loss of benefit is bounded such that $\OfferingUtility{\OfferingState} - \OfferingUtility{\OfferingState + \Trade} \leq \nicefrac{\beta d^2}{2}.$
To prevent these losses, the offering agent can avoid overshooting by decreasing the offer magnitude.
Alternatively, one can eliminate such trade directions. If the trade \(\Trade\) is not beneficial despite \( \langle \Trade, \nabla \OfferingUtility{\OfferingState} \rangle \geq 0 \), the offering agent is near-optimal along the line of \(\Trade\). Consequently, we can exclude trades with a component on this line and continue trading within the null space of the line.
\end{remark}

\begin{figure}[t]
\centering

\begin{subfigure}[t]{0.49\linewidth}%
    \centering
    \resizebox{\linewidth}{!}{\begin{tikzpicture}

\definecolor{darkgray176}{RGB}{176,176,176}
\definecolor{green}{RGB}{0,128,0}
\definecolor{purple}{RGB}{128,0,128}

\begin{axis}[
tick align=outside,
tick pos=left,
xmin=-4, xmax=4,
ymin=-4, ymax=4,
width=\linewidth,
height=0.9\linewidth,
axis lines=box,
axis line style={draw=black, very thick},
ticks=none,
xticklabels=\empty,
yticklabels=\empty,
clip=false
]

\path [draw=blue, fill=blue, line width=1.1pt]
(axis cs:2.20124611797498,1.10062305898749)
--(axis cs:2.0335410196625,0.932917960675006)
--(axis cs:2.00022360679775,0.9995527864045)
--(axis cs:0.000223606797749909,-0.00044721359550004)
--(axis cs:-0.000223606797749909,0.000447213595499929)
--(axis cs:1.99977639320225,1.0004472135955)
--(axis cs:1.9664589803375,1.06708203932499)
--cycle;

\path [draw=red, fill=red, line width=1.1pt]
(axis cs:1.10062305898749,2.20124611797498)
--(axis cs:1.06708203932499,1.9664589803375)
--(axis cs:1.0004472135955,1.99977639320225)
--(axis cs:0.000447213595499929,-0.000223606797749909)
--(axis cs:-0.00044721359550004,0.000223606797749909)
--(axis cs:0.9995527864045,2.00022360679775)
--(axis cs:0.932917960675006,2.0335410196625)
--cycle;

\path [draw=green, fill=green, line width=1.1pt]
(axis cs:1.16099689437998,0.580498447189992)
--(axis cs:1.02683281573,0.446334368540005)
--(axis cs:1.00022360679775,0.4995527864045)
--(axis cs:0.00022360679775002,-0.000447213595499929)
--(axis cs:-0.000223606797749909,0.000447213595499985)
--(axis cs:0.99977639320225,0.5004472135955)
--(axis cs:0.973167184270002,0.553665631459995)
--cycle;

\path [draw=purple, fill=purple, line width=1.1pt]
(axis cs:0.917879220613579,-0.917879220613579)
--(axis cs:0.748173593128807,-0.833026406871193)
--(axis cs:0.790246446609407,-0.790953553390593)
--(axis cs:-0.000353553390593286,-0.000353553390593286)
--(axis cs:0.000353553390593286,0.000353553390593286)
--(axis cs:0.790953553390593,-0.790246446609407)
--(axis cs:0.833026406871193,-0.748173593128807)
--cycle;

\path [draw=blue, fill=blue, opacity=0.15]
(axis cs:2,-4)
--(axis cs:4,-4)
--(axis cs:4,4)
--(axis cs:-2,4)
--cycle;

\path [draw=red, fill=red, opacity=0.15]
(axis cs:4,-2)
--(axis cs:4,4)
--(axis cs:-4,4)
--(axis cs:-4,2)
--cycle;

\addplot [
    very thick,
    gray,
    dashed,
    opacity=0.4
]
coordinates {(-4,0) (4,0)};

\addplot [
    very thick,
    gray,
    dashed,
    opacity=0.4
]
coordinates {(0,-4) (0,4)};

\draw (axis cs:2.15,1.20) node[
  scale=0.7,
  anchor=base west,
  text=blue
]{$\nabla f^A$};

\draw (axis cs:1.20,2.35) node[
  scale=0.7,
  anchor=base west,
  text=red
]{$-\nabla f^B$};

\draw (axis cs:1.25,0.35) node[
  scale=0.7,
  anchor=north west,
  text=green
]{$T_{\text{greedy}}$};

\draw (axis cs:1.25,-0.85) node[
  scale=0.7,
  anchor=north west,
  text=purple
]{$T_{\text{ST-CR}}$};

\end{axis}
\end{tikzpicture}}
    \caption{Unaligned gradients. The overlap between the halfspaces induced by $\nabla f^A$ and $-\nabla f^B$ is large, making greedy offers highly mutually beneficial.}
    \label{fig:unaligned}
\end{subfigure}\hfill%
\begin{subfigure}[t]{0.49\linewidth}%
    \centering
    \resizebox{\linewidth}{!}{\begin{tikzpicture}

\definecolor{darkgray176}{RGB}{176,176,176}
\definecolor{green}{RGB}{0,128,0}
\definecolor{purple}{RGB}{128,0,128}

\begin{axis}[
tick align=outside,
tick pos=left,
xmin=-4, xmax=4,
ymin=-4, ymax=4,
width=\linewidth,
height=0.9\linewidth,
axis lines=box,
axis line style={draw=black, very thick},
ticks=none,
xticklabels=\empty,
yticklabels=\empty,
clip=false
]

\path [draw=blue, fill=blue, line width=1.1pt]
(axis cs:2.20124611797498,1.10062305898749)
--(axis cs:2.0335410196625,0.932917960675006)
--(axis cs:2.00022360679775,0.9995527864045)
--(axis cs:0.000223606797749909,-0.00044721359550004)
--(axis cs:-0.000223606797749909,0.000447213595499929)
--(axis cs:1.99977639320225,1.0004472135955)
--(axis cs:1.9664589803375,1.06708203932499)
--cycle;

\path [draw=red, fill=red, line width=1.1pt]
(axis cs:-1.10062305898749,-2.20124611797498)
--(axis cs:-1.06708203932499,-1.9664589803375)
--(axis cs:-1.0004472135955,-1.99977639320225)
--(axis cs:-0.000447213595499929,0.000223606797749909)
--(axis cs:0.00044721359550004,-0.000223606797749909)
--(axis cs:-0.9995527864045,-2.00022360679775)
--(axis cs:-0.932917960675006,-2.0335410196625)
--cycle;

\path [draw=green, fill=green, line width=1.1pt]
(axis cs:1.16099689437998,0.580498447189992)
--(axis cs:1.02683281573,0.446334368540005)
--(axis cs:1.00022360679775,0.4995527864045)
--(axis cs:0.00022360679775002,-0.000447213595499929)
--(axis cs:-0.000223606797749909,0.000447213595499985)
--(axis cs:0.99977639320225,0.5004472135955)
--(axis cs:0.973167184270002,0.553665631459995)
--cycle;

\path [draw=purple, fill=purple, line width=1.1pt]
(axis cs:0.917879220613579,-0.917879220613579)
--(axis cs:0.748173593128807,-0.833026406871193)
--(axis cs:0.790246446609407,-0.790953553390593)
--(axis cs:-0.000353553390593286,-0.000353553390593286)
--(axis cs:0.000353553390593286,0.000353553390593286)
--(axis cs:0.790953553390593,-0.790246446609407)
--(axis cs:0.833026406871193,-0.748173593128807)
--cycle;

\path [draw=blue, fill=blue, opacity=0.15]
(axis cs:2,-4)
--(axis cs:4,-4)
--(axis cs:4,4)
--(axis cs:-2,4)
--cycle;

\path [draw=red, fill=red, opacity=0.15]
(axis cs:-4,-4)
--(axis cs:4,-4)
--(axis cs:4,-2)
--(axis cs:-4,2)
--cycle;

\addplot [
    very thick,
    gray,
    dashed,
    opacity=0.4
]
coordinates {(-4,0) (4,0)};

\addplot [
    very thick,
    gray,
    dashed,
    opacity=0.4
]
coordinates {(0,-4) (0,4)};

\draw (axis cs:2.15,1.20) node[
  scale=0.7,
  anchor=base west,
  text=blue
]{$\nabla f^A$};

\draw (axis cs:-1.25,-2.45) node[
  scale=0.7,
  anchor=base west,
  text=red
]{$-\nabla f^B$};

\draw (axis cs:1.25,0.35) node[
  scale=0.7,
  anchor=north west,
  text=green
]{$T_{\text{greedy}}$};

\draw (axis cs:1.25,-0.85) node[
  scale=0.7,
  anchor=north west,
  text=purple
]{$T_{\text{ST-CR}}$};

\end{axis}
\end{tikzpicture}}
    \caption{Aligned gradients. The intersection of beneficial directions becomes small and often lies nearly orthogonal to both $\nabla f^A$ and $-\nabla f^B$.}
    \label{fig:aligned}
\end{subfigure}

\caption{Illustration of mutually beneficial regions under different alignment regimes.}
\label{fig:alignment-summary}
\end{figure}

\begin{remark}[Performance in Unaligned Settings]
In settings where the agents’ preferences are highly unaligned (i.e., 
$\nabla f^A$ and $-\nabla f^B$ point in similar directions), the set of mutually beneficial directions is large and typically well aligned with the offering agent’s own gradient. In such cases, greedy algorithms naturally perform well, since moving in the direction of $\nabla f^A$ tends to benefit both agents simultaneously. By contrast, the orthogonal offers generated by ST-CR may lie in directions that yield little utility improvement for either agent, reducing the overall societal benefit relative to simpler greedy approaches. We acknowledge this limitation and emphasize that ST-CR is designed to perform best in aligned settings, where its cone-refinement mechanism is most 
effective at identifying the narrower set of mutually beneficial directions.
Fig.~\ref{fig:alignment-summary} provides a simple geometric illustration of the phenomenon. Fig.~\ref{fig:unaligned} shows that, when the agents’ utility gradients are highly unaligned, the region of mutually beneficial directions is large and overlaps substantially with the offering agent’s gradient, so greedy offers yield significant gains. Fig.~\ref{fig:aligned} shows that, when the gradients are aligned, the mutually beneficial region becomes narrow and may lie nearly orthogonal to both gradients, causing orthogonal offers generated by ST-CR to yield little immediate utility improvement. We demonstrate this phenomenon in our experiments.
\label{rem:unaligned}
\end{remark}

\subsection{Theoretical Guarantees of ST-CR}
\label{section:theory}
In this section, we examine ST-CR under two scenarios. First, we analyze the case in which a certain number of offers are rejected consecutively. Second, we investigate the limit behavior as the number of accepted offers approaches infinity.

We show that, if the responding agent has a smooth utility function, then after a finite number of consecutively rejected offers made by ST-CR, either the agents’ gradients (i.e., preferences) are closely aligned, or the responding agent is at a near-optimal state. As a corollary, the resulting state is \(\epsilon\)\nobreakdash-weakly Pareto optimal for a constant \(\epsilon > 0\) that depends on the smoothness and trade parameters.

Additionally, we show that, if the agents’ utility functions are strongly concave and the accepted trades have a bounded maximum angle with at least one agent’s gradient, then, under a diminishing trade magnitude schedule, ST-CR converges to a weak Pareto-optimal point as the number of accepted offers approaches infinity.

ST-CR uses two-point comparisons to infer the signs of the directional derivatives and subsequently refine the cone of potential gradients. 
As discussed previously, this approach can result in sign errors if the current state is near-optimal along the offer direction. 
When such an error happens, the cone is cut using the wrong side of the hyperplane. The wrong cuts are the main challenge in deriving theoretical guarantees. 

To analyze the effects of wrong cuts, we use the $\beta$-smoothness of $f^{\respondingAgent}$. If a sign error occurs for an offer \(\Trade\) of magnitude $d$, i.e., \( \nabla_{-\Trade} f^{\respondingAgent}({\respondingState})  \geq 0\), inferred as \( \nabla_{-\Trade} f^{\respondingAgent}({\respondingState})  < 0\), the magnitude of the directional derivative satisfies \( \| \nabla_{-\Trade} f^{\respondingAgent}({\respondingState})\|  \leq \nicefrac{\beta d}{ 2}\). We consider two cases. 

\textit{Case 1:} There is no offer direction with a large directional derivative. In this case, all halfspace cuts may happen in the wrong direction. However, the responding agent’s current state is near-optimal since its gradient has a bounded magnitude. For example, the purple gradient \textcolor{purple}{$\nabla f^{B}(\respondingState) ($\ding{117}$)$}  in Fig.~\ref{fig:cone_error} is not enclosed by the halfspace cuts but is bounded\footnote{In Fig.~\ref{fig:cone_error}, the distance of a point from the center depends on the angle between \(\tau\) and the associated gradient of the point and does not depend on the gradient’s magnitude.}.

\definecolor{darkgreen}{RGB}{15,120,15}
\begin{figure}[t]
    \centering
    \begin{tikzpicture}[outer sep=0pt, inner sep=0pt]
    \definecolor{darkgreen}{RGB}{15,120,15}
    \def\radius{3/2}
    \def\radiustwo{2.55/2}
    \def\radiusthree{2.12132034356/2}

    \draw[thick, gray] (0/2, 0/2) circle (\radius);
    \draw[thick, dashed, gray] (1.5/2,1.5/2) circle (\radiustwo);
    \draw[thick, dotted, gray] (1.5/2,1.5/2) circle (\radiusthree);

    \draw[<->, gray!50!white](-4/2,0) -- (4/2,0);
    \draw[<->, gray!50!white](0,-4/2) -- (0,4/2);
    \draw[ thick](-3.5/2,0) -- (3.5/2,0) node[pos=0, above, yshift=2pt] {\large$h(\Trade_1)$};
    \draw[ thick](0,-3.5/2) -- (0,3.5/2) node[pos=0, left, xshift=-2pt] {\large$h(\Trade_2)$};

    \coordinate (A) at (3/2, 0/2);
    \coordinate (B) at (0, 3/2);

    \begin{scope}
        \clip (-\radius,-\radius) rectangle (\radius,\radius);
        \clip (0/2, 0/2) -- (A) -- (B) -- cycle; %
        \fill[gray, opacity=0.2] (0/2, 0/2) circle (\radius); %
    \end{scope}
    \begin{scope}
        \clip (-\radius,-\radius) rectangle (\radius,\radius);
        \clip (A) arc (0:90:100*\radius) -- (2.121, 2.121) arc (0:90:\radius) -- (B)-- cycle; %
        \fill[gray, opacity=0.2] (0/2, 0/2) circle (\radius); %
    \end{scope}
    \node[fill=white, text=darkgreen, below right] at (0.3/2,2.8/2) {$\nabla f_{\text{hyp}}^{B}(\respondingState)$};
    \filldraw[darkgreen] (0.3/2,2.8/2) circle (1pt);

    \node[fill=white, text=darkgreen, below left] at (-0.3/2,2.8/2) {$\nabla f^{B}(\respondingState)$};
    \filldraw[darkgreen] (-0.3/2,2.8/2) circle (1pt);

    \node[fill=white, text=blue, above right] at (0.3/2,0.3/2) {$\nabla f_{\text{hyp}}^{B}(\respondingState)$};
    \node[draw=blue, cross out, minimum size=4pt, inner sep=0pt, line width=0.8pt] at (0.3/2,0.3/2) {};

    \node[fill=white, text=blue, below left] at (-0.2/2,-0.2/2) {$\nabla f^{B}(\respondingState)$};
    \node[draw=blue, cross out, minimum size=4pt, inner sep=0pt, line width=0.8pt] at (-0.2/2,-0.2/2) {};

    \node[fill=white, text=purple, below left] at (-1.5/2,-1.5/2) {$\nabla f^{B}(\respondingState)$};
    \node[draw=purple, diamond, fill=purple, minimum size=4pt, inner sep=0pt] at (-1.5/2,-1.5/2) {};

    \node[below left, fill=white, inner sep=0.2em] at (6, 1.6) (legend) {
        \begin{tikzpicture}
            \node[inner sep=0pt] at (0.2,1.3) {\tikz \draw[thick, gray] (0,0) -- ++(0.75,0);};
            \node[anchor=west] at (0.4,1.3) {{\begin{tabular}{c} Initial cone\end{tabular}}};
            \node[inner sep=0pt] at (0.2,0) {\tikz \draw[thick, dashed, gray] (0,0) -- ++(0.75,0);};
            \node[anchor=west] at (0.4,0) {{\begin{tabular}{c} New cone\\  $\sqrt{1 - \nicefrac{1}{2n}}$ \end{tabular}}};
            \node[inner sep=0pt] at (0.2,-1.3) {\tikz \draw[thick, dotted, gray] (0,0) -- ++(0.75,0);};
            \node[anchor=west] at (0.4,-1.3) {{\begin{tabular}{c} New cone\\  $\sqrt{1 - \nicefrac{1}{n}}$ \end{tabular}}};
        \end{tikzpicture}
    };

\end{tikzpicture}
    \caption{2D top-down view of 3D cone refinement using the plane with normal \(\tau\). The ellipses are the cross-sections of the cones. The points are the cross-sections of the true and hypothetical gradients. \(h(\Trade_{i})\) is the hyperplane generated by offer \(\Trade_{i}\). The shaded region represents all possible gradient directions after the halfspace cuts using \(h(\Trade_{i})\).}
    \label{fig:cone_error}
\end{figure}

\textit{Case 2:} There is an offer direction with a large directional derivative. 
In this case, we consider a ``hypothetical'' gradient that matches the inferred signs. 
It shares the same directional derivatives as the true gradient in the directions where the inferred signs are correct and has the opposite directional derivatives in the other directions.
The hypothetical gradient is enclosed in the cone constructed with the angle update rule $\OpeningAngle \leftarrow \sin^{-1}(\sin(\OpeningAngle)\sqrt{1 - \nicefrac{1}{n}})$~\cite{karabag2021smooth}. We show that, due to the existence of a trade direction with a large directional derivative, the maximum angle between the hypothetical and true gradients is bounded by a constant depending on the trade magnitude and the smoothness constant. For example, the green gradient \textcolor{darkgreen}{$\nabla f^{B}(\respondingState)$ (\ding{108})} in Fig.~\ref{fig:cone_error} has a small directional derivative along direction \(\Trade_{2}\) resulting in a wrong halfspace cut. However, it has a larger directional derivative along direction \(\Trade_{1}\). Due to this component, the angle between the \textcolor{darkgreen}{$\nabla f^{B}(\respondingState)$} and \textcolor{darkgreen}{$\nabla f^{B}_{\text{hyp}}(\respondingState)$} is limited. Similarly, the blue gradient \textcolor{blue}{$\nabla f^{B}(\respondingState)$(\ding{53})} has small directional derivatives along directions \(\Trade_{1}\) and \(\Trade_{2}\), but has a larger directional derivative along direction \(\tau\). Using the closeness of the hypothetical and true gradients, we ensure that the true gradient, e.g., \textcolor{darkgreen}{$\nabla f^{B}(\respondingState)$(\ding{108})} and \textcolor{blue}{$\nabla f^{B}(\respondingState)$(\ding{53})} in Fig.~\ref{fig:cone_error}, is enclosed in the cone constructed with the angle update rule $\OpeningAngle \leftarrow \sin^{-1}(\sin(\OpeningAngle)\sqrt{1 - \nicefrac{1}{2n}})$. Since the offering agent’s directional derivatives share the same signs as the inferred signs for the rejected offers (Procedure \ref{algo:CR} Lines \ref{algo:offering_agent_constraint} and \ref{algo:offering_agent_constraint_2}), we observe that the offering agent’s gradient is also enclosed in the cone \(C(\ConeCenter, \OpeningAngle)\) of potential gradients for the responding agent. Hence, the maximum angle between the agents’ gradients is $2 \OpeningAngle$. 

Theorem \ref{thm} shows that, after a finite number of consecutively rejected offers, the preferences of the agents are closely aligned, or the responding agent’s state is near-optimal. 

\begin{theorem} \label{thm}
Let \(\kappa \geq \sqrt{n-1}\) be a parameter satisfying 
\begin{equation} \sqrt{1 - \frac{1}{2\NumItems}}^{\left\lfloor \frac{\UnacceptedOffers-1}{\NumItems-1} \right\rfloor} =2\NumItems\sqrt{1-\left(\frac{ \GradientThreshold^2 - (n-1)}{\GradientThreshold^2 + (n-1)}\right)^2}.
    \label{eq:condition}
\end{equation}

If $\respondingUtilityWoArguments$ is $\Smoothness$-smooth, at least one of the following holds true after Procedure \ref{algo:CR} makes $\UnacceptedOffers$ rejected offers at state $(\OfferingState, \respondingState)$:

\begin{enumerate}
    \item The angle between the agents’ gradients satisfies 
    \small
    \begin{equation*} \label{eq:angle_bound}
        \angle(\nabla \OfferingUtility{\OfferingState}, \nabla \respondingUtility{\respondingState}) < 2 \sin^{-1}\left(\sqrt{1 - \frac{1}{2\NumItems}}^{\left\lfloor \frac{\UnacceptedOffers-n}{\NumItems-1} \right\rfloor}\right).
    \end{equation*}
    \normalsize
    
    \item The responding agent’s benefit for any trade $\Trade$ satisfies
    \begin{equation*} \label{eq:benefit_bound}
    \respondingUtility{\respondingState + \Trade} - \respondingUtility{\respondingState} \leq \SamplingDistance \GradientThreshold \sqrt{\NumItems} \Smoothness \|\Trade\| + {\Smoothness \|\Trade\|^2}/{2}. 
    \end{equation*}
    If $\respondingUtilityWoArguments$ is concave, the responding agent’s benefit for any trade $\Trade$ further satisfies
    \begin{equation*} \label{eq:benefit_bound_2}
    \respondingUtility{\respondingState + \Trade} - \respondingUtility{\respondingState} \leq \SamplingDistance \GradientThreshold \sqrt{\NumItems} \Smoothness \|\Trade\| . 
        \end{equation*}
\end{enumerate}
\end{theorem}
A \(\kappa\) satisfying \eqref{eq:condition} exists and can be computed with line search as the l.h.s. of \eqref{eq:condition} is in $(0, 1]$, and the r.h.s. is a continuous, monotone decreasing function of \(\kappa\) with values $(0,2n]$.

The offer magnitude \(d\) balances the potential benefit per trade and estimation error. As \(\SamplingDistance\) gets larger, the potential benefit increases while the guarantee in Theorem \ref{thm} gets looser.

\begin{corollary}[Weak Pareto Optimality] \label{cor:epsilon_pareto}
Restate the definitions and conditions of Theorem~\ref{thm}.  Let \( \delta \) be the maximum feasible trade magnitude (bounded by the size of the state space), i.e.,  \(\|\Trade\| \leq \delta \) for all feasible \(\Trade\). 

If (i) $\OfferingUtilityWoArguments$ and $\respondingUtilityWoArguments$ are concave, and (ii) $\OfferingUtilityWoArguments$ or $\respondingUtilityWoArguments$ is $L$-Lipschitz continuous, then the current state $(\OfferingState, \respondingState)$ is \(\epsilon = \max\{\epsilon_{\text{alignment}}, \epsilon_{\text{responding}}\}\)-weakly Pareto optimal where,

\begin{enumerate}
    \item in Case 1 of Theorem \ref{thm}, 
    \[
    \epsilon_{\text{alignment}} = \delta  L  \sin\left(2 \sin^{-1}\left(\sqrt{1 - \frac{1}{2\NumItems}}^{\left\lfloor \frac{\UnacceptedOffers-n}{\NumItems-1} \right\rfloor}\right)\right),
    \]
    \item in Case 2 of Theorem \ref{thm}, 
    \[
    \epsilon_{\text{responding}} = d \kappa \sqrt{n} \beta \delta.
    \]
    
\end{enumerate}

\end{corollary}

In Case~1, the agents’ gradients are closely aligned, meaning any mutually beneficial trade direction is nearly orthogonal to both gradients. This geometric constraint, along with concavity, leads to an \( \epsilon \)-bound on both agents’ utility improvements, which approaches zero as the angle between their gradients decreases. In Case~2, the responding agent is at a near-optimal state, so any feasible trade yields a limited utility gain. Together, these cases imply that no feasible trade can significantly improve both agents’ utilities, establishing \( \epsilon \)-weak Pareto optimality.

We remark that as $\UnacceptedOffers \to \infty$,     \[
    \epsilon \approx 2\delta  L \sqrt{1 - \frac{1}{2\NumItems}}^{\left\lfloor \frac{\UnacceptedOffers-n}{\NumItems-1} \right\rfloor}
    \] since $\sin \theta  \approx \theta$ for small values of $\theta$.

While Theorem~\ref{thm} characterizes the outcome after $k$ sequential rejected offers, it does not address the regime in which ST-CR continues to generate accepted trades indefinitely. In this setting, cone refinement no longer drives the analysis. Instead, the improvements made by accepted offers drive the analysis. Theorem~\ref{thm:asymptotic_convergence} establishes asymptotic guarantees for this case by showing that, under mild assumptions on the utility functions and accepted trades, the state of the agents approaches optimality for at least one of the agents as the number of accepted trades approaches infinity.

To obtain this result, we impose additional assumptions on the utility functions and the alignment of the accepted trades with the utility gradients.
Specifically, we assume a uniform angular alignment condition: there exists $\hat{\theta}\in[0,\pi/2)$ such that each accepted trade $\Trade$ at state $(S_{A, t}, S_{B,t})$, either $\angle(\Trade,\nabla f^A(S_{A,t})) \le \hat{\theta}$ or $\angle(-\Trade,\nabla f^B(S_{B,t})) \le \hat{\theta}$. This alignment condition guarantees that the trade direction has a nonvacuous alignment with at least one agent’s utility gradient.
Additionally, we assume that $f^A$ and $f^B$ are $\mu$-strongly concave and bounded above by $f^{\max}$. Strong concavity, combined with the alignment assumption, ensures that each accepted trade yields an improvement lower bounded by a certain value that depends on the gradient magnitude.
When the trade magnitudes $d_t := \|T_t\|$ are chosen such that they satisfy diminishing conditions, $ \sum_{t=0}^{\infty} d_t = \infty$ and $\sum_{t=0}^{\infty} d_t^2 < \infty$, as is common in the literature \cite{bertsekas2000gradient}, these bounded improvements ensure that either $\|\nabla f^A(S_{A,t})\| \to 0$ or $\|\nabla f^B(S_{A,t})\| \to 0$. Consequently, the state approaches the agent’s optimal point due to concavity, which is a sufficient condition for weak Pareto optimality. The full proof of this result is provided in Appendix Section~\ref{appendix:convergence}

\begin{theorem}[Asymptotic Convergence under Aligned Trades]
\label{thm:asymptotic_convergence}

Let $f^A,f^B:\mathbb{R}^d\to\mathbb{R}$ be $\mu$-strongly concave and $\beta$-smooth.

Assume the following:
\begin{enumerate}
    \item \textbf{Trade updates.}  
    At each accepted trade step $t$, a nonzero trade vector $T_t\in\mathbb{R}^d$ is chosen, with step size
    \[
    d_t := \|T_t\| > 0,
    \]
    and the agents update their states according to
    \[
    S_{A,t+1} = S_{A,t} + T_t,
    \qquad
    S_{B,t+1} = S_{B,t} - T_t.
    \]

    \item \textbf{Angular alignment.}  
    There exists $\hat{\theta}\in[0,\pi/2)$ such that, at each step $t$, the trade direction satisfies at least one of
    \[
    \angle(T_t,\nabla f^A(S_{A,t})) \leq \hat{\theta},
    \qquad\text{or}\qquad
    \angle(-T_t,\nabla f^B(S_{B,t})) \leq \hat{\theta}.
    \]

    \item \textbf{Step-size conditions.}  
    The step sizes satisfy the standard diminishing-steps conditions
    \[
    \sum_{t=0}^\infty d_t = \infty,
    \qquad
    \sum_{t=0}^\infty d_t^2 < \infty.
    \]

    \item \textbf{Bounded utilities.}  
    Each agent’s utility is uniformly bounded above: there exists $f^{\max}<\infty$ such that
    \[
    f^A(S_A) \le f^{\max},
    \qquad
    f^B(S_B) \le f^{\max}
    \quad\text{for all } S_A,S_B.
    \]
\end{enumerate}

\paragraph{Conclusion.}
If the assumptions hold, then there exists at least one agent
$X \in \{A,B\}$ for which the cumulative step sizes of aligned trades diverge:
\[
\sum_{t:\, \angle(T_t,\nabla f^X(S_{X,t})) \le \hat{\theta}} d_t = \infty.
\]
For that agent $X$, the state sequence $\{S_{X,t}\}$ converges to a finite limit, and
\[
\lim_{t\to\infty} \|\nabla f^X(S_{X,t})\| = 0.
\]
\end{theorem}

The asymptotic convergence result relies on the angular alignment condition for accepted trades. We now examine the feasibility of this assumption. Theorem~\ref{thm:angular_or_epsilon} shows that either the angular assumption is satisfied for some offer generated by ST-CR, or the agents are $\epsilon$-weak Pareto-optimal. In the latter case, the accepted trade is near-perpendicular for both agents. Such a situation can only happen when the agents share an aligned gradient direction, which indicates $\epsilon_{\text{alignment}}$-weak Pareto optimality. This case inherently relies on the initial cone being correctly constructed. If this construction is inaccurate, then similar to Theorem~\ref{thm}, we can certify that the responding agent’s gradient has a small magnitude, yielding $\epsilon_{\text{responding}}$-weak Pareto optimality. A full proof of Theorem~\ref{thm:angular_or_epsilon} is provided in Appendix Section~\ref{appendix:angular}

\begin{theorem}[Either angular certification or $\epsilon$-weak Pareto optimality]
\label{thm:angular_or_epsilon}
Consider one iteration of ST-CR at state $(S_A,S_B)$ with gradients
$\nabla f^A(S_A)$ and $\nabla f^B(S_B)$. 
Assume each $f^i$ is $\mu$-strongly concave, $\beta$-smooth, and $L$-Lipschitz in the feasible region. 
Let $\delta := \max \|\Trade\|$ denote the maximum feasible trade magnitude.

For every $\hat\theta \in \bigl(\cos^{-1}(1/\sqrt{n}), {\pi}/{2}\bigr)$, at least one of the following holds:
\begin{enumerate}
    \item \textbf{Angular assumption holds (certifiable alignment).}
    There exists an offer $\Trade \in \bar V$ such that
    \[
    \angle\!\bigl(\Trade,\nabla f^A(S_A)\bigr)\le \hat\theta
    \qquad\text{or}\qquad
    \angle\!\bigl(-\Trade,\nabla f^B(S_B)\bigr)\le \hat\theta.
    \]

    \item \textbf{$\epsilon$-weak Pareto optimality certificate.}
    The current state $(S_A,S_B)$ is $\epsilon$-weakly Pareto-optimal with
    \[
    \epsilon
    := \max\{\epsilon_{\text{alignment}}, \epsilon_{\text{responding}}\},
    \]
    where
    \[
    \epsilon_{\text{alignment}} = 
    \delta\,L\,
    \sin\!\left(
        \min\left\{
            2\sin^{-1}\!\bigl(\sqrt{n-1}\cos(\hat\theta)\bigr),
            \frac{\pi}{2}
        \right\}
    \right).
    \]
    \[
    \epsilon_{\text{responding}} = 
    \frac{d \delta \beta (n-1)\sqrt{n}}{2}
    \]
\end{enumerate}
\end{theorem}

The parameter $\hat\theta$ captures a trade-off between directional improvement and certification strength. Choosing a smaller $\hat\theta$ enforces offers that are more closely aligned with an agent’s gradient and thus potentially yields stronger immediate improvement, but it weakens the $\epsilon$-weak Pareto certificate in Case~2. Conversely, taking $\hat\theta$ closer to $\pi/2$ tightens the Case~2 bound but allows accepted directions to become nearly orthogonal to the gradient, which may reduce per-trade improvement. 

We note that this theorem demonstrates the existence of an offer that satisfies the angular condition. Identifying such an offer \emph{a priori} is not possible since the responding agent’s utility gradient is unknown. However, in practice one can sample uniformly from the orthogonal offer set to invoke the angular condition with high probability.

\section{Numerical Experiments}

We test ST-CR using randomized scenarios. Implementation details and code can be found at \url{https://github.com/suryakmurthy/Seq_Trading_Cmpr}.

\subsection{Scenario}
 We consider quadratic utility functions such that 
 \small
\begin{align*}
&\OfferingUtility{\OfferingState} = \OfferingState^{\top} \frac{((\rho + 1) \UtilityA_A + \UtilityA_B)}{\rho + 2} \OfferingState + 2\OfferingState^{\top} \frac{((\rho + 1) \Utilityb_A + \Utilityb_B)}{\rho + 2}, \\
        &\respondingUtility{\respondingState} = \respondingState^{\top}  \frac{( \UtilityA_A + (\rho + 1) \UtilityA_B)}{\rho + 2} \respondingState + 2\respondingState^{\top} \frac{( \Utilityb_A + (\rho + 1) \Utilityb_B)}{\rho + 2}
\end{align*}
\normalsize
where $\UtilityA_A$ and $\UtilityA_B$ are random negative semi-definite matrices, $\Utilityb_A$ and $\Utilityb_B$ are random vectors, and $\rho\in [0, \infty)$ is a mixing constant. Each random matrix is generated by creating a $n \times n$ matrix with entries uniformly distributed between 0 and 1 and then multiplying it by the negative of its transpose. $\Utilityb_A$ and $\Utilityb_B$ are generated by sampling integers uniformly between 1 and 200. Values of $\rho$ closer to 0 correspond to high alignment between the utility gradients, while larger values of $\rho$ indicate lower alignment.
To control for alignment between the agents’ utility gradients and enable a consistent comparison, each agent's state is initialized with 100 resources in each category.
We limit the offers to trade at most $5$ resources from each category ($d = 5\sqrt{n}$).
The testing includes discrete and continuous scenarios. 
In discrete trading scenarios, offers are limited to integer values, while in continuous trading scenarios, offers can take any real value.
We adapt ST-CR to discrete trading by rounding offers to integers and progressively shrinking the set of potential gradients using polytope-based updates until a smaller cone is obtained. Full details of discrete ST-CR are provided in \cite{murthy2024sequential}.
If any of the trading algorithms reach a state where one agent possesses all resources in a category, we remove the category from consideration and continue with the remaining categories.

\begin{figure}[ht!]
      
    \begin{subfigure}[b]{\columnwidth}
        \centering
        \begin{tikzpicture}
    \definecolor{crimson2143940}{RGB}{214,39,40}
    \definecolor{darkgray176}{RGB}{176,176,176}
    \definecolor{darkorange25512714}{RGB}{141,141,233}
    \definecolor{forestgreen4416044}{RGB}{255,127,14}
    \definecolor{lightgray204}{RGB}{204,204,204}
    \definecolor{mediumpurple148103189}{RGB}{105,47,47}
    \definecolor{steelblue31119180}{RGB}{31,119,180}
    \definecolor{sienna1408675}{RGB}{60, 80, 50}
    \definecolor{orchid227119194}{RGB}{85, 107, 47}
    \definecolor{gray127}{RGB}{135, 160, 120}

    \fill[steelblue31119180] (0.2, -1) rectangle (0.7, -0.7);
    \node[anchor=west] at (0.8, -0.85) {ST-CR without Heuristics};
    
    \fill[darkorange25512714] (0.2, -1.35) rectangle (0.7, -1.05);
    \node[anchor=west] at (0.8, -1.2) {ST-CR with Previous Trade Heuristic};

    \fill[forestgreen4416044] (0.2, -1.7) rectangle (0.7, -1.4);
    \node[anchor=west] at (0.8, -1.55) {Random Trades without Heuristics};

    \fill[crimson2143940] (0.2, -2.05) rectangle (0.7, -1.75);
    \node[anchor=west] at (0.8, -1.9) {Random Trades with Previous Trade Heuristic};

    \fill[mediumpurple148103189] (0.2, -2.4) rectangle (0.7, -2.1);
    \node[anchor=west] at (0.8, -2.25) {Random Trades with Momentum};

    \fill[gray127] (0.2, -2.75) rectangle (0.7, -2.45);
    \node[anchor=west] at (0.8, -2.60) {GCA with updates every 100 offers};

    \fill[orchid227119194] (0.2, -3.1) rectangle (0.7, -2.80);
    \node[anchor=west] at (0.8, -2.95) {GCA with updates every 10 offers};

    \fill[sienna1408675] (0.2, -3.45) rectangle (0.7, -3.15);
    \node[anchor=west] at (0.8, -3.30) {GCA with updates every offer};
    
\end{tikzpicture}
    \end{subfigure}

\par\bigskip
    
    \centering
    \begin{subfigure}[b]{0.46\columnwidth}
        \centering
\input{figures/results/gca_testing/query_benefit_curves_alignment_0_1_items_3_int}
        \caption{3 categories, $\rho = 0.1$}
        \label{fig:discrete_3_01_s}
    \end{subfigure}%
    \hspace{0.05\columnwidth} %
    \begin{subfigure}[b]{0.46\columnwidth}
        \centering
        \input{figures/results/gca_testing/query_benefit_curves_alignment_10_items_3_int}
        \caption{3 categories, $\rho = 10$}
        \label{fig:discrete_3_10_s}
    \end{subfigure}

    \par\bigskip
    
    \begin{subfigure}[b]{0.46\columnwidth}
        \centering
\input{figures/results/gca_testing/query_benefit_curves_alignment_0_1_items_3_int_offering}
        \caption{3 categories, $\rho = 0.1$}
        \label{fig:discrete_3_01_o}
    \end{subfigure}%
    \hspace{0.05\columnwidth} %
    \begin{subfigure}[b]{0.46\columnwidth}
        \centering
\input{figures/results/gca_testing/query_benefit_curves_alignment_10_items_3_offering}
        \caption{3 categories, $\rho = 10$}
        \label{fig:discrete_3_10_o}
    \end{subfigure}

    \par\bigskip
    
    \begin{subfigure}[b]{0.46\columnwidth}
        \centering
\input{figures/results/gca_testing/query_benefit_curves_alignment_0_1_items_3_int_responding}
        \caption{3 categories, $\rho = 0.1$}
        \label{fig:discrete_3_01_r}
    \end{subfigure}%
    \hspace{0.05\columnwidth} %
    \begin{subfigure}[b]{0.46\columnwidth}
        \centering
    \input{figures/results/gca_testing/query_benefit_curves_alignment_10_items_3_int_responding}
        \caption{3 categories, $\rho = 10$}
        \label{fig:discrete_3_10_r}
    \end{subfigure}
    \caption{Offer-benefit plots for discrete trading scenarios.}
    \label{fig:combined_discrete}
\end{figure}

\begin{figure}[t]
      
    \begin{subfigure}[b]{\columnwidth}
        \centering
        \begin{tikzpicture}
    \definecolor{crimson2143940}{RGB}{214,39,40}
    \definecolor{darkgray176}{RGB}{176,176,176}
    \definecolor{darkorange25512714}{RGB}{141,141,233}
    \definecolor{forestgreen4416044}{RGB}{255,127,14}
    \definecolor{lightgray204}{RGB}{204,204,204}
    \definecolor{mediumpurple148103189}{RGB}{105,47,47}
    \definecolor{steelblue31119180}{RGB}{31,119,180}

    \fill[steelblue31119180] (0.2, -1) rectangle (0.7, -0.7);
    \node[anchor=west] at (0.8, -0.85) {ST-CR without Heuristics};
    
    \fill[darkorange25512714] (0.2, -1.35) rectangle (0.7, -1.05);
    \node[anchor=west] at (0.8, -1.2) {ST-CR with Previous Trade Heuristic};

    \fill[forestgreen4416044] (0.2, -1.7) rectangle (0.7, -1.4);
    \node[anchor=west] at (0.8, -1.55) {Random Trades without Heuristics};

    \fill[crimson2143940] (0.2, -2.05) rectangle (0.7, -1.75);
    \node[anchor=west] at (0.8, -1.9) {Random Trades with Previous Trade Heuristic};

    \fill[mediumpurple148103189] (0.2, -2.4) rectangle (0.7, -2.1);
    \node[anchor=west] at (0.8, -2.25) {Random Trades with Momentum};
    
\end{tikzpicture}
    \end{subfigure}

\par\bigskip
    
    \centering
    \begin{subfigure}[b]{0.43\columnwidth}
        \centering
\input{figures/results/new_alignment/query_benefit_curves_alignment_1_1_items_3}
        \caption{3 categories, $\rho = 0.1$}
        \label{fig:cont_3_01}
    \end{subfigure}%
    \hspace{0.07\columnwidth} %
        \begin{subfigure}[b]{0.43\columnwidth}
        \centering
\input{figures/results/new_alignment/query_benefit_curves_alignment_11_items_3}
        \caption{3 categories, $\rho = 10$}
        \label{fig:cont_3_10}
    \end{subfigure}%
    \par\bigskip
    \begin{subfigure}[b]{0.43\columnwidth}
        \centering
\input{figures/results/new_alignment/query_benefit_curves_alignment_1_1_items_5}
        \caption{5 categories, $\rho = 0.1$}
        \label{fig:cont_5_01}
    \end{subfigure}
    \hspace{0.07\columnwidth} %
    \begin{subfigure}[b]{0.43\columnwidth}
        \centering
\input{figures/results/new_alignment/query_benefit_curves_alignment_11_items_5}
        \caption{5 categories, $\rho = 10$}
        \label{fig:cont_5_10}
    \end{subfigure}
    \caption{Offer-benefit plots for continuous trading scenarios.}
    \label{fig:combined_cont}
\end{figure}

\begin{figure}[t]
      
    \begin{subfigure}[b]{\columnwidth}
        \centering
        \begin{tikzpicture}
    \definecolor{crimson2143940}{RGB}{214,39,40}
    \definecolor{darkgray176}{RGB}{176,176,176}
    \definecolor{darkorange25512714}{RGB}{141,141,233}
    \definecolor{forestgreen4416044}{RGB}{255,127,14}
    \definecolor{lightgray204}{RGB}{204,204,204}
    \definecolor{mediumpurple148103189}{RGB}{105,47,47}
    \definecolor{steelblue31119180}{RGB}{31,119,180}

    \fill[steelblue31119180] (0.2, -1) rectangle (0.7, -0.7);
    \node[anchor=west] at (0.8, -0.85) {$d = 0.01$};
    
    \fill[darkorange25512714] (0.2, -1.35) rectangle (0.7, -1.05);
    \node[anchor=west] at (0.8, -1.2) {$d = 0.1$};

    \fill[forestgreen4416044] (0.2, -1.7) rectangle (0.7, -1.4);
    \node[anchor=west] at (0.8, -1.55) {$d = 1$};

    \fill[crimson2143940] (0.2, -2.05) rectangle (0.7, -1.75);
    \node[anchor=west] at (0.8, -1.9) {$d = 10$};

    \fill[mediumpurple148103189] (0.2, -2.4) rectangle (0.7, -2.1);
    \node[anchor=west] at (0.8, -2.25) {$d = 100$};
    
\end{tikzpicture}
    \end{subfigure}

\par\bigskip
    
    \centering
    \begin{subfigure}[b]{0.43\columnwidth}
        \centering
\input{figures/hyperparameter_study/5_items/query_benefit_curves_alignment_0_01_items_5_nolegend}
        \caption{5 categories, $\rho = 0.01$}
        \label{fig:hyp_5_001}
    \end{subfigure}%
    \hspace{0.07\columnwidth} %
        \begin{subfigure}[b]{0.43\columnwidth}
        \centering
\input{figures/hyperparameter_study/5_items/query_benefit_curves_alignment_0_1_items_5_nolegend}
        \caption{5 categories, $\rho = 0.1$}
        \label{fig:hyp_5_01}
    \end{subfigure}%
    \par\bigskip
    \begin{subfigure}[b]{0.43\columnwidth}
        \centering
\input{figures/hyperparameter_study/5_items/query_benefit_curves_alignment_1_items_5_nolegend}
        \caption{5 categories, $\rho = 1$}
        \label{fig:hyp_5_1}
    \end{subfigure}
    \hspace{0.07\columnwidth} %
    \begin{subfigure}[b]{0.43\columnwidth}
        \centering
\input{figures/hyperparameter_study/5_items/query_benefit_curves_alignment_10_items_5_nolegend}
        \caption{5 categories, $\rho = 10$}
        \label{fig:hyp_5_10}
    \end{subfigure}
    \caption{Hyperparameter study for parameter $d$.}
    \label{fig:combined_hyp}
\end{figure}

\subsection{Baselines}
We implement two sets of baseline algorithms to compare with the proposed method ST-CR. 
For all baselines, the offering agent only presents offers that improve its utility.
The first set of baselines is based on a random search approach. The random search algorithm uniformly samples offer directions and then scales the offer.
This approach is similar to the mediated simulated annealing with hill-climbers proposed by Klein et al.\ \cite{Klein_1}, which used random adjustments to previously accepted outcomes to improve societal benefit and has been adapted to different negotiation settings \cite{fujita2024automated}.
To enhance the random search, we include a heuristic that starts by offering the most recently accepted trade.
Another improvement we consider is incorporating momentum into random search. In the random search with momentum, the offering agent uses the random search baseline for the first trade.
In subsequent trades, the agent makes the previously accepted offer with random deviations, with the magnitude of deviations increasing as more offers are rejected.

 The second set of baselines is based on the Greedy Concession Algorithm (GCA) \cite{baarslag2015optimal}, which aims to maximize the offering agent's expected utility.
 We use GCA as a comparison to evaluate ST-CR's performance in terms of the offering agent's utility. GCA consists of two steps: 1) sorting possible offers based on the expected utilities for the offering agent and 2) updating beliefs about the responding agent’s preferences. 
 
 In step 2 of our implementation for GCA, we model the responding agent's utility as a weighted sum over the amounts of resources in each category. We uniformly sample $100$ possible sets of weights from the unit hypersphere, representing possible configurations of the responding agent’s objective function. The offering agent initially assigns equal probabilities to each set of weights. After a number of offers are rejected, we use these rejections to update the probability of each set. We test GCA’s performance with belief update intervals of $1$, $10$, and $100$ rejections. 
If a set of weights indicates that a rejected offer should have been accepted, we reduce the probability of that weight vector by a scaling factor of $0.1$.
The weight probabilities are further smoothed using a softmax function after each accepted trade to account for shifts in preferences.

\subsection{Results}
In Figs.~\ref{fig:combined_discrete} and \ref{fig:combined_cont}, we present the societal, offering agent's, and responding agent's benefits for discrete and continuous trading scenarios, respectively, as a function of the offer count. For each setting, we average the benefits over 500 random scenarios. In all plots, the average benefit of each algorithm is normalized by the maximum cumulative average benefit achieved across all algorithms. 

In discrete settings with integer-constrained offers, the number of possible offers increases exponentially with the number of resource categories. 
In addition, the number of sampled weights must be increased to cover the responding agent's potential preferences. 
This growth presents computational challenges to algorithms that search through potential offers and weights, such as GCA. 
For this reason, for the discrete setting, we only provide comparisons for 3 items. 

For the societal benefit shown in Fig.~\ref{fig:discrete_3_10_s}, ST-CR achieves similar societal benefits with fewer offers compared to the other baselines. This efficiency stems from ST-CR’s modeling of the responding agent’s preferences, allowing it to quickly identify mutually beneficial trades. 
In highly aligned scenarios, such as those shown in Fig.~\ref{fig:discrete_3_01_s}, this difference is more pronounced, with the random baseline requiring significantly more offers to match ST-CR. The random trading baseline without heuristics eventually achieves a higher societal benefit than ST-CR.
This is likely because the random algorithm’s offering strategy allows for a comprehensive exploration of the space of potential offers.

For the offering agent's benefit shown in Figs.~\ref{fig:discrete_3_01_o} and \ref{fig:discrete_3_10_o}, we observe that GCA eventually achieves a higher offering benefit than both ST-CR and the random baselines due to prioritizing offers that maximize the offering agent’s expected benefit. For the lower number of offers, ST-CR provides a performance comparable to GCA. ST-CR efficiently elicits the responding agent's preferences and realizes more accepted trades, while GCA makes offers to maximize the offering agent's expected benefit for a fixed set of beliefs, limiting exploration. 
As shown in Fig.~\ref{fig:discrete_3_01_o}, the performance of ST-CR becomes more prominent in highly aligned scenarios, where the range of mutually beneficial trades is narrower. For the responding agent's benefit shown in Figs.~\ref{fig:discrete_3_01_r} and \ref{fig:discrete_3_10_r}, we observe that ST-CR achieves high performance with fewer offers.

For continuous trading scenarios shown in Figs.~\ref{fig:cont_3_01}--\ref{fig:cont_5_10}, we observe that increasing the number of resource categories from 3 to 5 decreases the performance of all algorithms as the search space grows. 
As resource categories increase, the performance gap between ST-CR and the baselines widens for a fixed number of offers, indicating ST-CR’s suitability for complex trading scenarios with limited offer budgets.

In Figs.~\ref{fig:combined_discrete} and \ref{fig:combined_cont}, we observe that methods that utilize previous trade heuristics can improve societal benefit with the fewest offers. 
Since the agents' utility functions are smooth and offers have bounded magnitudes, previously accepted trades are likely to be accepted again. 
However, as trading progresses and agents approach the Pareto front, the chances of prior trades being accepted decrease due to overshooting, reducing the efficiency of algorithms with the heuristic.

\subsection{Effects of the Trade Magnitude}

The primary hyperparameter used by the ST-CR algorithm is the trade magnitude $d$. For this hyperparameter study, we consider a $5$-category continuous trading scenario. We use the same utility function format as the previous experiments, and initialize each agent's state with $100$ resources in each category. However, unlike the previous example, we vary the trade magnitudes among the values $\{0.01, 0.1, 1, 10, 100\}$. As in the previous experiments, we collect results averaged across 500 random scenarios.

Fig.~\ref{fig:combined_hyp} presents the average cumulative societal benefit across different values of $d$ and alignment parameters $\rho$. In low-alignment settings ($\rho = 1$ and $\rho = 10$; see Figs.~\ref{fig:hyp_5_1}--~\ref{fig:hyp_5_10}), we observe that $d = 100$ initially achieves the highest societal benefit. Larger $d$ induces larger trade steps, enabling ST-CR to move more rapidly toward the Pareto front and realize greater benefit per accepted trade. However, increasing $d$ does not always improve long-term outcomes. In Fig.~\ref{fig:hyp_5_10}, performance with $d = 100$ eventually falls behind $d = 10$, suggesting that excessively large trade magnitudes hinder fine-grained adjustments near the Pareto front and limit how closely the agent can approach optimal allocations.

In highly aligned settings ($\rho = 0.01$ and $\rho = 0.1$; see Figs.~\ref{fig:hyp_5_001}--~\ref{fig:hyp_5_01}), the performance gap between $d \leq 1$ and $d > 1$ is substantially smaller than in unaligned cases. When utilities are aligned, precise gradient refinement becomes more important, and large $d$ values can produce noisier or less accurate gradient estimates. As a result, aggressive trade steps provide less advantage and may slow convergence or reduce final cumulative benefit.

\section{User Study}

We conducted a user study to evaluate the performance of ST-CR in human-agent interactions\footnote{This study was approved by UT IRB \#6345.}. The objective of this study is to assess how human participants respond to algorithmic trade offers and to investigate the real-world applicability of ST-CR. Participants were recruited online through Prolific to ensure a diverse sample and a controlled testing environment. We give further details of the study in \cite{murthy2024sequential}. 

\subsection{Trading Scenario}

\subsubsection{Initial Setup and Offer Constraints}
The user study considers a trading scenario in which a human participant, denoted as \( B \), interacts with a computer agent, denoted as \( A \), to exchange resources across three categories. In this study, we use different fruits as categories to capture different human preferences. Both agents start with an initial allocation of 50 units in each category. The computer agent generates trade offers that are constrained to integer values, ensuring that the trades align with discrete decision-making settings.

\subsubsection{Utility Functions and Preferences}
We instruct participants to specify a target allocation that aligns with their preferences. The target allocation \( \mathbf{b_B} \) has values constrained to the range \([0, 100]\). We do not use the target allocations for any of the compared algorithms. Instead, we use the target allocation to compare performances after trading is complete.

We provide bonus compensation based on how closely the participants reached their desired resource allocation, incentivizing them to move toward their target vector \( \mathbf{b_B} \). To quantify this, we assigned each user a score based on the normalized improvement from their initial allocation \( \mathbf{S_B^0} \) to their final allocation \( \mathbf{S_B^T} \), defined as
\[
\text{score} = 1 - \frac{\|\mathbf{b_B} - \mathbf{S_B^T}\|_1}{\|\mathbf{b_B} - \mathbf{S_B^0}\|_1}.
\]
A score of 1 indicates a perfect match with the user’s target allocation, while a score of 0 implies no improvement over the initial allocation. If the final allocation was further from the target than the initial state (i.e., a negative score), the user was treated as having a score of 0 for compensation purposes.

To lower variance in the results due to different preferences, we limit the computer agent's potential target allocation $\mathbf{b_A}$ to the following set
\(\{  [66, 33, 33], [33, 66, 33],
                [33, 33, 66], [66, 66, 66], 
                     [33, 33, 33] \}\).
We represent the computer's utility using a quadratic function of the form: 
\[
f^A(\mathbf{S_A}) = -\mathbf{S_A}^{\top} I \mathbf{S_A} + 2 \mathbf{S_A}^{\top} \mathbf{b_A}
\]
where \( I \) is the identity matrix. Note that this concave quadratic function is maximized at the computer's target allocation $\mathbf{b_A}$.

While ST-CR does not model the user's utility function, we use the same quadratic function as a metric to compare performance between algorithms
\begin{equation}
\label{eq:utilitymodel}
    f^B(\mathbf{S_B}) = -\mathbf{S_B}^{\top} I \mathbf{S_B} + 2 \mathbf{S_B}^{\top} \mathbf{b_B}.
\end{equation}
We emphasize that this quadratic utility is a modeling assumption used for evaluation and may not perfectly capture the user’s true underlying preferences.

\subsubsection{Interaction and Trading Progression}
Trading follows a sequential interaction process in which the computer agent generates offers while the human participant evaluates and responds to these offers. 
Participants can accept or reject offers and may also provide natural language feedback in the form of counteroffers. 
To analyze these responses, we use the GPT API \cite{openai2024gpt4technicalreport} for sentiment analysis, determining whether a participant is accepting or rejecting an offer. 
Additionally, the GPT API is employed to parse counteroffers into numerical vectors compatible with our trading algorithms. 
ST-CR treats counteroffers as additional halfspace constraints that further refine the gradient cone. Full details of the modifications to ST-CR for incorporating counteroffers are provided in \cite{murthy2024sequential}.
We remark that this tool only provides parsing and sentiment analysis and is not used for offer generation. 

The scenario continues for 10 minutes or until the participant chooses to end the session. We limit the response time to 120 seconds for each trade offer.

\subsection{Baseline Algorithms}
To assess the effectiveness of the proposed algorithm, we compare its performance against two baseline algorithms: random trades with a previous trade heuristic, and GCA. Each participant engages in a single trading scenario with one algorithm, ensuring that the study design isolates individual interactions to reduce cognitive load. The selection of the algorithm and computer agent target is systematically rotated across participants to achieve a balanced representation of different trading behaviors and utility alignments.

\subsection{Results}

\begin{table}[h!]
\centering
\caption{Distribution of user-agent interactions by alignment angle and method.}
\begin{tabular}{lccc|c}
\hline
\textbf{Method} & \textbf{0--60°} & \textbf{60--120°} & \textbf{120--180°} & \textbf{Total} \\
\hline
ST-CR   & 9  & 25 & 15 & 49 \\
GCA     & 7  & 29 & 7  & 43 \\
Random  & 6  & 36 & 11 & 53 \\
\hline
\textbf{Total}  & 22 & 90 & 33 & 145 \\
\hline
\end{tabular}
\label{tab:angle-distribution}
\end{table}

\paragraph{Alignment-Based Binning of User-Agent Interactions}

To better understand how alignment between user and agent preferences impacts outcomes, we segment interactions into three angular bins based on the direction of each party’s goals. These bins reflect the level of alignment between the human agent's target and the computer agent’s target:

\begin{itemize}
    \item \textbf{0--60°}: High alignment — agents want similar \emph{changes} in each resource category, leading to direct competition for resources.
    \item \textbf{60--120°}: Moderate alignment — partial overlap in preferences.
    \item \textbf{120--180°}: Low alignment — agents want different, complementary changes in resources, often making trades more feasible.
\end{itemize}

To assign each interaction to one of these bins, we compute the angle between the direction vectors from a common initial state \( \mathbf{S}_0 \) to each agent’s respective target. This reflects the direction of desired change rather than absolute position. The alignment angle is computed as:

\[
\cos^{-1}\left( \frac{(\mathbf{b}_A - \mathbf{S}_0) \cdot (\mathbf{b}_B - \mathbf{S}_0)}{\| \mathbf{b}_A - \mathbf{S}_0 \| \cdot \| \mathbf{b}_B - \mathbf{S}_0 \|} \right)
\]
We use this angle to categorize each of the $145$ collected interactions. Table~\ref{tab:angle-distribution} summarizes the breakdown by algorithm and angular bin.

\begin{table}[h!]
\centering
\caption{Societal benefit values (mean $\pm$ standard deviation) by angular range for each algorithm. The utility function is assumed to be quadratic as shown in Equation \ref{eq:utilitymodel}.}
\scriptsize
\resizebox{\columnwidth}{!}{%
\begin{tabular}{lccc}
\hline
\textbf{Algorithm} & \textbf{0--60°} & \textbf{60--120°} & \textbf{120--180°} \\
\hline
\multicolumn{4}{l}{\textit{Societal Benefit Values}} \\
GCA            & $-900.57 \pm 506.61$ & $543.31 \pm 521.74$ & $1317.43 \pm 961.29$ \\
Random Trades  & $-491.00 \pm 534.79$ & $326.67 \pm 789.49$ & $1633.64 \pm 678.31$ \\
ST-CR          & $-454.22 \pm 429.70$ & $568.40 \pm 564.50$ & $1368.40 \pm 717.24$ \\
\hline
\end{tabular}
}
\label{tab:societal_benefit}
\end{table}

\paragraph{Benefit by Alignment}
Table~\ref{tab:societal_benefit} reports the cumulative societal benefit for each algorithm across the three angular alignment ranges. In the 0--60° range, where agents are highly aligned and seek similar resources, all methods yield negative cumulative benefit. This outcome reflects the inherent difficulty of negotiating when agents are in direct competition for the same items. Additionally, these negative values could be due to a mismatch between the assumed human utility model and the human's actual underlying preferences. In the 60--120° range, where agents exhibit partial alignment, all algorithms achieve moderate positive benefit, indicating that limited compromise is possible. The highest societal benefits are observed in the 120--180° range, where agents’ goals are unaligned and complementary trades are easier to identify.

ST-CR achieves the highest societal benefit in both the aligned (0--60°) and moderately aligned (60--120°) settings, outperforming both GCA and the random baseline. This demonstrates ST-CR’s strength in efficiently navigating trade-offs when the space of beneficial offers is narrow. In the highly unaligned (120--180°) case, however, ST-CR underperforms the random baseline. As we state in Remark~\ref{rem:unaligned}, this is likely due to its strategy of using orthogonal offers to refine the gradient cone, leading to neutral trades in strongly unaligned settings.

\begin{itemize}
    \item In \textbf{aligned settings} (0--60°), all algorithms yield negative cumulative benefit as a natural consequence of the overlapping goals. We remark that the negative utilities may be caused by the users' lack of rationality or $f^B(\mathbf{S_B})$ not being an accurate utility model.
    \item In \textbf{moderately aligned settings} (60--120°), benefit values are positive but modest, reflecting partial alignment between agent objectives.
    \item In \textbf{highly unaligned settings} (120--180°), all methods show the greatest societal gain, confirming that trades are most valuable when agents have different targets.
\end{itemize}

\begin{table}[h!]
\centering
\caption{Summary of trade metrics by algorithm with mean $\pm$ standard deviation for each algorithm.}
\resizebox{\columnwidth}{!}{%
\begin{tabular}{lcccc}
\hline
\textbf{Algorithm} & \textbf{Total Offers} & \textbf{Accepted Offers} & \textbf{Offers/Accepted} & \textbf{Time per Trade (s)} \\
\hline
ST-CR         & 28.31 $\pm$ 16.37 & 10.02 $\pm$ 6.47 & 2.41 $\pm$ 2.25 & 38.12 $\pm$ 41.88 \\
GCA           & 18.02 $\pm$ 15.37 & 4.09  $\pm$ 2.12 & 3.49 $\pm$ 4.05 & 70.26 $\pm$ 85.37 \\
Random        & 22.58 $\pm$ 17.73 & 5.70  $\pm$ 3.88 & 2.65 $\pm$ 3.16 & 62.37 $\pm$ 93.85 \\
\hline
\end{tabular}
}
\label{tab:efficiency}
\end{table}

\begin{table}[h!]
\centering
\caption{Fraction of accepted trades that decrease user benefit (mean $\pm$ standard deviation) by angular range for each algorithm.}
\scriptsize
\resizebox{\columnwidth}{!}{%
\begin{tabular}{lccc}
\hline
\textbf{Algorithm} & \textbf{0--60°} & \textbf{60--120°} & \textbf{120--180°} \\
\hline \\
GCA           & $0.938 \pm 0.095$ & $0.551 \pm 0.263$ & $0.228 \pm 0.355$ \\
Random Trades & $0.924 \pm 0.094$ & $0.433 \pm 0.292$ & $0.110 \pm 0.176$ \\
ST-CR         & \textbf{$0.668 \pm 0.165$} & \textbf{$0.319 \pm 0.179$} & $0.306 \pm 0.262$ \\
\hline
\end{tabular}
}
\label{tab:score_decrease_rate}
\end{table}

\paragraph{Trade Efficiency}
Table~\ref{tab:efficiency} summarizes the efficiency of each algorithm in terms of total offers, accepted trades, offers per accepted trade, and average time to reach agreement. ST-CR achieves the highest number of accepted trades (\(10.02 \pm 6.47\)) while requiring fewer offers (\(28.31 \pm 16.37\)) and less time (\(38.12 \pm 41.88\) seconds) per accepted trade than either baseline. This efficiency stems from ST-CR’s ability to model the responding agent’s preferences and identify mutually beneficial offers more rapidly. In contrast, both GCA and the random baseline exhibit higher variance in performance and, on average, take longer or require more offers to secure agreement. These trends are consistent with the results observed in our numerical experiments, where ST-CR consistently reaches a higher benefit with fewer interactions.

Table~\ref{tab:score_decrease_rate} reports the fraction of accepted trades that resulted in a score decrease for each algorithm. The results show that ST-CR consistently exhibits lower rates of accepted score-decreasing trades than both GCA and Random Trades across the highly aligned angular ranges. In the highly unaligned case (120°–180°), ST-CR does exhibit an elevated rate of score-decreasing acceptances relative to the other baselines, suggesting that greedy approaches can yield higher benefit in such cases.

\section{Conclusion}
We consider a sequential multi-issue trading problem with two agents.
We introduced sequential trading with cone refinement (ST-CR), an algorithm that leverages comparisons to identify mutually beneficial trades. 
We demonstrated ST-CR’s effectiveness in two-agent negotiations and provided theoretical guarantees that, after a finite number of rejected offers, the algorithm either finds a mutually beneficial trade or certifies that the current state is \(\epsilon\)\nobreakdash-weakly Pareto optimal---meaning no feasible trade can significantly improve both agents’ utilities.
We also conducted user studies to explore the benefits and limitations of ST-CR in scenarios with human-agent interactions.
ST-CR relies on the greedy rationality of agents, which may not hold in some scenarios. 
Future work could explore modifications to handle agents with more complex acceptance strategies or deceptive agents, further expanding ST-CR’s applicability.

\section*{Acknowledgments}
This work was supported partially by the Army Research Laboratory (ARL), under grant number W911NF-23-1-0317, and partially by the National Science Foundation (NSF), under grant numbers 1836900 and 2211432.

\begin{IEEEbiography}[{\includegraphics[width=1in,height=1.25in,clip,keepaspectratio]{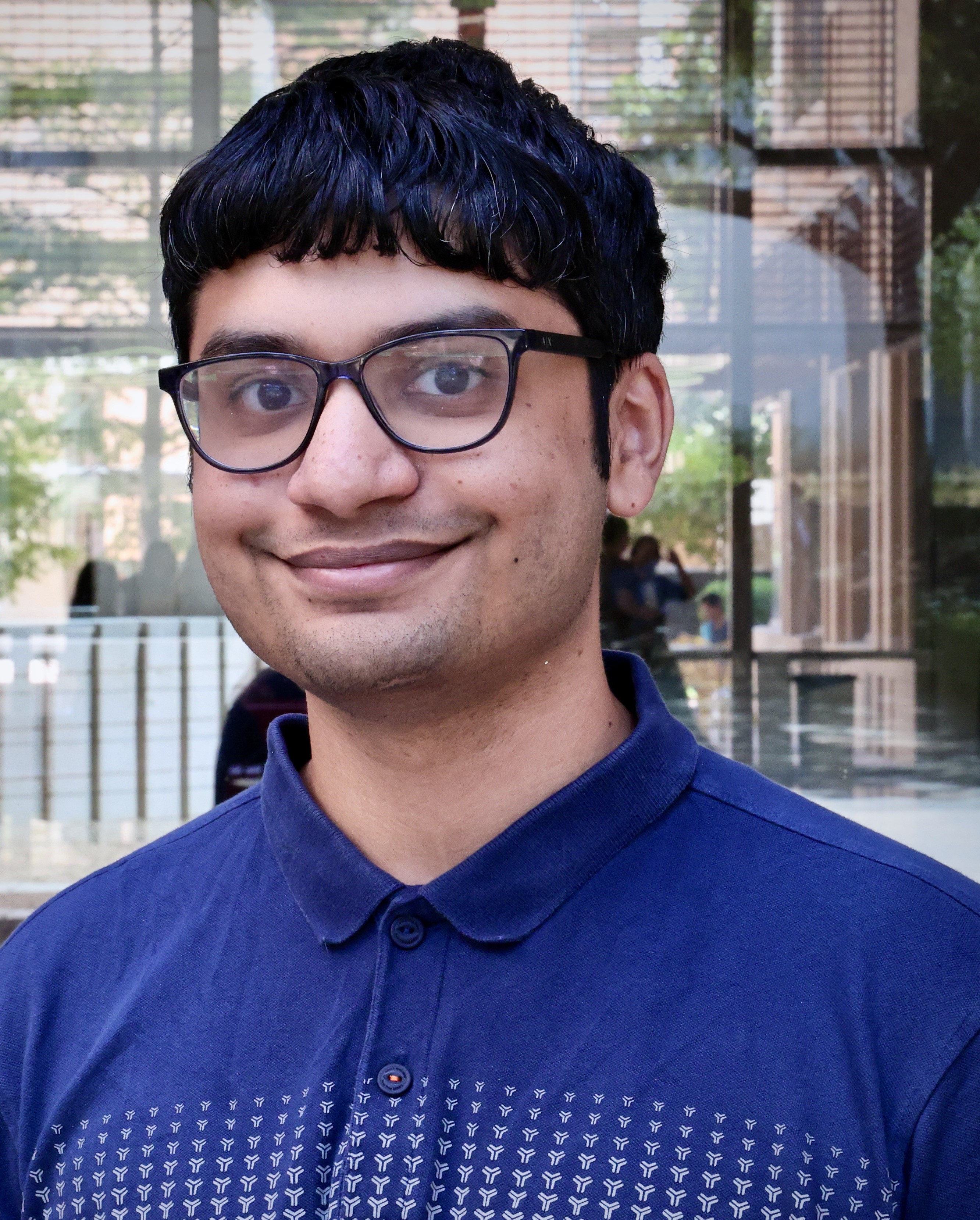}}]{Surya Murthy} is a PhD student in the Electrical and Computer Engineering department at the University of Texas at Austin. He received his B.S. degree in Computer Engineering from the University of Illinois at Urbana-Champaign. His research interests are multi-agent systems, multi-objective decision-making, and human-agent collaboration.
\end{IEEEbiography}
\begin{IEEEbiography}[{\includegraphics[width=1in,height=1.25in,clip,keepaspectratio]{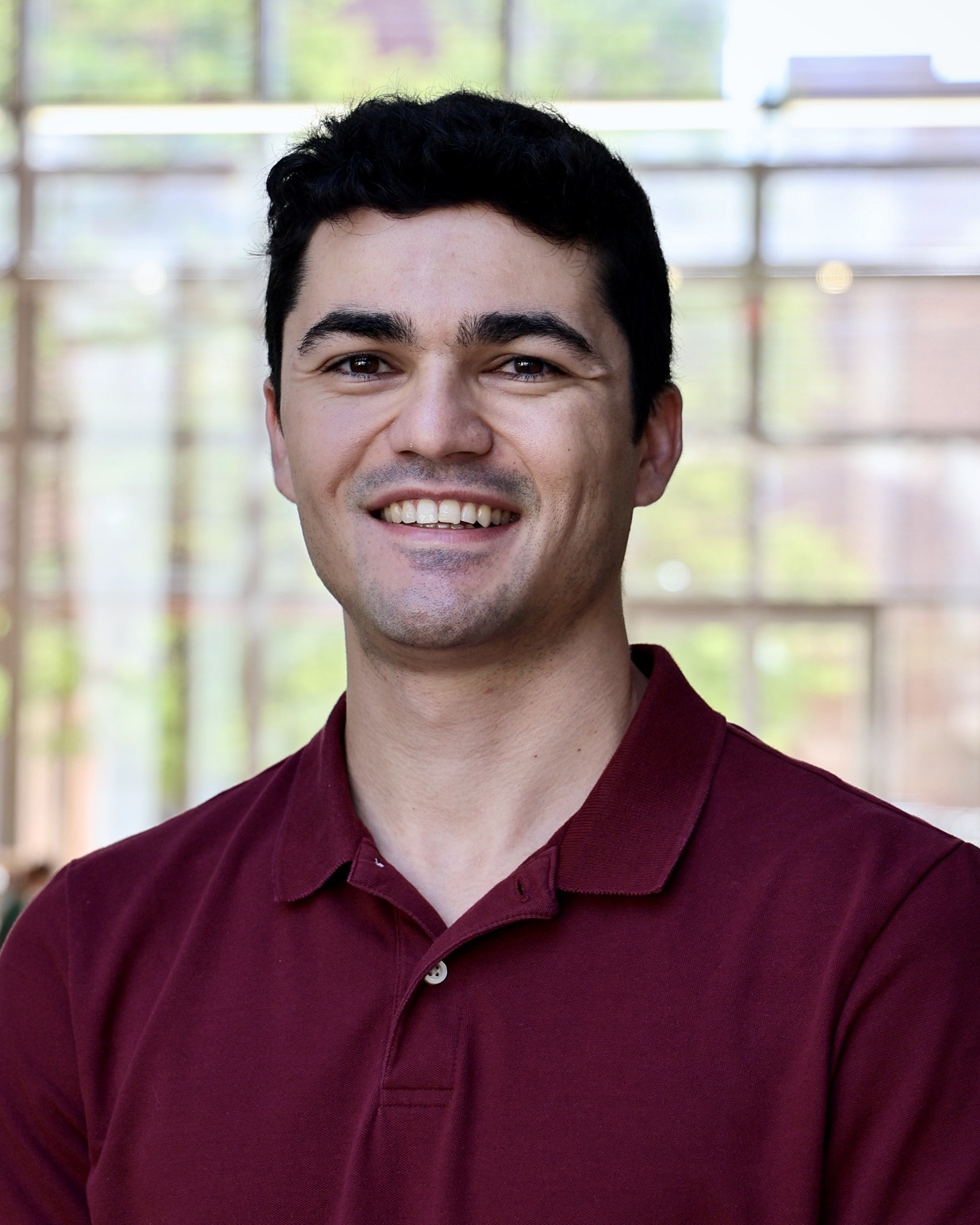}}]{Mustafa O. Karabag} is a postdoctoral fellow in the Oden Institute for Computational Engineering \& Sciences at the University of Texas at Austin. He received his Ph.D. degree from the University of Texas at Austin in 2023. His research focuses on developing theory and algorithms to control the information flow of autonomous systems to succeed in information-scarce or adversarial environments.
\end{IEEEbiography}
\begin{IEEEbiography}[{\includegraphics[width=1in,height=1.25in,clip,keepaspectratio]{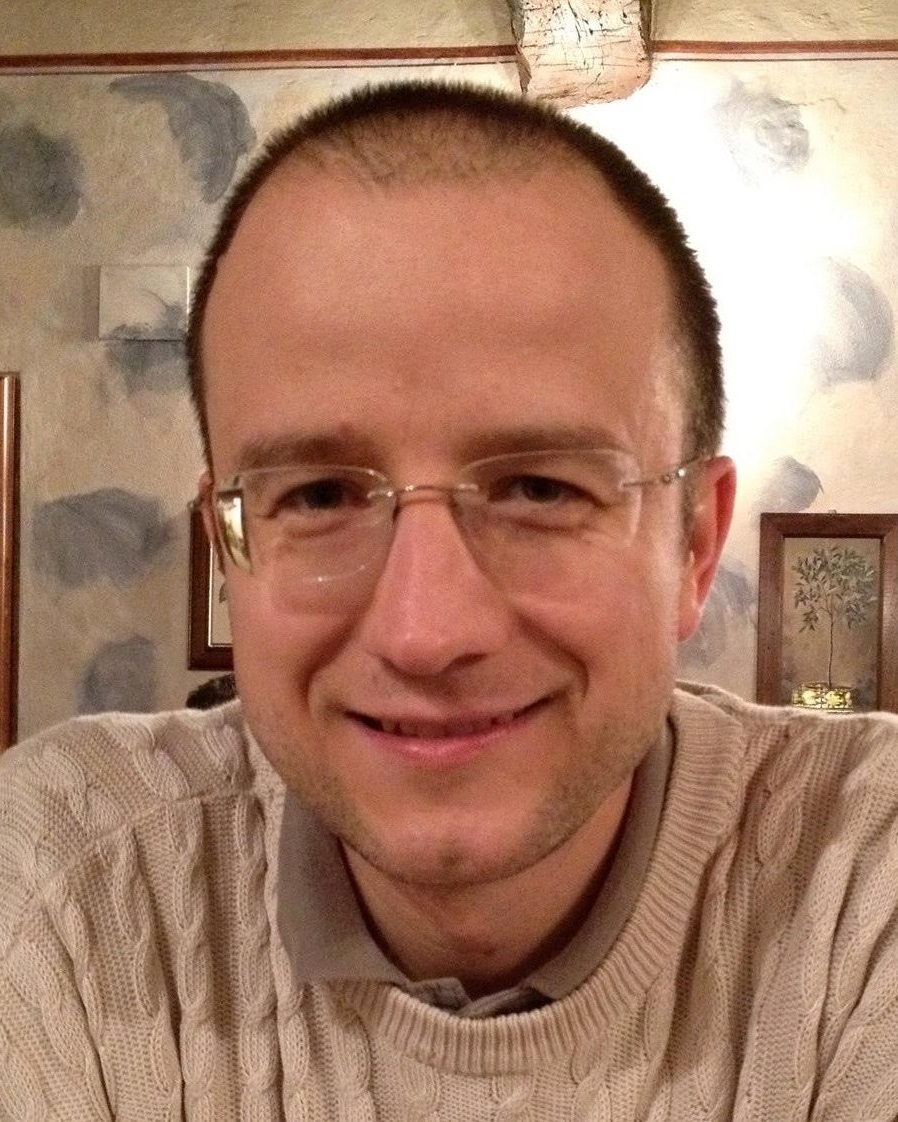}}]{Ufuk Topcu} received the Ph.D. degree from the University of California at Berkeley, Berkeley, CA, USA, in 2008. He joined the Department of Aerospace Engineering, University of Texas at Austin, Austin, TX, USA, in Fall 2015. He held research positions with the University of Pennsylvania, Philadelphia, PA, USA, and California Institute of Technology, Pasadena, CA, USA. His research focuses on the theoretical, algorithmic and computational aspects of design and verification of autonomous systems through novel connections between formal methods, learning theory, and controls.
\end{IEEEbiography}
\section*{References}
\bibliographystyle{unsrt}
\bibliography{bibleography}

\onecolumn

\section*{Appendix}
In this appendix, we present the theoretical guarantees of our algorithm through a case-by-case analysis. First, under smoothness assumptions, we show that if \(k\) consecutive offers are rejected, the current point can be certified as \(\epsilon\)-weakly Pareto-optimal. Second, when offers are accepted and satisfy an angular condition with respect to at least one agent’s objective, we prove that the algorithm asymptotically converges to a Pareto-stationary point. Finally, when this angular condition cannot be certified, we show that the current point must already be Pareto-stationary.

\subsection{Proofs for Theorem 1}

In this section, we provide a proof of Theorem \ref{thm}. Recall that \(\kappa \geq \sqrt{n-1}\) is a parameter satisfying 
\begin{equation} \sqrt{1 - \frac{1}{2\NumItems}}^{\left\lfloor \frac{\UnacceptedOffers-n}{\NumItems-1} \right\rfloor} =2\NumItems\sqrt{1-\left(\frac{ \GradientThreshold^2 - (n-1)}{\GradientThreshold^2 + (n-1)}\right)^2}.
    \label{eq:condition_repeat}
\end{equation}

\begin{theorem} \label{thm:appdx}
Let \(\kappa \geq \sqrt{n-1}\) be a parameter satisfying 
\begin{equation} \sqrt{1 - \frac{1}{2\NumItems}}^{\left\lfloor \frac{\UnacceptedOffers-1}{\NumItems-1} \right\rfloor} =2\NumItems\sqrt{1-\left(\frac{ \GradientThreshold^2 - (n-1)}{\GradientThreshold^2 + (n-1)}\right)^2}.
    \label{eq:condition}
\end{equation}

If $\respondingUtilityWoArguments$ is $\Smoothness$-smooth, at least one of the following holds true after Procedure \ref{algo:CR} makes $\UnacceptedOffers$ rejected offers at state $(\OfferingState, \respondingState)$:

\begin{enumerate}
    \item The angle between the agents' gradients satisfies 
    \small
    \begin{equation*} \label{eq:angle_bound}
        \angle(\nabla \OfferingUtility{\OfferingState}, \nabla \respondingUtility{\respondingState}) < 2 \sin^{-1}\left(\sqrt{1 - \frac{1}{2\NumItems}}^{\left\lfloor \frac{\UnacceptedOffers-n}{\NumItems-1} \right\rfloor}\right).
    \end{equation*}
    \normalsize
    
    \item The responding agent's benefit for any trade $\Trade$ satisfies
    \begin{equation*} \label{eq:benefit_bound}
    \respondingUtility{\respondingState + \Trade} - \respondingUtility{\respondingState} \leq \SamplingDistance \GradientThreshold \sqrt{\NumItems} \Smoothness \|\Trade\| + {\Smoothness \|\Trade\|^2}/{2}. 
    \end{equation*}
    If $\respondingUtilityWoArguments$ is concave, the responding agent's benefit for any trade $\Trade$ further satisfies
    \begin{equation*} \label{eq:benefit_bound_2}
    \respondingUtility{\respondingState + \Trade} - \respondingUtility{\respondingState} \leq \SamplingDistance \GradientThreshold \sqrt{\NumItems} \Smoothness \|\Trade\| . 
        \end{equation*}
\end{enumerate}
\end{theorem}

\begin{proof}[Proof of Theorem \ref{thm}]
Throughout the proof we use \(\Smoothness\)-smoothness of \(\respondingUtilityWoArguments\): For a trade \(\Trade\), we have $$|\respondingUtility{\respondingState - \Trade} - \respondingUtility{\respondingState} - \langle \nabla \respondingUtility{\respondingState}, -\Trade \rangle| \leq \frac{\beta \|\Trade\|^2}{2}.$$ 

Due to Cauchy-Schwartz inequality, we also have 
\begin{equation}
    \respondingUtility{\respondingState - \Trade} - \respondingUtility{\respondingState}\leq \lVert \nabla \respondingUtility{\respondingState}\rVert \|\Trade\| +   \frac{\beta \|\Trade\|^2}{2}. \label{ineq:nearopt}
\end{equation}

If $\respondingUtilityWoArguments$ is concave, we also have
\begin{equation}
    \respondingUtility{\respondingState - \Trade} - \respondingUtility{\respondingState}\leq \lVert \nabla \respondingUtility{\respondingState}\rVert \|\Trade\| . \label{ineq:concavity}
\end{equation}
due to the definition of concavity.

If \(\ResponseFunction{\Trade} = reject\), then \(\respondingUtility{\respondingState - \Trade} - \respondingUtility{\respondingState} \leq 0\), which implies that \(\langle \nabla \respondingUtility{\respondingState}, -\Trade \rangle \leq \frac{\beta \|\Trade\|^2}{2}\). For a trade \(T\) with \(\|\Trade\| = d\), we have \( \nabla_{-\Trade} f^{\respondingAgent}{\respondingState}  \leq \frac{\beta d}{ 2}\) . 

We first analyze Procedure \ref{algo:CR} between lines \ref{algo:stage_2_1_start} and \ref{algo:stage_2_1_end}, and show that $\nabla \respondingUtility{\respondingState} \in C=(Q, \pi/2)$ or \(\|\nabla \respondingUtility{\respondingState}\| \leq \frac{\kappa \sqrt{n} d \beta}{2}\). Let \(T_{1}, \ldots, T_{n}\) be the trades offered between lines 3 and 15. We note that \(Q = \sum_{i=1}^{n} T_{i}\). We explore two cases:

\textbf{Case 1: \( \forall i \in [n], \lVert\nabla_{-\Trade_{i}} \respondingUtility{\respondingState}\rVert \leq \frac{(n-1) d \beta}{2}\)} 

Since $\Trade_{i}$ are orthogonal and \(\kappa \geq \sqrt{n-1} \), we trivially have \(\lVert\nabla\respondingUtility{\respondingState}\rVert \leq \frac{\kappa \sqrt{n} d \beta}{2}\) which satisfies property 2 in Theorem 1 due to \eqref{ineq:nearopt} and \eqref{ineq:concavity} and concludes case 1.

\textbf{Case 2: \( \exists i \in [n], \lVert\nabla_{-\Trade_{i}} \respondingUtility{\respondingState}\rVert > \frac{(n-1) d \beta}{2}\)}

Let \(W_{--}\) be the set of \(i\) such that \(\lVert\nabla_{-\Trade_{i}} \respondingUtility{\respondingState}\rVert > \frac{(n-1) d \beta}{2}\). Because \(\lVert\nabla_{-\Trade_{i}} \respondingUtility{\respondingState}\rVert > \frac{ d \beta}{2}\) for all \(i \in W_{--}\), we must have \(\nabla_{-\Trade_{i}} \respondingUtility{\respondingState} < \frac{-(n-1) d \beta}{2}\) for all \(i \in W_{--}\). 

Also let \(W_{+}\) be the set of \(i\) such that \(\nabla_{-\Trade_{i}} \respondingUtility{\respondingState} \geq  0\). Note that \(\nabla_{-\Trade_{i}}\respondingUtility{\respondingState}  \leq \frac{\beta d}{ 2}\) for all \(i \in W_{+}\) as \(T_{i}\) is rejected, \(W_{--} \cap W_{+} = \emptyset\), and \(\nabla_{-\Trade_{i}} \respondingUtility{\respondingState} < 0\) for all \(i \in[n] \setminus (W_{+} \cup W_{--})\). 

We note that if $\nabla \respondingUtility{\respondingState} \not \in C=(Q, \pi/2)$, then we must have $\langle \nabla \respondingUtility{\respondingState}, Q \rangle < 0$, i.e., \(\sum_{i=1}^{n} \nabla_{-\Trade_{i}} \respondingUtility{\respondingState} d > 0\). We have
\begin{align}
\sum_{i=1}^{n} \nabla_{-\Trade_{i}} \respondingUtility{\respondingState} \cdot d 
&= \sum_{i \in W_{+}} \nabla_{-\Trade_{i}} \respondingUtility{\respondingState} \cdot d 
\nonumber \\
&\quad + \sum_{i \in W_{--}} \nabla_{-\Trade_{i}} \respondingUtility{\respondingState} \cdot d 
\nonumber \\
&\quad + \sum_{i \notin (W_{+} \cup W_{--})} \nabla_{-\Trade_{i}} \respondingUtility{\respondingState} \cdot d 
\nonumber \\
&\leq \sum_{i \in W_{+}} \nabla_{-\Trade_{i}} \respondingUtility{\respondingState} \cdot d 
\nonumber \\
&\quad + \sum_{i \in W_{--}} \nabla_{-\Trade_{i}} \respondingUtility{\respondingState} \cdot d
\label{eq:decomp_bound_1}
\end{align}

\noindent Note that \(|W_{--}| \geq 1\) and hence \(|W_{+}| \leq n - 1\). Thus,

\begin{align}
\sum_{i=1}^{n} \nabla_{-\Trade_{i}} \respondingUtility{\respondingState} d 
&\leq (n - 1) \cdot \frac{\beta d}{2} \cdot d 
     - (n - 1) \cdot \frac{\beta d}{2} \cdot d \nonumber \\
&\leq 0
\label{eq:decomp_bound_2}
\end{align}

which implies that \(\nabla \respondingUtility{\respondingState} \in C=(Q, \pi/2)\).

We now analyze between lines \ref{algo:stage_2_3_start} and \ref{algo:stage_2_2_end}. We inductively show that $\nabla \respondingUtility{\respondingState} \in C=(Q, \theta)$ holds or \(\|\nabla \respondingUtility{\respondingState}\| \leq \kappa \Smoothness \SamplingDistance \sqrt{n}\). 

Before we start the analysis, we note that after $k$ unaccepted trades, the lowest value \(\theta\) takes is \[\sin^{-1}\left(\left(\sqrt{1-\nicefrac{1}{2\NumItems}}\right)^{\left\lfloor \frac{\UnacceptedOffers - n}{\NumItems-1} \right\rfloor}\right).\] this is because the \(\theta\) is initialized to \(\pi/2\) after \(n\) unaccepted trades and gets scaled according to the rule    $\OpeningAngle \leftarrow \sin^{-1}(\sqrt{1-\frac{1}{2\NumItems}}\sin(\OpeningAngle))$ after every \(n-1\) unaccepted trades between lines lines \ref{algo:stage_2_3_start} and \ref{algo:stage_2_2_end}.

Assume that $\nabla \respondingUtility{\respondingState} \in C=(Q, \theta)$ which implies \(\nabla f^{\respondingAgent}_{\tau} (\respondingState) \geq 0\). As the base case, we have \(\nabla \respondingUtility{\respondingState} \in C=(Q, \pi/2)\).We note that the algorithm makes $n-1$ trade offers that are orthogonal to \(\tau\) and each other, i.e., the offers in \(V \setminus \{\tau\}\) to refine the cone.

\textbf{Case 2.1: $\lVert \nabla f^{\respondingAgent}_{-\Trade} (\respondingState)\rVert < \GradientThreshold \Smoothness \SamplingDistance, \  \forall \Trade \in V$}

We first analyze the case where all directions in \(V\) have a small directional derivative for the responding agent's utility function.

Since the directions in \(V\) are orthogonal, we trivially have \(\lVert\nabla\respondingUtility{\respondingState}\rVert \leq \kappa \Smoothness \SamplingDistance \sqrt{n}\), which satisfies property 2 in Theorem 1 due to \eqref{ineq:nearopt} and \eqref{ineq:concavity}, and concludes case 2.1.

\textbf{Case 2.2: \(\exists T \in V \) such that $\lVert \nabla_{-\Trade} f^{\respondingAgent}(\respondingState)\rVert \geq \GradientThreshold \Smoothness \SamplingDistance $.}

We now analyze the case where at least one of the directions in \(V\) has a directional derivative. For simplicity, let \(V \setminus \{\tau\} = \lbrace T_{1}, \ldots, T_{n-1} \rbrace\).

In this case, we show that the direction with a large enough directional derivative is sufficient to ensure that the new cone encloses the responding agent's gradient despite there being directions that have misinferred signs for the directional derivatives.

We first note that  $\nabla \respondingUtility{\respondingState} \in C=(Q, \theta)$ implies \(\nabla_{\tau} f^{\respondingAgent} (\respondingState) \geq 0\).

Without loss of generality let \({1}, \ldots, {l}\) be the directions such that \(\nabla_{-\Trade_i} f^{\respondingAgent} (\respondingState) > 0\) for all \(i \in 1, \ldots, l\) and \(\nabla _{\Trade_i} f^{\respondingAgent} (\respondingState) \leq 0\) for all \(i \in l+1, \ldots, n-1\). \(T_{1}, \ldots, T_{l}\) are the directions where the sign of the directional derivative is misinferred by the two-point comparison, and the cone is cut using the wrong side of the halfspace. Note that \(\nabla f^{\respondingAgent}_{-\Trade_i} (\respondingState) \leq \frac{\Smoothness d}{2}\) for all \(i \in 1, \ldots, l\). 

The responding agent's gradient \(\nabla f^{\respondingAgent} (\respondingState)\) is equal to
\begin{align}
\nabla f^{\respondingAgent} (\respondingState) 
&= 
\left\lVert \nabla_{\ConeCenter} f^{\respondingAgent} (\respondingState) \right\rVert 
\cdot \frac{\ConeCenter}{\lVert \ConeCenter \rVert} 
\nonumber \\
&\quad - \sum_{i = 1}^{l} 
\left\lVert \nabla_{-\Trade_i} f^{\respondingAgent} (\respondingState) \right\rVert 
\cdot \frac{\Trade_i}{\lVert \Trade_i \rVert}
\nonumber \\
&\quad - \sum_{i = l+1}^{n-1} 
\left\lVert \nabla_{-\Trade_i} f^{\respondingAgent} (\respondingState) \right\rVert 
\cdot \frac{\Trade_i}{\lVert \Trade_i \rVert}
\label{eq:real_gradient}
\end{align}
Consider a hypothetical gradient \(\nabla f^{\respondingAgent}_{\text{hyp}} (\respondingState)\) that is mirrored in the misinferred directions, i.e., \(\nabla f^{\respondingAgent}_{\text{hyp}} (\respondingState)\) is equal to 
\begin{align}
\nabla f^{\respondingAgent}_{\text{hyp}} (\respondingState)
&= 
\left\lVert \nabla_{\ConeCenter} f^{\respondingAgent} (\respondingState) \right\rVert 
\cdot \frac{\ConeCenter}{\lVert \ConeCenter \rVert} 
\nonumber \\
&\quad + \sum_{i = 1}^{l} 
\left\lVert \nabla_{\Trade_i} f^{\respondingAgent} (\respondingState) \right\rVert 
\cdot \frac{\Trade_i}{\lVert \Trade_i \rVert}
\nonumber \\
&\quad - \sum_{i = l+1}^{n-1} 
\left\lVert \nabla_{\Trade_i} f^{\respondingAgent} (\respondingState) \right\rVert 
\cdot \frac{\Trade_i}{\lVert \Trade_i \rVert}
\label{eq:hypothetical_gradient}
\end{align}

Lemma 2 of \cite{karabag2021smooth} shows that when updating the cone using the angle shrinkage rule
\begin{equation}
    \frac{\sin(\OpeningAngle')}{\sin(\OpeningAngle)} = \sqrt{1-\frac{1}{\NumItems}}
\end{equation}
is sufficient to ensure \(\nabla f^{\respondingAgent}_{\text{hyp}} (\respondingState) \in C=(\theta', \tau')\) where \(\tau'\) is equal to the new cone direction as in the lines \ref{algo:cone_center_update_1} and \ref{algo:cone_center_update_2} of Procedure \ref{algo:CR}. We next show that the angle between \(\nabla f^{\respondingAgent}_{\text{hyp}} (\respondingState)\) and \( \nabla f^{\respondingAgent} (\respondingState)\) is bounded above and using the angle shrinkage rule 
\begin{equation}
    \frac{\sin(\OpeningAngle'')}{\sin(\OpeningAngle)} =\sqrt{1-\frac{1}{2\NumItems}}
\end{equation}
is sufficient to ensure \(\nabla f^{\respondingAgent} (\respondingState) \in C=(\theta'', \tau')\).

The angle between\(\nabla f^{\respondingAgent}_{\text{hyp}} (\respondingState)\) and \( \nabla f^{\respondingAgent} (\respondingState)\) is 
\begin{align}
&\angle\big(
\nabla f^{\respondingAgent}_{\text{hyp}} (\respondingState), \ 
\nabla f^{\respondingAgent} (\respondingState)
\big) 
\nonumber \\
&\quad = \cos^{-1} \left(
\frac{
\left\langle 
\nabla f^{\respondingAgent}_{\text{hyp}} (\respondingState), 
\nabla f^{\respondingAgent} (\respondingState)
\right\rangle
}{
\left\lVert \nabla f^{\respondingAgent}_{\text{hyp}} (\respondingState) \right\rVert
\cdot 
\left\lVert \nabla f^{\respondingAgent} (\respondingState) \right\rVert
}
\right)
\label{eq:angle_between_gradients}
\end{align}

We note that the angle increases, i.e., the ratio \[\frac{\langle\nabla f^{\respondingAgent}_{\text{hyp}} (\respondingState), \nabla f^{\respondingAgent} (\respondingState)\rangle}{\lVert\nabla f^{\respondingAgent}_{\text{hyp}} (\respondingState)\rVert\lVert\nabla f^{\respondingAgent} (\respondingState)\rVert}\] decreases when \(\lVert \nabla f^{\respondingAgent}_{-\Trade_i} (\respondingState) \rVert \) for \(i=1, \ldots, l\) increases, and when \(\lVert\nabla f^{\respondingAgent}_{-\Trade_i} (\respondingState) \rVert\) for \(i=l+1, \ldots, n-1\) and  \(\lVert\nabla f^{\respondingAgent}_{\tau} (\respondingState)\rVert\) decreases. Consequently, the maximum angle is achieved only if \(\nabla f^{\respondingAgent}_{-\Trade_i} (\respondingState) = \frac{\beta d}{2}\) for \(i=1, \ldots, l\). To achieve maximum angle, the other directions in \(V\) must have small magnitudes for the directional derivatives. However, \(\exists T \in V \) such that $\lVert \nabla f^{\respondingAgent}_{-T}(\respondingState)\rVert \geq \GradientThreshold \Smoothness \SamplingDistance $. The directions \(T_{1}, \ldots, T_{l}\) does not satisfy this condition as \(\lVert \nabla f^{\respondingAgent}_{-T}(\respondingState)\rVert \leq 
 \frac{d \beta }{2} < \GradientThreshold\). Consequently, among all directions in \(V \setminus \lbrace T_{1}, \ldots, T_{l}\rbrace\), we set the magnitude of all but one directional derivative to \(0\) and one to \(\kappa \Smoothness \SamplingDistance\). Using this construction, we get
\begin{align}
&\angle\big(
\nabla f^{\respondingAgent}_{\text{hyp}} (\respondingState), 
\nabla f^{\respondingAgent} (\respondingState)
\big)
\nonumber \\
&\quad \leq \cos^{-1} \left( 
\frac{
\SamplingDistance^2 \Smoothness^2 (\GradientThreshold^2 - l)
}{
\SamplingDistance^2 \Smoothness^2 (\GradientThreshold^2 + l)
}
\right)
\nonumber \\
&\quad = \cos^{-1} \left( 
\frac{
\GradientThreshold^2 - l
}{
\GradientThreshold^2 + l
}
\right)
\label{eq:angle_bound_ratio}
\end{align}
We note that the right-hand side is maximized when \(l\) is maximized. Since \(\exists T \in V \) such that $\lVert \nabla f^{\respondingAgent}_{-T}(\respondingState)\rVert \geq \GradientThreshold \Smoothness \SamplingDistance $, and the directions \(T_{1}, \ldots, T_{l}\) does not satisfy this condition as \(\lVert \nabla f^{\respondingAgent}_{-T}(\respondingState)\rVert \leq 
 \frac{d \beta }{2} < \GradientThreshold \Smoothness \SamplingDistance\), we have \(l \leq n-1\). Setting \(l = n-1\) gives 
  \begin{equation}
     \angle(\nabla f^{\respondingAgent}_{\text{hyp}} (\respondingState), \nabla f^{\respondingAgent} (\respondingState)) \leq \cos^{-1}{ \left( \frac{  \GradientThreshold^2 - (n-1)}{\GradientThreshold^2 +(n-1)} \right)}.
\end{equation}

If \[\theta'' - \theta' \geq \cos^{-1}{ \left( \frac{  \GradientThreshold^2 - (n-1)}{ \GradientThreshold^2 - (n-1)} \right)},\] then \(\nabla f^{\respondingAgent} (\respondingState) \in C=(\theta'', \tau')\). This is because \(\nabla f^{\respondingAgent}_{\text{hyp}} (\respondingState) \in C=(\theta', \tau')\) and $$ \angle(\nabla f^{\respondingAgent}_{\text{hyp}} (\respondingState), \nabla f^{\respondingAgent} (\respondingState)) \leq \cos^{-1}{ \left( \frac{ \GradientThreshold^2 - (n-1)}{ \GradientThreshold^2 + (n-1)} \right)}.$$

Since \(\kappa\) satisfies 
\begin{equation*} \sqrt{1-\frac{1}{2\NumItems}}^{\left\lfloor \frac{\UnacceptedOffers-n}{\NumItems-1} \right\rfloor} = 4\NumItems\sqrt{1-\left(\frac{ \GradientThreshold^2 - (n-1)}{\GradientThreshold^2 + (n-1)}\right)^2},
\end{equation*} by Lemma \ref{lm:1}, we have 
    \begin{equation*}
\theta'' - \theta' \geq \cos^{-1}{ \left( \frac{ \GradientThreshold^2 - (n-1)}{ \GradientThreshold^2 + (n-1)} \right)}
    \end{equation*}\
    for all \(\theta \in \left[\sin^{-1}\left(\left(\sqrt{1-\frac{1}{2\NumItems}}\right)^{\left\lfloor \frac{\UnacceptedOffers - n}{\NumItems-1} \right\rfloor}\right), \frac{\pi}{2}\right]\). Since the the lowest value \(\theta\) takes is \[\sin^{-1}\left(\left(\sqrt{1-\frac{1}{2\NumItems}}\right)^{\left\lfloor \frac{\UnacceptedOffers - n}{\NumItems-1} \right\rfloor}\right),\] we have \[\theta'' - \theta' \geq \cos^{-1}{ \left( \frac{  \GradientThreshold^2 - (n-1)}{ \GradientThreshold^2 + (n-1)} \right)}\] after every cone refinement which implies that \(\nabla f^{\respondingAgent} (\respondingState) \in C=(\theta, \tau)\) after \(k\) unaccepted offers. In other words, the responding agent's gradient is contained within the cone maintained by the algorithm. 

    We note that after a rejected trade offer \(T\), each cut made to the cone discards the halfspace that is the polar cone of \(-T\) and preserves the halfspace that is the dual cone of \(T\). Since \(\langle \nabla \OfferingUtility{\OfferingState}, \Trade \rangle \geq 0\) for every made offer \(T\), and the new cones always preserve the remaining directions of the cone after cuts, the offering agents gradient is also contained in \(\nabla f^{\OfferingAgent} (\OfferingState) \in C=(\theta, \tau)\) for every cone refinement. 

    Since both \(\nabla f^{\respondingAgent} (\respondingState) \in C=(\theta, \tau)\) and \(\nabla f^{\OfferingAgent} (\OfferingState) \in C=(\theta, \tau)\), and the cone has angle \(\theta = \sin^{-1}\left(\left(\sqrt{1-\frac{1}{2\NumItems}}\right)^{\left\lfloor \frac{\UnacceptedOffers - n}{\NumItems-1} \right\rfloor}\right)\) after \(k\) unaccepted trades, we have    \begin{equation*} 
        \begin{split}
        \angle(\nabla \OfferingUtility{\OfferingState}, \nabla \respondingUtility{\respondingState}) <  2 \sin^{-1}\left(\left(\sqrt{1-\frac{1}{2\NumItems}}\right)^{\left\lfloor \frac{\UnacceptedOffers-n}{\NumItems-1} \right\rfloor}\right).
        \end{split}
    \end{equation*} which satisfies property 1 in Theorem \ref{thm} and concludes case 2.2.

\end{proof}

\begin{lemma}
    Let \(n\geq 2\), \(\gamma \leq 1\), \(\theta \in (0, \pi/2]\), \(\sin(\OpeningAngle') = \sqrt{1-\frac{1}{\NumItems}} \sin(\OpeningAngle)\), and  \(\sin(\OpeningAngle'') = \sqrt{1-\frac{1}{2\NumItems}} \sin(\OpeningAngle)\). If
    \begin{equation}
        \sin{(\OpeningAngle)} \sqrt{1-\frac{1}{2\NumItems}} \geq 4\NumItems\sqrt{1-\AngleIncrease^2},
        \label{eq:theta_2}
    \end{equation} then 
    \begin{equation}
        \theta'' - \OpeningAngle' \geq  + \cos^{-1}{(\AngleIncrease)}.
    \end{equation}
    \label{lm:1}
\end{lemma}

\begin{proof}[Proof of Lemma \ref{lm:1}]

We have
\begin{align*}
    \frac{\sin(\OpeningAngle' + \cos^{-1}(\AngleIncrease))}{\sin(\OpeningAngle)}
    &= \frac{\AngleIncrease \sin(\OpeningAngle')}{\sin(\OpeningAngle)} + \frac{\cos(\OpeningAngle')\sqrt{1-\AngleIncrease^2}}{\sin(\OpeningAngle)} 
    \\
    &= \AngleIncrease \sqrt{1-\frac{1}{\NumItems}} + \frac{cos(\OpeningAngle')\sqrt{1-\AngleIncrease^2}}{\sin(\OpeningAngle)}.
\end{align*}

Using $\AngleIncrease \leq 1$ and $\cos(\OpeningAngle') \leq 1$, we get

\begin{equation}
    \sin{(\OpeningAngle' + \cos^{-1}{\AngleIncrease})} \leq \sin{(\OpeningAngle)}\sqrt{1-\frac{1}{\NumItems}} + \sqrt{1 - \AngleIncrease^2}.
\end{equation}

Using     
\begin{equation*}
        \sin{(\OpeningAngle)} \sqrt{1-\frac{1}{2\NumItems}} \geq 4\NumItems\sqrt{1-\AngleIncrease^2}
    \end{equation*} we get
    
\begin{align}
\sin\left( \OpeningAngle' + \cos^{-1}(\AngleIncrease) \right)
&\leq \sin(\theta'') 
\cdot \sqrt{1 - \tfrac{1}{\NumItems}} 
\cdot \sqrt{ \left( 1 - \tfrac{1}{2\NumItems} \right)^{-1} } 
\nonumber \\
&\quad + \sin(\theta'') 
\cdot \sqrt{1 - \tfrac{1}{2\NumItems}} 
\cdot \tfrac{1}{4\NumItems}
\nonumber \\
&= \left(
\sqrt{ \frac{2\NumItems^2 - 2\NumItems}{2\NumItems^2 - \NumItems} } 
+ \frac{1}{4\NumItems}
\right) \cdot \sin(\theta'')
\label{eq:sin_bound_simplified}
\end{align}

Since $\NumItems \geq 2$, we have $ \sqrt{\frac{2\NumItems^2 - 2\NumItems}{2\NumItems^2 - \NumItems}} + \frac{1}{4\NumItems} < 1$. Therefore, we obtain

\begin{equation}
    \sin{(\OpeningAngle' + \cos^{-1}{\AngleIncrease})} \leq \left(\sqrt{\frac{2\NumItems^2 - 2\NumItems}{2\NumItems^2 - \NumItems}} + \frac{1}{4\NumItems}\right) \sin{(\theta'')} \leq \sin{(\theta'')}
\end{equation} which implies
\begin{equation}
    \theta'' \geq \OpeningAngle' + \cos^{-1}{(\AngleIncrease)}.
\end{equation}

\end{proof}

\subsection{\texorpdfstring{Corollary: Theorem~1 Implies \(\epsilon\)-Weak Pareto Optimality}{Corollary: Theorem~1 Implies ε-Weak Pareto Optimality}}

We now connect the conditions in Theorem~1 to approximate Pareto efficiency. Specifically, we show that the bounds provided in Theorem~1 imply \(\epsilon\)-weak Pareto optimality under certain conditions.

\paragraph{\(\epsilon\)-Weak Pareto Optimality}
A joint state \((\OfferingState^*, \respondingState^*)\) is \emph{\(\epsilon\)-weakly Pareto optimal} for some \(\epsilon > 0\) if there exists no feasible trade \( \Trade \in \mathcal{T} \) such that both agents improve by more than \(\epsilon\). That is, for all feasible \(\Trade\),
\[
f^A(\OfferingState^* + \Trade) - f^A(\OfferingState^*) \leq \epsilon \quad \text{or} 
\]\[f^B(\respondingState^* - \Trade) - f^B(\respondingState^*) \leq \epsilon.
\]
We now derive such bounds from the two cases considered in Theorem~1.

\paragraph*{Case 1: Gradients are Closely Aligned}
We begin by introducing the threshold angle \( \theta_c \), which serves as an upper bound on the angle between the agents’ gradients. Specifically, Theorem~1 analyzes the case where this angle is bounded as follows
\[
\angle(\nabla f^A(S_A), \nabla f^B(S_B)) \leq \theta_c.
\]
The value of \( \theta_c \) is defined as
\[
\theta_c = 2 \sin^{-1}\left(\left(\sqrt{1 - \frac{1}{2\NumItems}}\right)^{\left\lfloor \frac{\UnacceptedOffers - n}{\NumItems - 1} \right\rfloor}\right).
\]
We analyze the implications for the offering agent \( A \). Since $f^A$ is concave, we know that
\[
f^A(S_A + T) - f^A(S_A) \leq \langle \nabla f^A(S_A), T \rangle
\]
For a trade \( T \) that is mutually beneficial, it must lie within the intersection of the ascent halfspaces defined by \( \langle \nabla f^A(S_A), T \rangle > 0 \) and \( \langle \nabla f^B(S_B), -T \rangle > 0 \). In detail, consider a trade $T$ such that $\langle \nabla f^A(S_A), T \rangle < 0$ or $\langle \nabla f^B(S_B), -T \rangle < 0$. Due to the concavity of $f^A$ and $f^B$,  $f^A(S_A) > f^A(S_A + T)$ or similarly $f^B(S_B) > f^B(S_B - T)$. 

Note that $\angle(\nabla f^A(S_A), \nabla f^B(S_B)) \leq \theta_c$. Consequently, a trade $\Trade$ satisfying \( \langle \nabla f^A(S_A), T \rangle > 0 \) and \( \langle \nabla f^B(S_B), -T \rangle > 0 \), satisfies 
\[
\angle(T, \nabla f^A(S_A)) \geq \frac{\pi}{2} - \theta_c
\]
and 
\[
\angle(-T, \nabla f^B(S_B)) \geq \frac{\pi}{2} - \theta_c
\]

We have
\[
\langle \nabla f^A(S_A), T \rangle \leq \|\nabla f^A(S_A)\| \|T\|  \sin(\theta_c).
\]
\[
f^A(S_A + T) - f^A(S_A) \leq \|T\| \|\nabla f^A(S_A)\| \sin(\theta_c)
\]
If the function is $L$-Lipschitz continuous, assuming \( \max \|T\| \leq \delta \), we get
\[
f^A(S_A + T) - f^A(S_A) \leq \delta L \sin(\theta_c)
\]
This allows us to set the $\epsilon$ bound as
\[
\epsilon =  \delta L \sin(\theta_c).
\]
Same logic applies if $f^{B}$ is $L$-Lipschitz continuous instead of $f^{A}$. 
\paragraph*{Case 2: Responding Agent is Near-Optimal}
In Theorem~1, the responding agent’s utility gain under any trade \( T \) from state \( S_B \) is bounded by
\[
f^B(S_B - T) - f^B(S_B) \leq d \kappa \sqrt{n} \beta \|T\|,
\]
due to concavity where \( d \) is the sampling distance during cone refinement, \( \kappa \) is a constant depending on \( k \) and \( n \), and \( \beta \) is the smoothness constant of \( f^B \). Let \( \delta \) be the maximum feasible trade magnitude (bounded by the size of the state space), so that \( \max \|T\| \leq \delta \). Then
\[
f^B(S_B - T) - f^B(S_B) \leq d \kappa \sqrt{n} \beta \delta.
\]
This allows us to set the $\epsilon$ bound as
\[
\epsilon = d \kappa \sqrt{n} \delta.
\]

\subsection{Proof of Asymptotic Convergence}
\label{appendix:convergence}
\begin{theorem}[Asymptotic Convergence under Aligned Trades]
\label{thm:asymptotic_convergence}

Let $f^A,f^B:\mathbb{R}^d\to\mathbb{R}$ be $\mu$-strongly concave and $\beta$-smooth.

\paragraph{Assumptions.}
\begin{enumerate}
    \item \textbf{Trade updates.}  
    At each accepted trade step $t$, a nonzero trade vector $T_t\in\mathbb{R}^d$ is chosen, with step size
    \[
    d_t := \|T_t\| > 0,
    \]
    and the agents update their states according to
    \[
    S_{A,t+1} = S_{A,t} + T_t,
    \qquad
    S_{B,t+1} = S_{B,t} - T_t.
    \]

    \item \textbf{Angular alignment.}  
    There exists $\hat{\theta}\in[0,\pi/2)$ such that, at each step $t$, the trade direction satisfies at least one of
    \[
    \angle(T_t,\nabla f^A(S_{A,t})) \leq \hat{\theta},
    \qquad\text{or}\qquad
    \angle(-T_t,\nabla f^B(S_{B,t})) \leq \hat{\theta}.
    \]

    \item \textbf{Step-size conditions.}  
    The step sizes satisfy the standard diminishing-steps conditions
    \[
    \sum_{t=0}^\infty d_t = \infty,
    \qquad
    \sum_{t=0}^\infty d_t^2 < \infty.
    \]

    \item \textbf{Bounded utilities.}  
    Each agent’s utility is uniformly bounded above: there exists $f^{\max}<\infty$ such that
    \[
    f^A(S_A) \le f^{\max},
    \qquad
    f^B(S_B) \le f^{\max}
    \quad\text{for all } S_A,S_B.
    \]
\end{enumerate}

\paragraph{Conclusion.}
Under these assumptions, there exists at least one agent
$X \in \{A,B\}$ for which the cumulative step sizes of aligned trades diverge:
\[
\sum_{t:\, \angle(T_t,\nabla f^X(S_{X,t})) \le \hat{\theta}} d_t = \infty.
\]
For that agent $X$, the state sequence $\{S_{X,t}\}$ converges to a finite limit, and
\[
\lim_{t\to\infty} \|\nabla f^X(S_{X,t})\| = 0.
\]
\end{theorem}

\subsubsection{Preliminaries}

The convergence proof relies on a result concerning sequences that exhibit approximate ascent with summable perturbations. We restate below a version of Lemma~1 from Bertsekas and Tsitsiklis~\cite{bertsekas2000gradient}, adapted to the notation used in this manuscript and adjusted for maximization instead of minimization.

\begin{lemma}[Bertsekas and Tsitsiklis -- ascent form]
\label{lemma:b&t}
Let $\{Y_t\}$, $\{W_t\}$, and $\{Z_t\}$ be sequences of real numbers with $W_t \ge 0$ for all $t$. Suppose that
\[
Y_{t+1} \ge Y_t + W_t - Z_t, \quad t = 0,1,2,\ldots
\]
and that the series $\sum_{t=0}^{\infty} Z_t$ converges. Then either
\[
Y_t \to +\infty,
\]
or else $Y_t$ converges to a finite value and
\[
\sum_{t=0}^{\infty} W_t < \infty.
\]
\end{lemma}

This lemma implies that if a sequence admits a descent relation up to a summable perturbation, then either the sequence diverges to $-\infty$ or the cumulative descent term must be finite.

\subsubsection{Convergence Proof}
\begin{proof}
Fix an agent $X\in\{A,B\}$ such that
\[
\sum_{t:\,\angle(\pm T_t,\nabla f^X(S_{X,t}))\le\hat\theta} d_t = \infty,
\]
where the sign is chosen so that $\pm T_t$ is the update applied to agent $X$.

\paragraph*{Step 1: One-step ascent bound.}
By $\mu$-strong concavity, for any $x,d$:
\[
f^X(S_{X,t+1}) \ge f^X(S_{X,t})
+ \langle\nabla f^X(S_{X,t}),\pm T_t\rangle
+ \frac{\mu}{2} d_t^2.
\]
Whenever $\angle(\pm T_t,\nabla f^X(S_{X,t}))\le\hat\theta$,
\[
\langle\nabla f^X(S_{X,t}),\pm T_t\rangle
\ge d_t\cos(\hat\theta)\|\nabla f^X(S_{X,t})\|.
\]

\paragraph*{Step 2: Approximate ascent with summable error.}
Define
\[
\begin{aligned}
Y_t &:= f^X(S_{X,t}), \\
W_t &:= d_t \cos(\hat\theta)\,\|\nabla f^X(S_{X,t})\|
       \mathbf{1}_{\{t\in\mathcal T_X\}}, \\
Z_t &:= -\tfrac{\mu}{2} d_t^2 .
\end{aligned}
\]
Then
\[
Y_{t+1} \ge Y_t + W_t - Z_t.
\]
Since $\sum_t Z_t<\infty$ and $Y_t$ is bounded above,
Lemma~\ref{lemma:b&t} implies that $Y_t$ converges and
\[
\sum_{t=0}^\infty W_t < \infty.
\]

\paragraph*{Step 3: Vanishing gradient.}
If $\|\nabla f^X(S_{X,t})\|$ does not converge to zero, then there exists
$\varepsilon>0$ such that
$\|\nabla f^X(S_{X,t})\|\ge\varepsilon$ infinitely often on $\mathcal T_X$.
Since $\sum_{t\in\mathcal T_X} d_t=\infty$, this contradicts the finiteness of
$\sum_t W_t$. Hence
\[
\|\nabla f^X(S_{X,t})\|\to 0.
\]
\end{proof}
\subsection{When the Angular Assumption Cannot Be Certified}
\label{appendix:angular}

This section characterizes the regime in which the angular ascent
assumption used in the convergence analysis cannot be certified.
We show that failure of the angular bound arises in cases that admit
explicit certificates of $\epsilon$-weak Pareto optimality.

Let
\[
g_A := \nabla f^A(S_A),
\qquad
g_B := \nabla f^B(S_B)
\]
denote the agents’ gradients, and let $\tau$ denote the axis of the
responding agent’s feasibility cone during refinement.

We assume throughout:
\begin{itemize}
\item \textbf{(Differentiability)} Each $f^i$ is differentiable on the feasible region.
\item \textbf{(Concavity/strong concavity)} Each $f^i$ is concave (and $\mu$-strongly concave, if desired).
\item \textbf{(Smoothness)} Each $f^i$ is $\beta$-smooth.
\item \textbf{(Lipschitzness)} Each $f^i$ is $L$-Lipschitz on the feasible region (equivalently $\|\nabla f^i(x)\|\le L$ for all feasible $x$).
\item \textbf{(Feasible step size)} Feasible trades satisfy $\max\|T\| = \delta$.
\item \textbf{(Target angular condition)} The target angular condition is
$\angle(T,g_A)\le \hat\theta$ or $\angle(-T,g_B)\le \hat\theta$ for some offered trade $T$.
\end{itemize}

\begin{theorem}[Either angular certification or $\epsilon$-weak Pareto optimality]
\label{thm:angular_or_epsilon_appdx}
Consider one iteration of ST-CR at state $(S_A,S_B)$ with gradients
$\nabla f^A(S_A)$ and $\nabla f^B(S_B)$. 
Assume each $f^i$ is $\mu$-strongly concave, $\beta$-smooth, and $L$-Lipschitz in the feasible region. 
Let $\delta := \max \|T\|$ denote the maximum feasible trade magnitude.

For every $\hat\theta \in \bigl(\cos^{-1}(1/\sqrt{n}), {\pi}/{2}\bigr)$, at least one of the following holds:
\begin{enumerate}
    \item \textbf{Angular assumption holds (certifiable alignment).}
    There exists an offer $T \in \bar V$ such that
    \[
    \angle\!\bigl(T,\nabla f^A(S_A)\bigr)\le \hat\theta
    \qquad\text{or}\qquad
    \angle\!\bigl(-T,\nabla f^B(S_B)\bigr)\le \hat\theta.
    \]

    \item \textbf{$\epsilon$-weak Pareto optimality certificate.}
    The current state $(S_A,S_B)$ is $\epsilon$-weakly Pareto optimal with
    \[
    \epsilon
    := \max\{\epsilon_{\text{alignment}}, \epsilon_{\text{responding}}\},
    \]
    where
    \[
    \epsilon_{\text{alignment}} = 
    \delta\,L\,
    \sin\!\left(
        \min\left\{
            2\sin^{-1}\!\bigl(\sqrt{n-1}\cos(\hat\theta)\bigr),
            \frac{\pi}{2}
        \right\}
    \right).
    \]
    \[
    \epsilon_{\text{responding}} = 
    \frac{d \delta \beta (n-1)\sqrt{n}}{2}
    \]
\end{enumerate}
\end{theorem}
\begin{proof}
We show that if the angular condition cannot be certified, then the current state is
$\epsilon$-weakly Pareto optimal with the stated value.

\paragraph*{Step 1: Initialization always gives a uniform angle bound for $\nabla f^A(S_A)$.}

During initialization, ST-CR makes $n$ mutually orthogonal offers
$T_1,\dots,T_n \in \mathbb{R}^n$.
Fix any nonzero gradient $\nabla f^A(S_A) \in \mathbb{R}^n$.
Since the $T_i$ are mutually orthogonal and nonzero, the normalized vectors
\[
\frac{T_i}{\|T_i\|}, \quad i=1,\dots,n,
\]
form an orthonormal set (and hence an orthonormal basis) of $\mathbb{R}^n$. Hence,
\[
\sum_{i=1}^n
\left\langle
\nabla f^A(S_A), \frac{T_i}{\|T_i\|}
\right\rangle^2
\;=\;
\|\nabla f^A(S_A)\|^2.
\]
Dividing both sides by $\|\nabla f^A(S_A)\|^2$ gives
\[
\sum_{i=1}^n
\left(
\frac{
\left\langle
\nabla f^A(S_A), \frac{T_i}{\|T_i\|}
\right\rangle
}{
\|\nabla f^A(S_A)\|
}
\right)^2
\;=\;
1.
\]
Since the sum of the $n$ nonnegative terms equals $1$, they cannot all be strictly less than $1/n$. Indeed, if
\[
\left(
\frac{
\left\langle
\nabla f^A(S_A), \frac{T_i}{\|T_i\|}
\right\rangle
}{
\|\nabla f^A(S_A)\|
}
\right)^2
<
\frac{1}{n}
\quad
\text{for all } i=1,\dots,n,
\]
then summing over $i$ would give
\[
\sum_{i=1}^n
\left(
\frac{
\left\langle
\nabla f^A(S_A), \frac{T_i}{\|T_i\|}
\right\rangle
}{
\|\nabla f^A(S_A)\|
}
\right)^2
<
n \cdot \frac{1}{n}
=
1,
\]
contradicting the equality above. Therefore, there exists an index $i$ such that
\[
\left(
\frac{
\left\langle
\nabla f^A(S_A), \frac{T_i}{\|T_i\|}
\right\rangle
}{
\|\nabla f^A(S_A)\|
}
\right)^2
\;\ge\;
\frac{1}{n},
\]
which implies
\[
\frac{
\left|
\left\langle
\nabla f^A(S_A), \frac{T_i}{\|T_i\|}
\right\rangle
\right|
}{
\|\nabla f^A(S_A)\|
}
\;\ge\;
\frac{1}{\sqrt{n}}.
\]
By definition of the angle,
\[
\cos\angle(\nabla f^A(S_A), T_i)
=
\frac{
\left\langle
\nabla f^A(S_A), \frac{T_i}{\|T_i\|}
\right\rangle
}{
\|\nabla f^A(S_A)\|
}.
\]
Hence,
\[
|\cos\angle(\nabla f^A(S_A), T_i)|
\;\ge\;
\frac{1}{\sqrt{n}}.
\]
Since the offers are only accepted if they improve an agent's utility, we choose the sign of $T_i$ so that
\[
\cos\angle(\nabla f^A(S_A), T_i)
\;\ge\;
\frac{1}{\sqrt{n}},
\]
which implies
\[
\angle(\nabla f^A(S_A), T_i)
\;\le\;
\cos^{-1}\!\left(\frac{1}{\sqrt{n}}\right).
\]

\paragraph*{Step 2: Refinement angle identities (offering agent).}

Define the $(n-1)$–dimensional refinement hyperplane orthogonal to the cone axis by
\[
H := \{ v \in \mathbb{R}^n : \langle v,\tau\rangle = 0 \}.
\]
During refinement, the offers $T_1,\dots,T_{n-1}$ lie in $H$ and are mutually orthogonal.
Assume $\nabla f^A(S_A) \neq 0$.

Denote the orthogonal projection of $\nabla f^A(S_A)$ onto $H$ by
\[
\overline{\nabla f^A(S_A)}
:=
\nabla f^A(S_A) - \frac{\langle \nabla f^A(S_A),\tau\rangle}{\|\tau\|^2}\tau.
\]

Since the normalized vectors $T_i/\|T_i\|$ form an orthonormal set (and hence an orthonormal basis) of $H$, applying the argument from Step~1 within $H$ yields an index $i$ such that
\[
\cos\angle\!\Bigl(\overline{\nabla f^A(S_A)},\, T_i\Bigr)
\;\ge\;
\frac{1}{\sqrt{n-1}},
\]
and therefore
\[
\angle\!\Bigl(\overline{\nabla f^A(S_A)},\, T_i\Bigr)
\;\le\;
\cos^{-1}\!\left(\frac{1}{\sqrt{n-1}}\right).
\]

Moreover, because $T_i \in H$, we have $\langle \tau, T_i\rangle = 0$, and hence
\[
\begin{aligned}
\bigl\langle \nabla f^A(S_A), T_i\bigr\rangle
&=
\left\langle
\overline{\nabla f^A(S_A)}
+ \frac{\langle \nabla f^A(S_A),\tau\rangle}{\|\tau\|^2}\tau,
\, T_i
\right\rangle \\
&=
\bigl\langle \overline{\nabla f^A(S_A)}, T_i\bigr\rangle .
\end{aligned}
\]
Thus,
\[
\begin{aligned}
\cos\angle\!\bigl(\nabla f^A(S_A),T_i\bigr)
&=
\frac{\bigl\langle \nabla f^A(S_A),T_i\bigr\rangle}
     {\|\nabla f^A(S_A)\|\,\|T_i\|} \\
&=
\frac{\bigl\langle \overline{\nabla f^A(S_A)},T_i\bigr\rangle}
     {\|\nabla f^A(S_A)\|\,\|T_i\|} \\
&=
\frac{\bigl\|\overline{\nabla f^A(S_A)}\bigr\|}
     {\|\nabla f^A(S_A)\|}
\,\cos\angle\!\Bigl(\overline{\nabla f^A(S_A)},T_i\Bigr).
\end{aligned}
\]

Since $\overline{\nabla f^A(S_A)}$ is the orthogonal projection of $\nabla f^A(S_A)$ onto $H$,
\[
\frac{\bigl\|\overline{\nabla f^A(S_A)}\bigr\|}{\|\nabla f^A(S_A)\|}
=
\cos\angle\!\Bigl(\nabla f^A(S_A),\overline{\nabla f^A(S_A)}\Bigr)
=
\sin\angle\!\bigl(\nabla f^A(S_A),\tau\bigr).
\]
Therefore,
\begin{equation}\label{eq:refine-lowerbound-A}
\cos\angle\!\bigl(\nabla f^A(S_A),T_i\bigr)
\;\ge\;
\sin\angle\!\bigl(\nabla f^A(S_A),\tau\bigr)\,
\frac{1}{\sqrt{n-1}}.
\end{equation}

In particular, if
\[
\sin\angle\!\bigl(\nabla f^A(S_A),\tau\bigr)\,
\frac{1}{\sqrt{n-1}}
\;\ge\;
\cos\hat\theta,
\]
then \eqref{eq:refine-lowerbound-A} implies
\[
\cos\angle\!\bigl(\nabla f^A(S_A),T_i\bigr)\ge\cos\hat\theta,
\qquad
\text{i.e.}
\qquad
\angle\!\bigl(\nabla f^A(S_A),T_i\bigr)\le \hat\theta.
\]
Hence, if the angular condition cannot be certified for agent $A$, then necessarily
\begin{equation}\label{eq:starA}
\begin{aligned}
\frac{\sin\angle\!\bigl(\nabla f^A(S_A),\tau\bigr)}
     {\sqrt{n-1}}
&< \cos\hat\theta \\
&\Longleftrightarrow
\angle\!\bigl(\nabla f^A(S_A),\tau\bigr)
< \arcsin\!\bigl(\sqrt{n-1}\cos\hat\theta\bigr).
\end{aligned}
\tag{$\star_A$}
\end{equation}

We note that the condition
\[
\sin\angle\!\bigl(\nabla f^A(S_A),\tau\bigr)\,\frac{1}{\sqrt{n-1}} \;\ge\; \cos\hat\theta
\]
is satisfiable only if
\[
\begin{aligned}
\cos\hat\theta
&\le \frac{1}{\sqrt{n-1}}, \\
\text{equivalently}\quad
\hat\theta
&\ge \cos^{-1}\!\left(\frac{1}{\sqrt{n-1}}\right).
\end{aligned}
\]
since $\sin\angle\!\bigl(\nabla f^A(S_A),\tau\bigr)\le 1$.

Applying the same argument to the responding agent $B$ yields, if the angular condition cannot be certified for agent $B$,
\begin{equation}\label{eq:starB}
\angle\!\bigl(\nabla f^B(S_B),\tau\bigr)
\;<\;
\sin^{-1}\!\bigl(\sqrt{n-1}\cos\hat\theta\bigr).
\tag{$\star_B$}
\end{equation}

\paragraph*{Step 3: Deriving $\epsilon$ bounds.}
Suppose the angular condition cannot be certified for either agent. Then $(\star_A)$ and $(\star_B)$ both hold, and by the triangle inequality,
\[
\begin{aligned}
\angle\!\bigl(\nabla f^A(S_A),\nabla f^B(S_B)\bigr)
&\le \angle\!\bigl(\nabla f^A(S_A),\tau\bigr)
   + \angle\!\bigl(\nabla f^B(S_B),\tau\bigr) \\
&< 2\sin^{-1}\!\bigl(\sqrt{n-1}\cos(\hat\theta)\bigr) \\
&=: \theta_c .
\end{aligned}
\]
Moreover, Step~1 shows that for each agent there exists an offer
satisfying the angular condition with 
$\hat\theta = \cos^{-1}(1/\sqrt{n})$.
Hence, if the algorithm proceeds past initialization without
certifying the angular condition for either agent, it follows from
the initialization rule that both gradients must assign the same sign
pattern to the orthogonal offer directions. In particular, the
gradients lie in the same orthant.
Consequently, we obtain the geometric bound
\[
\angle\!\bigl(\nabla f^A(S_A),\nabla f^B(S_B)\bigr)
\le \frac{\pi}{2}.
\]

Assume that this angular upper bound holds, we then define
\[
\tilde\theta_c \;:=\; \min\left\{\theta_c,\frac{\pi}{2}\right\}.
\]
Then
\[
\theta:=\angle\!\bigl(\nabla f^A(S_A),\nabla f^B(S_B)\bigr)\le \tilde\theta_c.
\]

Now consider any feasible trade $T$ that is first-order improving for both agents:
\[
\left\langle \nabla f^A(S_A),T\right\rangle>0,
\qquad
\left\langle \nabla f^B(S_B),-T\right\rangle>0,
\]
equivalently $\langle \nabla f^A(S_A),T\rangle>0$ and $\langle \nabla f^B(S_B),T\rangle<0$.
Any such $T$ satisfies
\[
\angle\!\bigl(\nabla f^A(S_A),T\bigr)\ \ge\ \frac{\pi}{2}-\theta,
\qquad
\angle\!\bigl(\nabla f^B(S_B),-T\bigr)\ \ge\ \frac{\pi}{2}-\theta,
\]
and therefore
\[
\begin{aligned}
\left\langle \nabla f^A(S_A),T\right\rangle
&=
\|\nabla f^A(S_A)\|\,\|T\|
   \cos\angle\!\bigl(\nabla f^A(S_A),T\bigr) \\
&\le
\|\nabla f^A(S_A)\|\,\|T\|\,\sin(\theta) \\
&\le
\|\nabla f^A(S_A)\|\,\|T\|\,\sin(\tilde\theta_c).
\end{aligned}
\]

Since $f^A$ is concave,
\[
\begin{aligned}
f^A(S_A+T)-f^A(S_A)
&\le \left\langle \nabla f^A(S_A),T\right\rangle \\
&\le \|\nabla f^A(S_A)\|\,\|T\|\,
      \sin(\tilde\theta_c).
\end{aligned}
\]
By $L$-Lipschitzness we have $\|\nabla f^A(S_A)\|\le L$, and by feasibility $\|T\|\le \delta$, hence
\[
f^A(S_A+T)-f^A(S_A)\le \delta\,L\,\sin(\tilde\theta_c).
\]
Thus no feasible trade can improve both agents by more than
\[
\begin{aligned}
\epsilon_{\text{alignment}}
&:= \delta\,L\,\sin(\tilde\theta_c) \\
&= \delta\,L\,
   \sin\!\left(
      \min\left\{
         2\sin^{-1}\!\bigl(\sqrt{n-1}\cos(\hat\theta)\bigr),
         \frac{\pi}{2}
      \right\}
   \right).
\end{aligned}
\]
which certifies $\epsilon$-weak Pareto optimality.

Now we consider the case where the initialization trades lead to an incorrect orthant being selected for the gradient cone. The proof of Theorem \ref{thm} shows that this can only happen if the directional derivatives in all trade directions $T_i$ satisfy
\[
\forall i \in [n], \lVert\nabla_{-\Trade_{i}} \respondingUtility{\respondingState}\rVert \leq \frac{(n-1) d \beta}{2}
\]

This means that the gradient magnitude for the responding agent satisfies
\[
\|\nabla f^B (S_B)\| \leq \frac{(n-1) d \beta \sqrt{n}}{2}
\]

Due to the concavity of $f^B$, we obtain 
\[
\begin{aligned}
f^B(S_B - T) - f^B(S_B)
&\le \|f^B(S_B)\|\,\|T\| \\
&\le \frac{d\,\delta\,\beta\,(n-1)\sqrt{n}}{2} \\
&=: \epsilon_{\text{responding}} .
\end{aligned}
\]

We therefore obtain $\epsilon$-weak Pareto-Optimality where:
\[
\epsilon = \max\{\epsilon_\text{alignment}, \epsilon_{\text{responding}}\}
\]
\end{proof}

\subsection{Cone Update in Discrete Settings}
\begin{algorithm}[H]
  \SetAlgoNlRelativeSize{0}   
  \SetAlgoLined
  \setcounter{AlgoLine}{0}
  \KwIn{$\OfferingState, f_{\OfferingAgent}, R_{\respondingAgent}, d, \Theta^{*}$}
  \KwOut{$\Trade, \ConeCenter, \OpeningAngle, n$}
  \SetKwRepeat{Do}{do}{while}
  \tcp{Initialize quadrant using the same approach as non-integer constrained algorithm}

  $\ConeCenter \leftarrow \Quadrant$ \;
  $\OpeningAngle \leftarrow \cos^{-1}{\frac{\pi}{2}}$ \;
  $V \leftarrow \{\ConeCenter\}$ \;
  \tcp{Initialize basis vectors for hyperplane with normal $\tau$. We use QR decomposition to obtain the set of basis vectors.}
  $V_{\text{b}} \leftarrow \{v_1, \ldots v_{n-1}\}$, $M = [v_1, \ldots, v_{n-1}]$ \;
  
  \Do{$\ResponseFunction{\Trade} \neq \text{accept or }  \OpeningAngle \geq \Theta^{*}$}{
    $V \leftarrow V \cup \{\frac{\Trade}{\lVert \Trade \rVert} \}$\;
    
    \tcp{Calculate New Cone}
    
    \If{$|V| \geq n$}{
      \tcp{Calculate New Cone}
      Set $e_1^\top = [1, 0, \ldots, 0], e_2^\top = [0, 1, \ldots, 0], \ldots e_{\NumItems-1}^\top = [0, 0, \ldots, 1]$\;
      
      $H_{\text{cone}} \leftarrow \{x\in \Reals^{\NumItems-1}: \pm e_i^T x \geq -|\tan{\theta}| \quad \text{for } i \leftarrow 1, 2, \ldots, n -1 \}$\; \label{algo:hypercube}
      
       For each trade $T \in V \setminus \tau$, set $g_{i} = \frac{\langle \Trade, v_i \rangle }{\| v_i \|^2}, \forall v_i \in V_{b}.$ \;
       
      $H_{\text{offers}} \leftarrow \{ x\in \Reals^{\NumItems-1}: g^{\top} x \geq \cos^{-1}(\angle(\Trade, \ConeCenter)), \text{for all } \Trade \in V \setminus \tau\}$\; \label{algo:int_offers}
      
      $x_1, x_2  \leftarrow \argmax_{(x_1', x_2') \in H_{\text{cone}}\cap H_{\text{offers}}}{\lVert x_1' - x_2' \rVert}$ \;\label{algo:farthest_points}
      
      $c \leftarrow \frac{x_1 + x_2}{2}$ \; \label{algo:hypersphere_1}
      $r \leftarrow \frac{\sqrt{3}}{2}\lVert x_1 - x_2 \rVert$ \; \label{algo:hypersphere_2}
      
      $\ConeCenter' \leftarrow \ConeCenter \cos(\tan^{-1}(\lVert c \rVert)) +  M c\sin(\tan^{-1}(\lVert c \rVert))$\; \label{algo:cone_start_int}
      
      $\ClosePoint \leftarrow \CircleCenter - \CircleCenter \frac{\CircleCenter}{\lVert \CircleCenter \rVert}$ \;
      
      $\FarPoint \leftarrow \CircleCenter + \CircleRadius \frac{\CircleCenter}{\lVert \CircleCenter \rVert}$ \;
        
      $\OpeningAngle' \leftarrow \frac{1}{2}\cos^{-1}{\left(\frac{4\CircleRadius^2 - (1 + \lVert \ClosePoint \rVert^2) - (1 + \lVert \FarPoint \rVert^2)}{-2\sqrt{(1 + \lVert \ClosePoint \rVert^2)}\sqrt{(1 + \lVert \FarPoint \rVert^2)}}\right)}$ \; \label{algo:cone_end_int}
      \tcp{If the semi-vertical angle is smaller, update the cone}
      \If{$\OpeningAngle' \leq \OpeningAngle$}{
        $\ConeCenter \leftarrow \ConeCenter'$\; \label{algo:cone_update_int_start}
        
        $\OpeningAngle \leftarrow \OpeningAngle'$\;
        
        $V \leftarrow \{\ConeCenter\}$\; \label{algo:cone_update_int_end}
        
      }
      \Else{
      \tcp{If not, generate a new offer that bisects the farthest corner points}
         $\HyperplaneA \leftarrow \frac{x_1 - x_2}{ \lVert x_1 - x_2 \rVert}$ \; \label{algo:int_hyperplane_start}
         
        $\HyperplaneB \leftarrow \frac{\lVert x_1 \rVert^2  -\lVert x_2 \rVert^2 }{2\lVert x_1 - x_2 \rVert}$ \; \label{algo:int_hyperplane_end}
        
        $\Trade \leftarrow M \HyperplaneA \cos(\tan^{-1}(\left(\frac{|\HyperplaneB|}{\lVert \HyperplaneA \rVert}\right))) + \ConeCenter \sin(\tan^{-1}(\left(\frac{|\HyperplaneB|}{\lVert \HyperplaneA \rVert}\right)))$ \; \label{algo:int_trade_start}
        
        $\Trade \leftarrow \Trade_{\text{int}}$ such that $\Trade_{\text{int}} \in \mathbb{Z}^n$ and $\Trade_{\text{int}} \leftarrow \argmin_{\Trade_{\text{int}}}{\angle(\Trade_{\text{int}}, \Trade)}, \|\Trade_{\text{int}}\| \leq d$\; \label{algo:int_trade_end}
      }
    }
    \Else{
    \tcp{The first n-1 trades follow the pattern established in the non-integer constrained algorithm}
        $\Trade \leftarrow \Trade_{i}$ such that $\Trade_i \perp \Trade_j$ for all $\Trade_j \in V$,  $\lVert \Trade \rVert \leq \SamplingDistance$ and $\langle \nabla \OfferingUtility{\OfferingState}, \Trade \rangle \geq 0$\;
        
        $\Trade \leftarrow \Trade_{\text{int}}$ such that $\Trade_{\text{int}} \in \mathbb{Z}^n$ and $\Trade_{\text{int}} \leftarrow \argmin_{\Trade_{\text{int}}}{\angle(\Trade_{\text{int}}, \Trade)}, \|\Trade_{\text{int}}\| \leq d$,\; 
    }
  }
  \If{$\ResponseFunction{\Trade} \neq \text{accept}$}{return $-1, \ConeCenter, \OpeningAngle$ \;}
  \caption{Cone Refinement with Integer Constraints}
  \label{algo:stcr_int}
\end{algorithm}
\newpage
\begin{figure}[t]
    \begin{minipage}{\linewidth}
        \centering
        \resizebox{0.2\linewidth}{!}{\begin{tikzpicture}[scale=2, transform shape, shift={(1,-1)}]
    \def\sidelength{2}
    
    \coordinate (A) at (0,0);
    \coordinate (B) at (\sidelength,0);
    \coordinate (C) at (\sidelength/2,{sqrt(3)*\sidelength/2});
    
    \draw[thick] (A) -- (B) -- (C) -- cycle;
    
    \coordinate (M) at ($(A)!0.5!(B)$);
    \filldraw[black] (M) circle (1pt) node[below left] {$c'$};
    
    \draw[thick, red] (M) circle (\sidelength/2);
    \draw[thick, blue] (M) circle ({sqrt(3) * \sidelength/2});
    
    \draw[<->, thick, blue] (M) -- ++({sqrt(3) * \sidelength/2}, 0) node[right] {$\frac{\sqrt{3}}{2}$};
    \draw[<->, thick, red] (M) -- ++(\sidelength/2, 0) node[midway, below] {$1$};
\end{tikzpicture}}
        \caption{Using the half-distance between the two farthest points on a polytope as the radius $r$ of the hypersphere can fail to include all possible gradient directions when the center is chosen as the midpoint between the two farthest points. Instead, by scaling the radius $r$ by a factor of $\sqrt{3}$, we can ensure that all possible gradients are included. Note that a point that is more than $\sqrt{3}r$ from the center would be one of the two farthest points by the law of cosines.}
        \label{no_scaling}
    \end{minipage}%
\end{figure}

As we discussed in previous sections, making discrete offers can lead to off-center or non-orthogonal halfspace cuts when refining the cone of potential gradients.
We consider scenarios where offers are limited to integer values.
In these scenarios, the space of potential gradients cannot be readily enclosed using ST-CR's cone update rules.
Instead, we explicitly model the space of potential gradients after the rejected offers as a polytope and calculate the cone that encloses the space.

When modeling the space of potential gradients, we consider the $\NumItems - 1$ dimensional hyperplane with the normal vector $\ConeCenter$ and orthogonal basis vectors $\BasisSet = \{v_1, v_2, .... v_{n-1}\}$ a unit distance from the cone's vertex.
We represent the current cone \( C(\ConeCenter, \OpeningAngle) \) of potential gradients as a hypercube defined by \( 2(\NumItems-1) \) halfspace constraints:
\begin{equation*}
    H_{\text{cone}} = \left\{ x \in \mathbb{R}^{\NumItems-1} : \pm e_i^{\top} x \geq -\tan{\theta} \quad \text{for } i = 1, 2, \ldots, \NumItems-1 \right\},
\end{equation*}
where $e_1^\top = [1, 0, \ldots, 0], e_2^\top = [0, 1, \ldots, 0], \ldots, e_{\NumItems-1}^\top = [0, 0, \ldots, 1]$
(Algorithm \ref{algo:stcr_int}, Line \ref{algo:hypercube}). We chose a hypercube representation because it allows us to easily calculate the corner points of the space of potential gradients after the halfspace cuts.
Given a set of $k$ rejected offers $V_{\text{rejected}}$, the halfspaces corresponding to the rejected offers are (Algorithm \ref{algo:stcr_int} Line \ref{algo:int_offers})
\begin{equation*}
    H_{\text{offers}} = \{ x\in \Reals^{\NumItems-1}: g^{\top} x \geq -\|\Trade\|\cos(\angle(\Trade, \ConeCenter)), \Trade \in V_{\text{rejected}}\}
\end{equation*}
where 
\begin{equation*}
    g_{i} = \frac{\langle \Trade, v_i \rangle }{\| v_i \|^2}, \forall v_i \in V_{b}.
\end{equation*}
We can define the space of potential gradients as a polytope $P = \{ x\in \Reals^{\NumItems-1}: x\in H_{\text{cone}} \cap H_{\text{offer}}\}$ defined by the intersection of the rejected offer halfspaces and the hypercube corresponding to the current cone.
We use the farthest corner points of $P$ to enclose the space of potential gradients in a new hypersphere and corresponding cone.

We now determine if the polytope $\polytope$ can be enclosed by a smaller cone.
We first calculate the two farthest corner points ($x_1, x_2$) of $\polytope$ (Algorithm \ref{algo:stcr_int} Line \ref{algo:farthest_points}). 
We then enclose $\polytope$ with the hypersphere $\Circle(\CircleCenter, \CircleRadius)$ with center $\CircleCenter = \frac{x_1 +x_2}{2} $ and radius $ \CircleRadius = \frac{\sqrt{3}}{2} \lVert x_1 - x_2 \rVert.$  (Algorithm \ref{algo:stcr_int}, Lines \ref{algo:hypersphere_1} - \ref{algo:hypersphere_2}).
In \cref{no_scaling}, we observe that the diameter being the distance between the two farthest points excludes possible gradient directions. We address this issue by scaling the radius by a factor of $\sqrt{3}$. 
Next, we determine the cone $C(\ConeCenter', \OpeningAngle')$ that encloses the hypersphere $\Circle(\CircleCenter, \CircleRadius)$. 
We begin by finding the $n$-dimensional vector corresponding to the hypersphere center $v_{center} = M \CircleCenter$ where $M \in \mathbb{R}^{n \times n - 1} $ is a matrix such that $M = [v_1, \ldots, v_{n-1}]$ .
We then obtain $\ConeCenter'$ by rotating $\ConeCenter$ in the direction of $v_{center}$ by $\tan^{-1}(\lVert c\rVert)$ (Algorithm \ref{algo:stcr_int}, Lines \ref{algo:cone_start_int} - \ref{algo:cone_end_int})
\begin{equation*}
    \ConeCenter' = \ConeCenter \cos(\tan^{-1}(\lVert c \rVert)) + Mc  \sin(\tan^{-1}(\lVert c \rVert)).
\end{equation*}
To find the semi-vertical angle $\OpeningAngle$, we calculate the angle between closest $\ClosePoint = \CircleCenter - \nicefrac{\CircleRadius\CircleCenter'}{\lVert \CircleCenter \rVert}$ and farthest $ \FarPoint = \CircleCenter + \nicefrac{\CircleRadius\CircleCenter}{\lVert \CircleCenter \rVert}$ points on the hypersphere from the origin. Since we know the distance between $\ClosePoint$ and $\FarPoint$ and the distance between these points and the cone vertex, we can determine $\OpeningAngle'$

 \begin{equation*}
     \OpeningAngle' = \frac{1}{2}\cos^{-1}{\left(\frac{ (1 + \lVert \ClosePoint \rVert^2) + (1 + \lVert \FarPoint \rVert^2) - 4\CircleRadius'^2}{2\sqrt{(1 + \lVert \ClosePoint \rVert^2)}\sqrt{(1 + \lVert \FarPoint \rVert^2)}}\right)}. 
 \end{equation*}

If the current angle $\OpeningAngle$ is smaller than the new angle $\OpeningAngle'$ after $\NumItems-1$ offers, we continue refining the cone.
We use the hyperplane $\HyperplaneA^{T} x = \HyperplaneB$ that bisects the two farthest corner points of the polytope as a basis for the next offer, ensuring the farthest corner points of the new polytope are closer together than those in the previous one (Algorithm \ref{algo:stcr_int}, Lines \ref{algo:int_hyperplane_start}-\ref{algo:int_hyperplane_end}). We then obtain the offer $\Trade$ in $\NumItems$ dimensions $M \HyperplaneA$ and rotate the offer in the direction of $\tau$ by an angle $\tan^{-1}\left(\nicefrac{|\HyperplaneB|}{\lVert \HyperplaneA \rVert}\right)$ (Algorithm \ref{algo:stcr_int}, Lines \ref{algo:int_trade_start}-\ref{algo:int_trade_end})

\begin{equation*}
      \Trade =  M \HyperplaneA \cos\left(\tan^{-1}\left(\frac{|\HyperplaneB|}{\lVert \HyperplaneA \rVert}\right)\right) + \ConeCenter \sin\left(\tan^{-1}\left(\frac{|\HyperplaneB|}{\lVert \HyperplaneA \rVert}\right)\right).
\end{equation*}
Repeating this process shrinks the size of the polyhedron.
Once $\OpeningAngle' < \OpeningAngle$, ST-CR updates the cone and repeats the refinement process (Algorithm \ref{algo:stcr_int}, Lines \ref{algo:cone_update_int_start}-\ref{algo:cone_update_int_end}).

\newpage
\subsection{Numerical Experiment Baseline Psudocode}
\begin{figure}[H]

\begin{minipage}[t]{0.48\textwidth}
\begin{algorithm}[H]
 \SetAlgoNlRelativeSize{0}   
 \SetAlgoLined
 \setcounter{AlgoLine}{0}
  \KwIn{$\OfferingState, \respondingState, f^{\OfferingAgent}$, $d$, $n$}
  \KwOut{$\Trade$}
  
  \tcp{Sample a random trade direction}
  \For{$i \leftarrow 1$ \KwTo $n$}{
    $v_i \leftarrow \mathcal{N}(0, 1)$ \;
    
    \tcp{Sample each component from a standard normal distribution}
  }
  $\mathbf{v} \leftarrow \frac{\mathbf{v}}{\|\mathbf{v}\|}$ \;
  
  \tcp{Normalize to get a unit vector}
  
  $T \leftarrow T \text{ such that } \frac{\Trade}{\lVert\Trade\rVert} = \mathbf{v}, \respondingState - \Trade > 0, \OfferingState + \Trade > 0, \text{ and } \lVert\Trade\rVert \leq d$ \;

  \If{$f^{\OfferingAgent}(\OfferingState + \Trade) \leq 0$}{
    $\Trade \leftarrow -1 * \Trade $\;
  }
  return $\Trade$ \;
  
  \caption{Obtain a Uniform Random Trade (URT)}
\end{algorithm}

\begin{algorithm}[H]
    \SetAlgoNlRelativeSize{0}   
    \SetAlgoLined
    \setcounter{AlgoLine}{0}
    \KwIn{$\OfferingState, f_{\OfferingAgent}, R_{\respondingAgent}, \Theta^{*}, \SamplingDistance$, $m$, $n$}
    $q \leftarrow 0$\;
    
    \While{$q \leq m$}{
        \tcp{Offer previous trade}
        \If{$\ResponseFunction{\Trade_{\text{prev}}} = \text{accept}$}{ %
            $\OfferingState \leftarrow \OfferingState + \Trade_{\text{prev}}$ \;
            $\respondingState \leftarrow \respondingState - \Trade_{\text{prev}}$ \;
            continue \;
        }
        \tcp{Find trade Random Trading}
        $\Trade, q_{\Trade} \leftarrow \text{ RT }(\OfferingState, \respondingState, f_{\OfferingAgent}, R_{\respondingAgent}, d, m, n)$ \; %
        
        $q \leftarrow q + q_{\Trade}$ \;
        
        \If{$T \neq -1$}{ %
            $\OfferingState \leftarrow \OfferingState + \Trade$ \;
            $\respondingState \leftarrow \respondingState - \Trade$ \;
            continue \;
        }
        \Else{
             break \; \tcp{End loop if no valid trade found}
        }
    }
    \caption{Random Trading Baseline Progression}
\end{algorithm}

\begin{algorithm}[H]
  \SetAlgoNlRelativeSize{0}   
  \SetAlgoLined
  \setcounter{AlgoLine}{0}
  \SetKwRepeat{Do}{do}{while}
  \KwIn{$\OfferingState, \respondingState f^{\OfferingAgent}, R_{\respondingAgent}, d, m, n$}
  \KwOut{$\Trade, q_{\Trade}$}
  $\Trade \leftarrow \text{URT}(\OfferingState, \respondingState, f^{\OfferingAgent}, d, n)$ \;
  \tcp{Continue sampling until the offer is accepted or query limit is reached}
  $q_{\Trade} \leftarrow 0$ \;
  
  \Do{$\ResponseFunction{\Trade} \neq \text{accept}$ and $ q < m$}{
    $\Trade \leftarrow \text{URT}(\OfferingState, f^{\OfferingAgent})$ \;
    
    Offer $\Trade$, $q_{\Trade} \leftarrow q_{\Trade} + 1$ \;
  }
  \If{$\ResponseFunction{\Trade} \neq \text{accept}$}{
    return $-1, q_{\Trade}$ \;
  }
  \Else{
   return $\Trade, q_{\Trade}$ \;
  }
  \caption{Random Trading Baseline (RT)}
\end{algorithm}
\end{minipage}
\hfill
\begin{minipage}[t]{0.48\textwidth}
\begin{algorithm}[H]
    \SetAlgoNlRelativeSize{0}   
    \SetAlgoLined
    \setcounter{AlgoLine}{0}
    \KwIn{$\OfferingState, \respondingState, f_{\OfferingAgent}, R_{\respondingAgent}, \Theta^{*}, \SamplingDistance, \SamplingDistance_{\text{interval}}, \SamplingDistance_{\text{max}}, m, n$}
    $q \leftarrow 0$ \;
    \While{$q \leq m$}{
        \tcp{Offer previous trade}
        \If{$q = 0$}{\tcp{Find initial trade using Random Trading}
        $\Trade, q_{\Trade} \leftarrow \text{RT}(\OfferingState, \respondingState, f^{\OfferingAgent}, R_{\respondingAgent}, m, d)$\;
        }
        \Else{\tcp{Find subsequent trades using Random Trading with Momentum}
        $\Trade, q_{\Trade} \leftarrow \text{ RTM }(\OfferingState, \respondingState, f_{\OfferingAgent}, R_{\respondingAgent}, d, \SamplingDistance_{\text{interval}}, \SamplingDistance_{\text{max}}, m - q, \Trade_{\text{prev}}, n)$ \;}
        $q \leftarrow q + q_{\Trade}$\; 
        \If{$T \neq -1$}{ %
            $\OfferingState \leftarrow \OfferingState + \Trade$ \;
            $\respondingState \leftarrow \respondingState - \Trade$ \;
            $\Trade_{\text{prev}}\leftarrow \Trade$ \;
        }
        \Else{
            break \;
        }
        
    }
    \caption{Random Trading With Momentum Progression}
\end{algorithm}

\begin{algorithm}[H]
    \SetAlgoNlRelativeSize{0}   
    \SetAlgoLined
    \setcounter{AlgoLine}{0}
    \KwIn{$\OfferingState, \respondingState, f^{\OfferingAgent}, R_{\respondingAgent}, d, d_{\text{interval}, n}, \SamplingDistance_{\text{max}}, m, \Trade_{\text{prev}}, n$}
    \KwOut{$\Trade, q_{\Trade}$}
    \tcp{Deviation scaling factor}
    
    $d_\text{dev} \leftarrow 0$, $q_{\Trade} \leftarrow 0$, $\Trade \leftarrow \Trade_{\text{prev}}$ \;
    
   Offer $\Trade$, $q_{\Trade} \leftarrow q_{\Trade} + 1$ \;
   \While{$\ResponseFunction{\Trade} \neq \text{accept}$ and $ q_{\Trade} < m$}{
    \tcp{Increase deviation scaling factor by some constant.}
    $d_\text{dev} \leftarrow \min{(d_\text{dev} + d_{\text{interval}}, d_{\text{max}})}$ \;
    
    \tcp{Uniformly Sample Adjustments}
    
    \For{$i \leftarrow 1$ \KwTo $n$}{
        $v_i \leftarrow \mathcal{N}(0, 1)$
      }
      
    $\mathbf{v} \leftarrow \frac{\mathbf{v}}{\|\mathbf{v}\|}$, \;
    $\mathbf{v} \leftarrow d_\text{dev} * \mathbf{v}$, \;
    $\Trade_{\text{dev}} \leftarrow \nicefrac{\Trade_{\text{prev}}}{\|\Trade_{\text{prev}}\|} + \mathbf{v}$ \;
    
    $\Trade \leftarrow \Trade \text{ such that } \nicefrac{\Trade}{\lVert\Trade\rVert} = \nicefrac{\Trade_{\text{dev}}}{\lVert\Trade_{\text{dev}}\rVert}, \respondingState - \Trade > 0, \OfferingState + \Trade > 0, \text{ and } \lVert\Trade\rVert \leq d$ \;
    Offer $\Trade$, $q_{\Trade} \leftarrow q_{\Trade} + 1$
   }
   \If{$\ResponseFunction{\Trade} \neq \text{accept}$}{
      return $-1, q_{\Trade}$ \;
   }
   \Else{
    return $\Trade, q_{\Trade}$ \;
   }
   
    \caption{Random Trading With Momentum (RTM)}
\end{algorithm}
\end{minipage}
\end{figure}
\subsection{Greedy Concession Algorithm Baseline Psudocode}
\begin{figure}[H]
\begin{minipage}[t]{0.48\textwidth}
\begin{algorithm}[H]
    \SetAlgoNlRelativeSize{0}   
    \SetAlgoLined
    \setcounter{AlgoLine}{0}
    \KwIn{$\OfferingState, \respondingState, f_{\OfferingAgent}, R_{\respondingAgent}, \SamplingDistance$, $m$, $n_{\text{weights}}$, $\lambda$, $q_\text{int}, n$}
    \tcp{GCA Arguments: The scaling factor ($\lambda$), Number of sampled weights ($n_{\text{weights}}$), belief update interval ($q_\text{int}$)}
    $q \leftarrow 0$\;

    \tcp{Initialize Weights and Belief Probabilities}
    $W = \{w_i = \frac{\mathcal{N}(0, 1)^{n}}{\|\mathcal{N}(0, 1)^{n}\|} \mid i = 1, \ldots, n_{\text{weights}}\}$\;

    $P(w_i) = \frac{1}{n_{\text{weights}}}$ \;

    \While{$q \leq m$}{
        
        \tcp{Find trade GCA}
        $\Trade, q_{\Trade}, P \leftarrow \text{ GCA }(\OfferingState, \respondingState, f_{\OfferingAgent}, R_{\respondingAgent}, \SamplingDistance, m, 
        q_\text{int}, \lambda, W, P, n)$ \; %
        
        $q \leftarrow q + q_{\Trade}$ \;
        
        \If{$T \neq -1$}{ %
            $\OfferingState \leftarrow \OfferingState + \Trade$ \;
            
            $\respondingState \leftarrow \respondingState - \Trade$ \;
            
            $P \leftarrow \text{Softmax}(P)$ \;
        }
        \Else{
            break \;
            
            \tcp{End loop if no valid trade found}
        
        }
    }
    \caption{Greedy Concession Algorithm Trade Progression}
\end{algorithm}

\begin{algorithm}[H]
    \SetAlgoNlRelativeSize{0}   
    \SetAlgoLined
    \setcounter{AlgoLine}{0}
    \KwIn{$\OfferingState, \respondingState, f_{\OfferingAgent}, R_{\respondingAgent}, \SamplingDistance, m, q_\text{int}, \lambda, W, P, n$}
    
    \tcp{Arguments: The set of possible weights $W$, the belief distribution over the weights $P$, the set of feasible offers $\Trade_{\text{total}}$, scaling factor ($\lambda$), number of sampled weights ($n_{\text{weights}}$)}
    \KwOut{$\Trade, q_{\Trade}, P$}

    $q_{\Trade} \leftarrow 0$\;
    
    $T_{\text{total}} \leftarrow \{\Trade \in \mathbb{Z}^n \mid \OfferingState + \Trade > 0, \respondingState - \Trade > 0, \|\Trade\| \leq \SamplingDistance\}$\;
    
    $\Trade_{\text{sorted}} \leftarrow \text{GCA\_Sort}(\OfferingState, W, P, T_{\text{total}})$\;
    
    $T_{\text{rej}} \leftarrow \{\}$\;
    
    \While{$q_{\Trade} \leq m$}{
        $\Trade \leftarrow \Trade_{\text{sorted}}[0]$\;
        
        $\Trade_{\text{sorted}} \leftarrow \Trade_{\text{sorted}} \setminus \{\Trade\}$\;
        
        \If{$f_{\OfferingAgent}(\OfferingState + \Trade) - f_{\OfferingAgent}(\OfferingState) > 0$}{
            $q_{\Trade} \leftarrow q_{\Trade} + 1$\;
            
            \If{$\ResponseFunction{\Trade} \neq \text{accept}$}{
                $T_{\text{rej}} \leftarrow T_{\text{rej}} \cup \{\Trade\}$\;
                
                \tcp{Update belief weights every $q_\text{int}$ offers}\;
                
                \If{$q_{\Trade} \bmod q_\text{int} = 0$}{
                    $P \leftarrow \text{BU}(\respondingState, W, P, T_{\text{rej}})$\;
                    
                    $\Trade_{\text{sorted}} \leftarrow \text{GCA\_Sort}(\OfferingState, W, P, \Trade_{\text{sorted}})$\;
                }
            }
            \Else{
                \Return{$\Trade, q_{\Trade}, P$}
            }
        }
        \Else{
            \Return{$-1, q_{\Trade}, P$}
        }
    }
    
    \Return{$-1, q_{\Trade}, P$}\;
    \caption{Greedy Concession Algorithm (GCA)}
\end{algorithm}

\end{minipage}
\hfill
\begin{minipage}[t]{0.48\textwidth}
\begin{algorithm}[H]
    \SetAlgoNlRelativeSize{0}   
    \SetAlgoLined
    \setcounter{AlgoLine}{0}
    \KwIn{$\OfferingState, W, P, \Trade_{\text{total}}, f_{\OfferingAgent}, n$}
    \tcp{Arguments: The set of possible weights $W$, the belief distribution over the weights $P$, and the set of feasible offers $\Trade_{\text{total}}$}
    \For{$\Trade \in \Trade_{\text{total}}$}{
        $p_\text{accept} = 0$ \;
        
        \For{$w_i \in W$}{
            \If{$\langle S_B - \Trade, w_i \rangle - \langle S_B, w_i \rangle > 0$}{
                $p_\text{accept} \leftarrow p_\text{accept} + P(w_i)$
            }
        }
        $E(f(\OfferingState + \Trade) - f(\OfferingState)) = p_\text{accept} \cdot (f(\OfferingState + \Trade) - f(\OfferingState))$
    }
    \tcp{Sort offers in descending order of expected utility}
    $\Trade_{\text{sorted}} = \text{Sort}(\Trade_{\text{total}}, \text{by } E(f(\OfferingState + \Trade) - f(\OfferingState)) \text{ in descending order})$ \;
    
    return $\Trade_{\text{sorted}}$ \;
    \caption{Greedy Concession Algorithm Sorting (GCA\_Sort)}
\end{algorithm}

\begin{algorithm}[H]
    \SetAlgoNlRelativeSize{0}   
    \SetAlgoLined
    \setcounter{AlgoLine}{0}
    \KwIn{$\respondingState, W, P, T_{\text{rej}}$}
    \tcp{Arguments: The set of possible weights $W$, the belief distribution over the weights $P$, the set of feasible offers $\Trade_{\text{total}}$, and the scaling factor ($\lambda$)}
    \For{$T \in T_{\text{rej}}$}{
        \For{$w_i \in W$}{
            \If{$\langle S_B - T, w_i \rangle - \langle S_B, w_i \rangle > 0$}{
                $P(w_i) \leftarrow \lambda P(w_i)$ \;
                \tcp{Reduce the probability of weights that do not match rejected offers.}
            }
        }
    }
    $P \leftarrow \frac{P}{\sum_{i = 1}^{\lVert W \rVert}{P(w_i)}}$ \; 
    
    return $P$ \;
    \caption{Greedy Concession Algorithm Belief Update (BU)}
\end{algorithm}
\end{minipage}
\newpage
\end{figure}

\subsection{ST-CR Adjustments for Numerical Experiments}
We make several adjustments when implementing ST-CR in continuous and discrete scenarios. We detail the modifications we made to ST-CR for our numerical experiments.

\subsubsection{Cone Initialization in Sequential Trading}
When making sequential trades, ST-CR can bypass stage 2.1 by using the previous cone to initialize the cone for the next trade. 
Since we assume that the responding agent's utility function is smooth and the offers have a bounded magnitude, the change in the responding agent's gradient direction is bounded.
Given a cone $C(\tau, \theta)$ from the previously accepted trade $\Trade$, we initialize a new cone $C(\tau', \theta')$ where
\begin{equation*}
    \tau' = \tau, \theta' = \theta + b \|\Trade\|
\end{equation*}
where $b$ is a constant corresponding to the increase in the cone's angle. 
Note that, as the smoothness constant $\beta$, decreases $b$ can be set lower. 
For example, in scenarios with linear utility functions, $b = 0$ is sufficient to enclose all possible gradient directions after a trade.
We scale the increment by $\| \Trade \|$ since larger trades can lead to a larger change in the gradient direction.
If $\theta' > \pi/2$, the cone has exceeded the size of the initial cone from stage 2.1. In such cases, we repeat stage 2.1 to narrow the cone to a single quadrant.

\subsubsection{Ensuring Beneficial Trades}
As we discussed in prior sections, ST-CR selects offer directions that align with the offering agent's gradient direction ensuring that \( \langle T, \nabla f^{A}(S_{A})\rangle \geq 0 \). 
For non-linear utility functions, this method can lead to non-beneficial trades for the offering agent if \(\lVert\nabla f^{A}_{\Trade}(\OfferingState)\rVert\) is small, overshooting the optimal trade in the direction of \(\Trade\). 
In our implementation, we resolve this issue by reducing the offer magnitude to prevent overshooting.
In discrete cases, offers are constrained to integer values of resources, and reducing the offer magnitude may not prevent overshooting.
In these cases, the ST-CR reduces the offer magnitude as much as possible and accepts a bounded loss of benefit.

\subsubsection{Improving Self-Interested Behavior}
ST-CR uses $n-1$ rejected orthogonal offers to refine the gradient cone.
If an offer is accepted, the agents transition to a new state.
When making the $n-1$ offers, the offering agent prioritizes offers that are most beneficial to itself, ensuring that accepted offers will have a larger benefit for the offering agent.

\subsubsection{Cone Update in Discrete Trading}
As we have established, a risk of two-point comparisons is potentially making incorrect halfspace cuts.
This is also a risk in discrete trading scenarios, where offers are constrained to integer values.
In the worst case, incorrect cuts can lead to an empty polyhedron of potential gradients, where the halfspace constraints of the current cone and the rejected offers do not have any overlapping region.
In such cases, we increase the current cone's angle until the polyhedron becomes non-empty.
By increasing the cone's angle of opening, we enclose the full space of potential gradients.

\subsubsection{Generating new Trades in Discrete Trading}
As we stated in the discrete trading section, the non-orthogonal or off-center halfspace cuts caused by discrete offers can result in ST-CR needing more than $n-1$ offers to update the gradient cone.
In such cases, we use the halfspace that bisects the farthest corner points of the polyhedron of potential gradients as a basis for the next offer.
However, these offers must also abide by the integer constraints and may change direction when rounding occurs. 
In the worst case, the rounded offer will no longer bisect the farthest corner points, and will not result in a reduction of the space of potential gradients.
If the rounded offer will not reduce the gradient space, ST-CR increases the offer's magnitude $d$ before rounding. 
By increasing $d$, the difference between the original offer and the rounded offer becomes smaller.
This can conflict with the feasibility of the offers if the offering or responding agents do not have the resources to meet the scaled-up offers.
If ST-CR cannot find feasible integer trades that can refine the cone, it stops trading.
\subsection{Reproducibility of Numerical Experiments}
All of the code for our numerical experiments is provided in the attached folder.
Our numerical experiments were conducted on Python version 3.9.7.
We provide Python package requirements in the \textit{requirements.txt} file.
The repository contains the file \textit{run\_algo\_tests.py}, which will run all of the tests we discussed in the numerical experiments section.
The results presented in the paper were obtained on a Dell XPS 13 laptop with an 11th Gen Intel(R) Core(TM) i7-1185G7 @ 3.00GHz processor and 16.0 GB of RAM.
To ensure reproducibility, we used a random seed of \textit{10} when generating our results.
Our numerical experiments also limited the offers to trade at most $5$ resources from each category ($\text{max\_trade\_value} = 5$) and set ST-CR's angle threshold $\OpeningAngle^*$ to $0.00001$ ($\text{theta\_closeness} = 0.00001$).
For the random trading with momentum baseline, we set the interval of deviation increase to $d_{\text{interval}} = 0.05$ ($\text{deviation\_interval} = 0.05$) and the maximum deviation magnitude to $d_{\text{max}} = 5$ ($\text{max\_deviation\_magnitude} = 5$).
For the GCA baseline, we set the belief shrinking factor as $\text{shrinking\_factor} = 0.1$, the number of sampled weights as  $\text{num\_sampled\_weights} = 100$, and the Softmax temperature as $\text{softmax\_temperature}=0.02$.
Using this random seed and these hyperparameters will reproduce the presented results.
We provide Python and package versions in the repository's requirements.txt file.

We also provide the file \textit{gpt\_integration\_test.py} for running trading scenarios with humans using GPT. 
We provide full transcripts of our GPT examples in chat\_folders/paper
We note that, since the output of GPT models changes with each run, we can not exactly reproduce our examples.
Finally, we note that, when running the GPT tests, users will need to provide their own OpenAI API keys.
\newpage
\subsection{Integrating ST-CR and Language Models}

We consider discrete trading scenarios between ST-CR and a human trader where offers are limited to integer values.
In these example scenarios, ST-CR and the human trade apples, bananas, and oranges.
ST-CR acts as the offering agent $A$ and the human is the responding agent $B$.
The state of each agent is $[x_1, x_2, x_3]$ where $x_1$ is the number of apples, $x_2$ is the number of bananas, and $x_3$ is the number of oranges.
Instead of having an explicit utility function, the human negotiator begins each scenario by inputting a target state $\Utilityb_{B}$ it would like to reach through trading.
To show the non-trivial sections of ST-CR, we use the target state to initialize the responding agent's gradient quadrant. Therefore, we assume that stage 2.1 of ST-CR has already been performed and begin offering from stage 2.2.
ST-CR is also given a target state $\Utilityb_{A}$ that is generated by sampling integers uniformly between 1 and 100.
Using these two states, the utility functions for both agents are $ \OfferingUtility{\OfferingState} = -\OfferingState^{\top}  I \OfferingState + 2\OfferingState^{\top} \Utilityb_A $ 
 and $\respondingUtility{\respondingState} = -\respondingState^{\top}  I \respondingState + 2\respondingState^{\top} \Utilityb_B$
where $I$ is the $n \times n$ identity matrix.
The human negotiator is not required to follow this utility function. However, approximating it enables the testing program to estimate the human's benefit following a trade.
When making offers ST-CR uses the format
\begin{center}
\begin{minipage}{\linewidth}
\begin{quote}
\textit{\noindent
\textbf{ST-CR's Trade Offer:}\\
\hspace*{1em}ST-CR Receives: 3 apples\\
\hspace*{1em}User Receives: 2 oranges, 4 bananas.}
\end{quote}
\end{minipage}
\end{center}
Using a standardized offer format allows the language model to easily interpret the human's response.

We utilize language models to perform sentiment analysis on human responses to determine if a human is accepting or rejecting an offer.
To do this, we first replace the identifiers in the previous offer with two named individuals to allow the language model to easily interpret who is receiving what resources. In our code, we replace "User" with "Alice" and "ST-CR" agent with "Bob".
We then provide the language model with the previous offer and the human's response, then give it the prompt
\begin{quote}
\textit{In a trading scenario, Alice has been given the following trade offer \_\_\_\_\_, Alice has given the following response: \_\_\_\_\_. Given Alice's response, is she accepting or rejecting the offer. Please answer with 1 for accept, otherwise answer with 0. If the text doesn't provide information on whether Alice accepting or rejecting Bob's trade offer, respond with 0.
}
\end{quote}
Using the language model's response to this prompt, we determine if the human is accepting or rejecting the offer.

Beyond acceptance or rejection, humans can provide additional information to ST-CR through counteroffers, adjustments to prior offers, or general preferences.
If sentiment analysis determines that the human user is not accepting the trade offer, we prompt the language model to determine if the user is providing any additional information
\begin{quote}
    \textit{Given the following conversation history, is Alice providing an explicit counteroffer, a general preference, or neither? Please respond 1 for a counteroffer, 2 for a preference, 0 for neither. (only answer with 0, 1, and 2). If you are unsure, respond with 0.}
\end{quote}
Adjustments to prior offers are readily represented as counteroffers, so we use the same logic to handle both cases.
If the language model believes that the human user is providing a counteroffer, we give GPT the prompt shown in \cref{fig:alice_prompt}. Using this approach, we obtain a counteroffer in a standardized form, which is readily integrated into ST-CR.

\begin{figure}[htpb]
    \centering
    \begin{gptbox}
You are a language model assisting with negotiation scenarios.

In a trading scenario, Alice has been given the following trade offer: \_\_\_\_\_. Alice has given the following response: \_\_\_\_\_. Based on Alice's feedback, provide a counteroffer that accurately reflects her request. The counteroffer should be formatted as: \\
\texttt{Alice gives: a apples, b oranges, c bananas, Alice receives: d apples, e oranges, f bananas.} \\

Guidelines:
\begin{itemize}
    \item If Alice asks to give an item (e.g., "Can I give you X bananas?"), set the number of that item she gives to that amount.
    \item If Alice asks to receive an item, set the number of that item she receives to the requested amount. If the previous offer had her giving that, ensure that her new offer reflects the current request accurately.
    \item If the previous offer has Alice giving an item, and Alice's response asks to receive the same item, then set the number of that item she gives to 0.
    \item If the previous offer has Alice receiving an item, and Alice's response asks to give the same item, then set the number of that item she receives to 0.
    \item If Alice is not giving any items in the new trade, replace the 'Alice gives' list with \texttt{Nothing.} If Alice is not receiving any items in the new trade, replace the 'Alice receives' list with \texttt{Nothing.}
    \item Do not respond with placeholders or ambiguous terms like ‘some’, ‘more’, ‘fewer’, etc. Infer specific numerical values (e.g., ‘1 apple’ instead of ‘some apples’).
    \item Ensure the trade offer is specific and formatted as requested.
\end{itemize}
\end{gptbox}
    \caption{Prompt for Parsing Alice’s Counteroffer}
    \label{fig:alice_prompt}
\end{figure}

If the language model believes that the human is providing a general preference, we utilize the GPT-4.0 function calling feature to get a linear relationship of the form $\langle w_1, v \rangle \geq\langle w_2, v \rangle$ where $w_1$ and $w_2$ represent states and $v$ is the vector of values for each category. For example, "I prefer 2 apples over 3 oranges and a banana" can be represented as $\langle[2, 0, 0], [3, 1, 1]\rangle \geq \langle[0, 1, 3], [3, 1, 1]\rangle$.
"I prefer apples over oranges" can be represented as $\langle[1, 0, 0], [2, 1, 0]\rangle \geq \langle[0, 1, 0], [2, 1, 0]\rangle$.
"I don't want oranges" can be represented as $\langle[0, -1, 0], [1, -1, 1]\rangle \geq \langle[0, 0, 0], [1, -1, 1]\rangle$.
We use $w_1$ and $w_2$ as comparison states in ST-CR and do not use the values $v$ generated by GPT.
This linear relationship is represented as a counteroffer that increases the resources in $w_1$ and decreases the resources in $w_2$.
For example, $\langle[0, -1, 0], [1, -1, 1]\rangle \geq \langle[0, 0, 0], [1, -1, 1]\rangle$ can be represented as:
\begin{center}
\begin{minipage}{\linewidth}
\begin{quote}
\textit{\noindent \textbf{ST-CR's Trade Offer:}\\
    \hspace*{1em}ST-CR Receives: 1 orange.\\
    \hspace*{1em}User Receives: Nothing.}
\end{quote}
\end{minipage}
\end{center}

One challenge when working with language models is ensuring accuracy in their responses.
As we will later show, without an underlying algorithm, language models can make mistakes when interpreting human feedback.
A simple way to resolve this issue is by having the language model verify its responses.
After the language model parses the human's response, we feed the human's response and the generated counteroffer back into the language model using the following prompt
\begin{quote}
    \textit{In a trading scenario, Alice has been given the following trade offer \_\_\_\_\_, Alice has given the following response: \_\_\_\_\_. Given Alice's feedback, does the following counteroffer: \_\_\_\_\_ accurately reflect her feedback? Please respond with only yes or no. If the counteroffer includes ambiguous or non-numerical values for the items (e.g. 'some apples'), please respond with no."
}
\end{quote}
If the language model claims that the counteroffer is inaccurate, the language model generates a new one, repeating this process until the counteroffer is deemed accurate.
Once verified, ST-CR assesses the counteroffer's benefit to the offering agent and presents it to the human as a new offer.
Allowing the human to reject the parsed counteroffer adds an extra layer of verification.

\subsection{Example ST-CR Transcript (Stay)}
In \cref{fig:chat-transcript-stcr}, we present an example transcript of ST-CR with the previous trade heuristic trading with a human negotiator. 
Lines \ref{line:explicit_counteroffer_1_start}-\ref{line:explicit_counteroffer_1_end}, show an interaction where ST-CR makes an offer, and the human negotiator responds with an adjustment to the offer.
The language model successfully interprets the adjustment as a counteroffer, but it is not beneficial for ST-CR.
Therefore, ST-CR integrates the counteroffer into the gradient cone but does not present it to the user.
Lines \ref{line:explicit_counteroffer_2_start}-\ref{line:explicit_counteroffer_2_end} show an example where the counteroffer is beneficial to ST-CR.
The counteroffer is presented to the user, who then accepts the offer.
Just as in the previous examples, ST-CR successfully parses the response into a counteroffer, which it uses to refine the gradient cone. 

\subsection{Language Model Performance without ST-CR}

To obtain a baseline for human trading using natural language, we implemented a method that uses the language model GPT-4.0 to generate trade offers without any underlying model. 
When a trading scenario begins, we give GPT the prompt in \cref{fig:GPT_Prompt} to motivate trading and structure its responses.
The prompt provides the language model with its utility function which has a target state of 25 apples, 50 bananas, and 75 oranges.
The model presents a trade to the human user, who then provides a natural language response.
Just as with ST-CR, if the human accepts the trade, the model transitions to a new state.
If the human rejects the offer, the model must generate a new offer.
This process repeats until the human decides to end the trading process.

In \cref{fig:chat-transcript-success-pure}, we observe that GPT without an underlying model can successfully generate mutually beneficial trades and adjust its offers according to parsed human feedback.
However, we note that language models can exhibit inconsistency and make mistakes. 
For example, in \cref{fig:chat-transcript-fail-1}, we observe that the GPT model proposes trades that are not beneficial to itself. 
This suggests that the GPT model may not obey the self-interested and rational constraint or may wrongly interpret its utility function.
In \cref{fig:chat-transcript-fail-2}, we highlight a specific example where the GPT model repeatedly offers similar trades despite prior rejection.
These offers point in similar directions, which are not beneficial to the human negotiator.
Our algorithm, which uses orthogonal offers to refine the search space, would consider offers in different directions, leading to a more efficient use of the offer budget.
These observations collectively suggest that, without an underlying algorithm, language models may lack the structure to effectively negotiate with humans and may require extensive prompt tuning or chain-of-thought reasoning to accomplish such tasks. 
To address these limitations, we integrate ST-CR, leveraging language models primarily for sentiment analysis and feedback parsing.
\newpage

\begin{figure}[H]
\centering
\begin{lstlisting}[numbers=left, numberstyle=\tiny\color{gray}]
(*@\textcolor{system}{System:}@*)
    (*@\textcolor{system}{Start of Trading}@*)
    (*@\textcolor{system}{User Target Items: 60 apples, 70 bananas, 30 oranges}@*)
    (*@\textcolor{system}{ST-CR Utility Function Q: [[-1.,-0.,-0.], [-0.,-1.,-0.], [-0.,-0.,-1.]]}@*)
    (*@\textcolor{system}{ST-CR Utility Function u: [33. 33. 33.]}@*)
    (*@\textcolor{system}{Initial Human Items: 50 apples, 50 bananas, 50 oranges}@*)
    (*@\textcolor{system}{Initial ST-CR Items: 50 apples, 50 bananas, 50 oranges}@*)
(*@\textcolor{computer}{ST-CR's Trade Offer:}@*)
    (*@\textcolor{computer}{ST-CR receives: Nothing}@*)
    (*@\textcolor{computer}{User receives: 5.0 apples}@*)
(*@\textcolor{user}{User: How about I give you 5 oranges for 10 bananas}@*)
(*@\textcolor{computer_internal}{Parsed User Counteroffer:}@*)
    (*@\textcolor{computer_internal}{ST-CR receives: 5.0 oranges}@*)
    (*@\textcolor{computer_internal}{User receives: 10.0 bananas}@*)
(*@\textcolor{computer_internal}{ST-CR:  Counteroffer Direction is Beneficial}@*)
(*@\textcolor{computer}{ST-CR's Trade Offer:}@*)
    (*@\textcolor{computer}{ST-CR receives: 5.0 oranges}@*)
    (*@\textcolor{computer}{User receives: 10.0 bananas}@*)
(*@\textcolor{user}{User: Are you giving me 10 bananas?}@*)
(*@\textcolor{computer_internal}{Parsed User Counteroffer:}@*)
(*@\textcolor{system}{Human State Unchanged}@*)
(*@\textcolor{computer}{ST-CR's Trade Offer:}@*)
    (*@\textcolor{computer}{ST-CR receives: Nothing}@*)
    (*@\textcolor{computer}{User receives: 5.0 bananas}@*)
(*@\textcolor{user}{User: Give me 10 bananas and I will give you 5 oranges}@*) (*@\label{line:explicit_counteroffer_1_start}@*)
(*@\textcolor{computer_internal}{Parsed User Counteroffer:}@*)
    (*@\textcolor{computer_internal}{ST-CR receives: 5.0 oranges}@*)
    (*@\textcolor{computer_internal}{User receives: 10.0 bananas}@*)
(*@\textcolor{computer_internal}{ST-CR:  Counteroffer Direction is Beneficial}@*)
(*@\textcolor{computer}{ST-CR's Trade Offer:}@*)
    (*@\textcolor{computer}{ST-CR receives: 5.0 oranges}@*)
    (*@\textcolor{computer}{User receives: 10.0 bananas}@*)
(*@\textcolor{user}{User: yes}@*)
(*@\textcolor{system}{Offer Accepted!}@*)
(*@\label{line:explicit_counteroffer_1_end}@*)
(*@\textcolor{system}{New Human State: 50.0 apples, 60.0 bananas, 45.0 oranges}@*)
(*@\textcolor{system}{New ST-CR State: 50.0 apples, 40.0 bananas, 55.0 oranges}@*)
(*@\textcolor{system}{Estimated Human Benefit: 475.0}@*)
(*@\textcolor{system}{ST-CR Benefit: 45.0}@*)
(*@\textcolor{computer}{ST-CR's Trade Offer:}@*)
    (*@\textcolor{computer}{ST-CR receives: 5.0 bananas}@*)
    (*@\textcolor{computer}{User receives: 3.0 oranges}@*)
(*@\label{line:explicit_counteroffer_2_start}@*)
(*@\textcolor{user}{User: I will give you 5 oranges for 10 apples}@*)
(*@\textcolor{computer_internal}{Parsed User Counteroffer:}@*)
    (*@\textcolor{computer_internal}{ST-CR receives: 5.0 oranges}@*)
    (*@\textcolor{computer_internal}{User receives: 10.0 apples}@*)
(*@\textcolor{computer_internal}{ST-CR:  Counteroffer Direction is not Beneficial. Incorporating feedback into Gradient Cone}@*)
(*@\textcolor{computer}{ST-CR: I understand that you would prefer the following offer:}@*)
    (*@\textcolor{computer}{ST-CR receives: 5.0 oranges}@*)
    (*@\textcolor{computer}{User receives: 10.0 apples}@*)
(*@\textcolor{computer}{This trade is not beneficial for me. I will keep your preferences in mind for future trades.}@*)
(*@\label{line:explicit_counteroffer_2_end}@*)
(*@\textcolor{computer}{Please consider the following offer:}@*)
(*@\textcolor{computer}{ST-CR's Trade Offer:}@*)
    (*@\textcolor{computer}{ST-CR receives: Nothing}@*)
    (*@\textcolor{computer}{User receives: 5.0 apples}@*)
(*@\textcolor{user}{User: yes}@*)
(*@\textcolor{system}{Offer Accepted!}@*)
(*@\textcolor{system}{New Human State: 55.0 apples, 60.0 bananas, 45.0 oranges}@*)
(*@\textcolor{system}{New ST-CR State: 45.0 apples, 40.0 bananas, 55.0 oranges}@*)
(*@\textcolor{system}{Estimated Human Benefit: 75.0}@*)
(*@\textcolor{system}{ST-CR Benefit: 145.0}@*)
\end{lstlisting}
\caption{Transcript of ST-CR interacting with a human negotiator. The human negotiator has a target state of 60 apples, 70 bananas, and 30 oranges. Blue text denotes trade offers that ST-CR presents to the user. The red text is the user's response. The orange text corresponds to ST-CR's internal decision-making and is not presented to the user. Green text is system information that denotes initialization and trade progression.}
\label{fig:chat-transcript-stcr}
\end{figure}

\begin{figure}[H]
    \centering
    \begin{gptbox}
System: You are a computer negotiation agent that aims to maximize your utility function by trading apples, oranges, and bananas with a human negotiator. Your utility function is: $-\text{num\_apples}^2 + 2 * 25 * \text{num\_apples} - \text{num\_bananas}^2 + 2 * 50 * \text{num\_bananas} - \text{num\_oranges}^2 + 2 * 75 * \text{num\_oranges}$. You should only make offers if they improve your utility. You start with 50 apples, oranges, and bananas. You will present trade offers to the human, and they will provide responses. Please only respond with trade offers in the same format as the following example (the ordering and number of each fruit can vary depending on the offer). Please use "GPT" to refer to yourself and "User" to refer to the negotiating opponent: \\
"GPT's Trade Offer: \\
    GPT receives: 5 apples, 2 bananas \\
    User receives: 3 oranges "\\
Negotiation will continue from a new state even after a trade offer is accepted by the human user. Please start the negotiation by providing an initial trade offer for the User.
\end{gptbox}
    \caption{Prompt Given to GPT Trading Baseline}
    \label{fig:GPT_Prompt}
\end{figure}
\begin{figure}[H]
\centering
\begin{subfigure}{0.45\linewidth}
\centering
\begin{subfigure}{\linewidth}
\centering
\begin{lstlisting}
(*@\textcolor{system}{System:}@*)
    (*@\textcolor{system}{GPT Utility Function Q: [[-1,-0,-0],[-0,-1,-0],[-0,-0,-1]]}@*)
    (*@\textcolor{system}{GPT Utility Function u: [50, 100, 150]}@*)
(*@\textcolor{computer}{GPT's Trade Offer:}@*) 
    (*@\textcolor{computer}{GPT receives: 10 apples, 5 bananas}@*)
    (*@\textcolor{computer}{User receives: 6 oranges}@*)
(*@\textcolor{user}{User: No, I can give you 5 oranges if you give me 5 apples}@*)
(*@\textcolor{computer}{GPT's Trade Offer:}@*) 
    (*@\textcolor{computer}{GPT receives: 5 oranges}@*)
    (*@\textcolor{computer}{User receives: 5 apples}@*)
(*@\textcolor{user}{User: Yes}@*)
(*@\textcolor{system}{System:}@*)
    (*@\textcolor{system}{Offer Accepted!}@*) 
    (*@\textcolor{system}{New User State: 55 apples, 50 bananas, 45 oranges,}@*) 
    (*@\textcolor{system}{New GPT State: 45 apples, 50 bananas, 55 oranges,}@*) 
    (*@\textcolor{system}{Estimated User Benefit: 450}@*) 
    (*@\textcolor{system}{GPT's Benefit: 450}@*)
\end{lstlisting}
\caption{Example transcript of GPT making a Successful Trade. The user has a target state of 75 apples, 50 bananas, and 25 oranges.}
\label{fig:chat-transcript-success-pure}
\end{subfigure}

\vspace{10 pt}

\begin{subfigure}{\linewidth}
\centering
\begin{lstlisting}
(*@\textcolor{system}{Start of Trading}@*)
    (*@\textcolor{system}{GPT Utility Function Q: [[-1,-0,-0],[-0,-1,-0],[-0,-0,-1]]}@*)
    (*@\textcolor{system}{GPT Utility Function u: [ 50 100 150]}@*)
(*@\textcolor{computer}{GPT's Trade Offer: }@*) 
    (*@\textcolor{computer}{GPT receives: 5 apples, 3 oranges}@*)
    (*@\textcolor{computer}{User receives: 3 bananas}@*)
(*@\textcolor{user}{User: No, how about you give me 5 oranges and I give you 5 apples?}@*)
(*@\textcolor{computer}{GPT's Trade Offer: }@*) 
    (*@\textcolor{computer}{GPT receives: 5 apples}@*)  
    (*@\textcolor{computer}{User receives: 5 oranges}@*)
(*@\textcolor{user}{User: Yes}@*)
(*@\textcolor{system}{System:}@*)
    (*@\textcolor{system}{Offer Accepted!}@*) 
    (*@\textcolor{system}{New User State: 45 apples, 50 bananas, 55 oranges}@*)
    (*@\textcolor{system}{New GPT State: 55 apples, 50 bananas, 45 oranges}@*) 
    (*@\textcolor{system}{Estimated User Benefit: 450}@*) 
    (*@\textcolor{system}{GPT's Benefit: -550}@*)
\end{lstlisting}
\caption{Example Transcript of GPT making a Non-Beneficial Trade. The user has a target state of 25 apples, 50 bananas, and 75 oranges}
\label{fig:chat-transcript-fail-1}
\end{subfigure}
\end{subfigure}\hfill
\begin{subfigure}{0.45\linewidth}
\centering
\begin{lstlisting}
(*@\textcolor{system}{System:}@*)
    (*@\textcolor{system}{GPT Utility Function Q: [[-1,-0,-0],[-0,-1,-0],[-0,-0,-1]]}@*)
    (*@\textcolor{system}{GPT Utility Function u: [50, 100, 150]}@*)
(*@\textcolor{computer}{GPT's Trade Offer:}@*)
    (*@\textcolor{computer}{GPT receives: 5 apples, 5 oranges}@*)
    (*@\textcolor{computer}{User receives: 3 bananas}@*)
(*@\textcolor{user}{User: No thanks}@*)
(*@\textcolor{computer}{GPT's Trade Offer:}@*)
    (*@\textcolor{computer}{GPT receives: 4 oranges}@*)
    (*@\textcolor{computer}{User receives: 4 bananas, 1 apple}@*)
(*@\textcolor{user}{User: Nah}@*)
(*@\textcolor{computer}{GPT's Trade Offer:}@*)
    (*@\textcolor{computer}{GPT receives: 3 oranges}@*)
    (*@\textcolor{computer}{User receives: 3 bananas, 2 apples}@*)
(*@\textcolor{user}{User: No}@*)
(*@\textcolor{computer}{GPT's Trade Offer:}@*)
    (*@\textcolor{computer}{GPT receives: 3 oranges}@*)
    (*@\textcolor{computer}{User receives: 2 bananas, 3 apples}@*)
(*@\textcolor{user}{User: No}@*)
(*@\textcolor{computer}{GPT's Trade Offer:}@*)
    (*@\textcolor{computer}{GPT receives: 2 oranges}@*)
    (*@\textcolor{computer}{User receives: 4 bananas, 3 apples}@*)
(*@\textcolor{user}{User: No}@*)
(*@\textcolor{computer}{GPT's Trade Offer:}@*)
    (*@\textcolor{computer}{GPT receives: 2 oranges}@*)
    (*@\textcolor{computer}{User receives: 5 bananas, 1 apple}@*)
(*@\textcolor{user}{User: Stopped Trading}@*) 
\end{lstlisting}
\caption{Example Transcript of GPT repeating rejected trade offers. The user has a target state of 55 apples, 50 bananas, and 55 oranges}
\label{fig:chat-transcript-fail-2}
\end{subfigure}
\caption{Example transcripts of GPT negotiating with a human agent without an underlying algorithm. In these examples, GPT is the offering agent, and the human user is the responding agent.}
\end{figure}
\newpage

\subsection{User Study Details}

\begin{figure}
    \centering
    \includegraphics[width=0.75\linewidth]{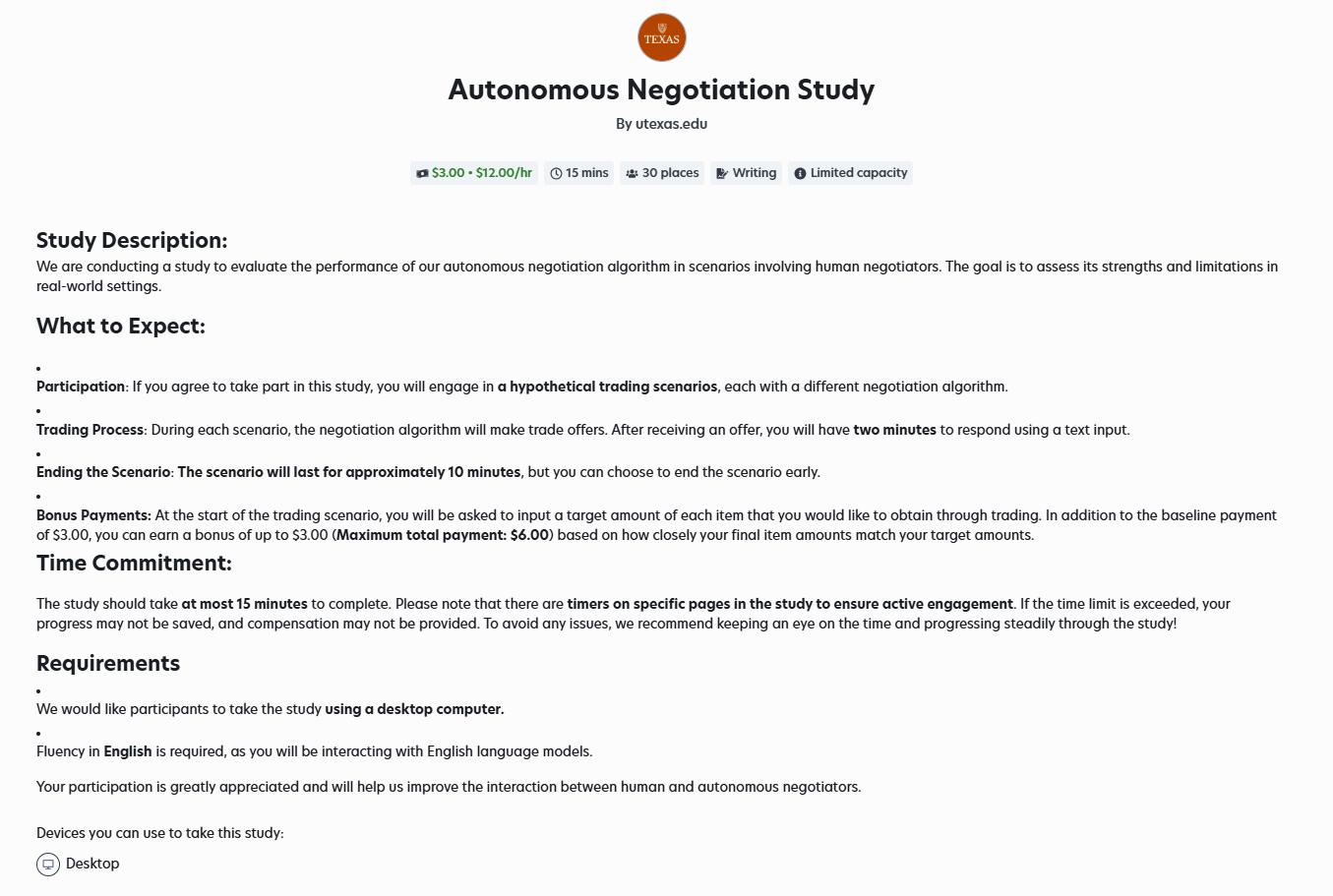}
    \caption{User Study Recruitment Interface on Prolific}
    \label{fig:prolificInstructions}
\end{figure}

Participants in the user studies were recruited using Prolific. All participants provided informed consent before participation and were compensated at a rate consistent with Prolific guidelines and institutional policy. \cref{fig:prolificInstructions} shows the recruitment interface. The full website that the users interacted with can be found at \url{https://github.com/suryakmurthy/Seq_Trading_Cmpr}.

The users were given bonus compensation of up to \$3 based on how closely they reached their desired resource allocation, incentivizing them to move toward their target vector \( \mathbf{b_B} \). To quantify this, each user was assigned a score based on the normalized improvement from their initial allocation \( \mathbf{S_B^0} \) to their final allocation \( \mathbf{S_B^T} \), defined as
\(
3 \left( 1 - \frac{\|\mathbf{b_B} - \mathbf{S_B^T}\|_1}{\|\mathbf{b_B} - \mathbf{S_B^0}\|_1} \right).
\)
A score of 1 indicates a perfect match with the user’s target allocation, while a score of 0 implies no improvement over the initial allocation. If the final allocation was further from the target than the initial state (i.e., a negative score), the user was treated as having a score of 0 for compensation purposes.

\subsection{User Study Additional Results}

\begin{table}[h!]
\centering
\scriptsize
\resizebox{0.5\columnwidth}{!}{%
\begin{tabular}{lccc}
\hline
\textbf{Algorithm} & \textbf{0--60°} & \textbf{60--120°} & \textbf{120--180°} \\
\hline
\multicolumn{4}{l}{\textit{Computer Benefit Values}} \\
GCA            & $614.86 \pm 332.15$  & $650.93 \pm 254.53$  & $588.29 \pm 360.19$ \\
Random Trades  & $377.17 \pm 260.98$  & $536.58 \pm 278.58$  & $608.64 \pm 215.06$ \\
ST-CR          & $320.22 \pm 261.58$  & $504.12 \pm 223.35$  & $518.47 \pm 286.74$ \\
\hline
\end{tabular}
}
\caption{Computer benefit values (mean $\pm$ standard deviation) by angular range for each algorithm.}
\label{tab:computer_benefit}
\end{table}

\begin{table}[h!]
\centering
\scriptsize
\resizebox{0.5\columnwidth}{!}{%
\begin{tabular}{lccc}
\hline
\textbf{Algorithm} & \textbf{0--60°} & \textbf{60--120°} & \textbf{120--180°} \\
\hline
\multicolumn{4}{l}{\textit{Human Benefit Values}} \\
GCA            & $-1515.43 \pm 828.26$ & $-107.62 \pm 586.21$ & $729.14 \pm 665.24$ \\
Random Trades  & $-868.17 \pm 776.10$  & $-209.92 \pm 847.76$ & $1025.00 \pm 608.80$ \\
ST-CR          & $-774.44 \pm 633.19$  & $64.28 \pm 541.32$   & $849.93 \pm 633.67$ \\
\hline
\end{tabular}
}
\caption{Estimated human benefit values (mean $\pm$ standard deviation) by angular range for each algorithm.}
\label{tab:human_benefit}
\end{table}

In \cref{tab:human_benefit} and \cref{tab:computer_benefit}, we consider cumulative benefits for the computer and human agents respectively. Across all alignment conditions, GCA consistently yields the highest benefit for the computer agent, while ST-CR provides the most favorable outcomes for human participants, especially by mitigating losses in low-alignment scenarios and achieving positive gains in medium alignment. In terms of societal benefit (the combined utility of both agents), ST-CR performs best overall in low and medium alignment settings, whereas Random Trades slightly outperforms others in high-alignment scenarios. These results highlight ST-CR’s strength in balancing trade-offs, particularly when agent preferences are misaligned.

\end{document}